\documentclass[twocolumn]{aastex63}

\begin{document}

\newcommand{\kms}{km~s$^{-1}$} \newcommand{\msun}{$M_{\odot}$} 
\newcommand{\rsun}{$R_{\odot}$} \newcommand{\teff}{$T_{\rm eff}$} 
\newcommand{\logg}{$\log{g}$} \newcommand{\mas}{mas~yr$^{-1}$}

\title{The 100 pc White Dwarf Sample in the SDSS Footprint}

\author{Mukremin Kilic} \affiliation{Homer L. Dodge Department of Physics and 
Astronomy, University of Oklahoma, 440 W. Brooks St., Norman, OK, 73019 USA}

\author{P. Bergeron} \affiliation{D\'epartement de Physique, Universit\'e de Montr\'eal,
C.P. 6128, Succ. Centre-Ville, Montr\'eal, QC H3C 3J7, Canada}

\author{Alekzander Kosakowski} \affiliation{Homer L. Dodge Department of Physics and 
Astronomy, University of Oklahoma, 440 W. Brooks St., Norman, OK, 73019 USA}

\author{Warren R.\ Brown} \affiliation{Smithsonian Astrophysical Observatory, 60
Garden Street, Cambridge, MA 02138 USA}

\author{Marcel A. Ag{\"u}eros} \affiliation{Department of Astronomy, Columbia University,
550 West 120th Street, New York, NY 10027, USA}

\author{Simon Blouin} \affiliation{Los Alamos National Laboratory, P.O. Box 1663,
Mail Stop P365, Los Alamos, NM 87545, USA}

\email{kilic@ou.edu}

\shorttitle{The 100 pc SDSS WD sample} 
\shortauthors{Kilic et al.}

\begin{abstract}

We present follow-up spectroscopy of 711 white dwarfs within 100 pc, and present
a detailed model atmosphere analysis of the 100 pc white dwarf sample in the
SDSS footprint. Our spectroscopic follow-up is complete for 83\% of the white
dwarfs hotter than 6000~K, where the atmospheric composition can be constrained
reliably.  We identify 1508 DA white dwarfs with pure hydrogen atmospheres. The
DA mass distribution has an extremely narrow peak at $0.59~M_{\odot}$, and
reveals a shoulder from relatively massive white dwarfs with
$M=0.7$-$0.9~M_{\odot}$. Comparing this distribution with binary population
synthesis models, we find that the contribution from single stars that form
through mergers cannot explain the over-abundance of massive white dwarfs. In
addition, the mass distribution of cool DAs shows a near absence of
$M>1~M_{\odot}$ white dwarfs. The pile-up of 0.7-$0.9~M_{\odot}$ and the
disappearance of $M>1~M_{\odot}$ white dwarfs is consistent with the effects of
core crystallization. Even though the evolutionary models predict the location
of the pile-up correctly, the delay from the latent heat of crystallization by
itself is insufficient to create a significant pile-up, and additional cooling
delays from related effects like phase separation are necessary. We also discuss
the population of infrared-faint (ultracool) white dwarfs, and demonstrate for
the first time the existence of a well defined sequence in color and magnitude.
Curiously, this sequence is connected to a region in the color-magnitude
diagrams where the number of helium-dominated atmosphere white dwarfs is low.
This suggests that the infrared-faint white dwarfs likely have mixed H/He
atmospheres.

\end{abstract}


\section{Introduction}

It took astronomers more than half a century to identify the first three white dwarfs and understand their nature \citep{holberg09}.
Starting with the detection of the companion of Sirius in 1862, only two other white dwarfs, 40 Eridani B and Van Maanen 2 were
identified by the 1920s. Interestingly, these white dwarfs reveal a relatively broad mass range, with $M = 0.97 \pm 0.01, 0.51 \pm 0.04$, and $0.68 \pm 0.02~M_{\odot}$, respectively \citep{holberg16}. 

Thanks to large scale proper motion \citep[e.g.][]{luyten76} and
imaging surveys \citep[e.g.,][]{green86}, the number of confirmed white dwarfs reached $\sim2\times10^3$ in the late nineties
\citep{mccook87,mccook99}. The next decade saw a ten-fold increase, thanks to the Sloan Digital Sky Survey spectroscopy
\citep{eisenstein06,kleinman13, kepler19}, and this decade saw another ten-fold increase thanks to the Gaia Data Release 2 
astrometry \citep{gaia18,gentile19}. Before Gaia, the only volume-limited and nearly-complete white dwarf sample extended only up
to 20 pc, and included 139 systems \citep[see][and references therein]{holberg16,hollands18}.

Volume-limited surveys provide unbiased estimates of the luminosity and mass functions of white dwarfs, which map
the star formation history, the initial-final mass relation, the effects of close binary evolution,
and the changes in the white dwarf cooling rate as a function of temperature and age.  Studying the mass distribution of
298 DA white dwarfs with $T_{\rm eff}\geq 13,000$ K in the magnitude limited Palomar-Green Survey \citep{green86},
\citet{liebert05} found a primary peak at $0.57~M_{\odot}$ along with a smaller peak from low-mass white
dwarfs at $0.4~M_{\odot}$, and a broad high-mass shoulder centered at $0.78~M_{\odot}$. 

Magnitude-limited surveys are biased against the intrinsically fainter and more massive white dwarfs. Correcting
for this bias, \citet{liebert05} found that the broad high-mass component accounts for the 22\% of the hot DA white
dwarfs in their sample, and they suggested binary mergers as a potential explanation for their numbers.
White dwarf mass distributions obtained from the SDSS spectroscopy also show similar peaks \citep[for example see
Fig. 10 in][]{kleinman13}, but the contribution from massive white dwarfs is underestimated given the SDSS spectroscopic
survey selection. 

\citet{kilic18} attempted to constrain the mass distribution using the 100 pc volume-limited white dwarf sample from Gaia.
Based on pure hydrogen atmosphere model fits, they found a bimodal mass distribution with peaks at 0.6 and $0.8~M_{\odot}$,
and estimated a contribution of $\sim11$\% from massive white dwarfs. They attributed these to the remnants of
main-sequence and post-main-sequence mergers. \citet{elbadry18} also found an excess of massive white dwarfs
in the SDSS DA white dwarf sample, and attributed this excess to a flattened initial-final mass relation that favors the
formation of massive white dwarfs. On the other hand, \citet{hollands18} did not find any evidence of an overabundance
of massive white dwarfs among the 139 systems in the 20 pc Gaia white dwarf sample.

To constrain the frequency of massive white dwarfs in the solar neighborhood, we initiated a spectroscopic follow-up
survey of the 100 pc Gaia white dwarf sample within the SDSS footprint. Here we present the results from this spectroscopic
survey. Section 2 describes our sample selection and follow-up observations, whereas section 3 presents our model
atmosphere analysis highlighting the model fits for different spectral types.  We discuss the properties of our 100 pc white
dwarf sample, including the DA white dwarf mass distribution and the infrared-faint (IR-faint) white dwarfs in section 4. We conclude in section 5.

\section{Survey Definition}

\subsection{Sample Selection}
\label{sample}

\begin{deluxetable*}{cccrcrrccc} 
\tablecolumns{10} \tablewidth{0pt} 
%\tabletypesize{\scriptsize}
\tablecaption{The 100 pc White Dwarf Sample in the SDSS Footprint.\label{tabdata}} 
\tablehead{\colhead{Object} & \colhead{Gaia Source ID} & \colhead{RA} & \colhead{DEC} &
\colhead{$\varpi$} & \colhead{$\mu_{\rm RA}$} & \colhead{$\mu_{\rm DEC}$} & \colhead{$G$}
& \colhead{$G_{\rm BP}$} & \colhead{$G_{\rm RP}$}\\
 &  & ($^{\circ}$) & ($^{\circ}$) & (mas) & (mas yr$^{-1}$) &  (mas yr$^{-1}$) & (mag) & (mag) & (mag)}
\startdata  
J0000$-$0403 & 2447815253423324544 & 0.04710 & $-$4.05414 & 22.31 $\pm$ 0.22 & $-$117.5 $\pm$ 0.4& 21.6 $\pm$ 0.2 & 17.80 & 18.15 & 17.25 \\
J0000+0132 & 2738626591386423424 & 0.17875 & 1.53930 & 14.95 $\pm$ 0.10 & 65.6 $\pm$ 0.2& $-$28.4 $\pm$ 0.1 & 16.26 & 16.29 & 16.16 \\
J0001+3237 & 2874216647336589568 & 0.26821 & 32.61768 & 10.15 $\pm$ 0.30 & $-$11.3 $\pm$ 0.4& $-$76.9 $\pm$ 0.3 & 19.20 & 19.53 & 18.71 \\
J0001$-$1111 & 2422442334689173376 & 0.34738 & $-$11.19932 & 13.51 $\pm$ 0.29 & 90.3 $\pm$ 0.4& $-$114.6 $\pm$ 0.3 & 18.30 & 18.47 & 17.90 \\
J0001+3559 & 2877080497170502144 & 0.48816 & 35.99629 & 11.66 $\pm$ 0.28 & 38.5 $\pm$ 0.3& $-$57.9 $\pm$ 0.2 & 18.85 & 19.02 & 18.35 \\
J0003$-$0111 & 2449594087142467712 & 0.81988 & $-$1.18834 & 14.07 $\pm$ 0.51 & 99.8 $\pm$ 0.6& $-$12.6 $\pm$ 0.4 & 18.81 & 19.21 & 18.20 \\
J0003$-$0426 & 2444731325169827200 & 0.88831 & $-$4.44813 & 11.62 $\pm$ 0.31 & $-$100.3 $\pm$ 0.6& $-$64.2 $\pm$ 0.4 & 18.57 & 18.78 & 18.16 \\
J0004$-$0340 & 2447889401738675072 & 1.04394 & $-$3.66918 & 21.08 $\pm$ 0.10 & 218.7 $\pm$ 0.2& $-$52.9 $\pm$ 0.1 & 16.74 & 16.90 & 16.40 \\
J0004+0838 & 2746843589674667264 & 1.06322 & 8.64444 & 14.22 $\pm$ 0.24 & 67.6 $\pm$ 0.6& $-$52.8 $\pm$ 0.3 & 18.43 & 18.77 & 17.95 \\
J0004+1430 & 2768919442402016896 & 1.12819 & 14.50006 & 11.63 $\pm$ 0.49 & 33.4 $\pm$ 1.1& 98.5 $\pm$ 0.8 & 19.37 & 19.78 & 18.68                 
\enddata
\tablecomments{The object names, RA, and DEC are based on the Gaia DR2 epoch of 2015.5.
This table is available in its entirety in machine-readable format in the online journal. A portion is shown here
for guidance regarding its form and content.}
\end{deluxetable*}

\begin{figure}
\includegraphics[width=3.2in, bb=18 144 592 718]{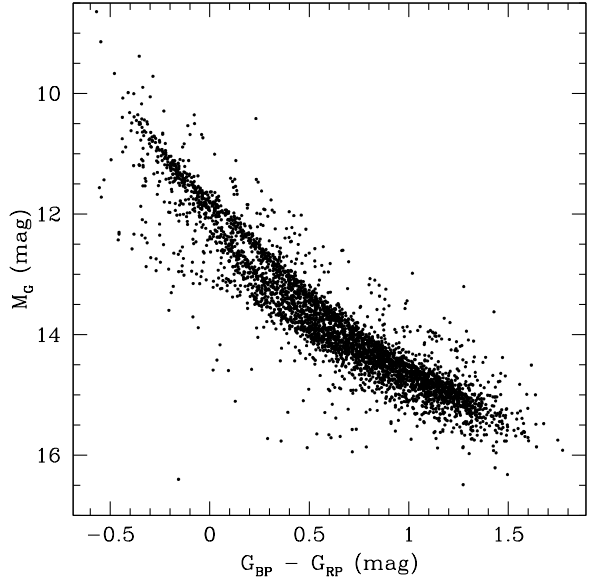} 
\caption{A color-magnitude diagram showing the 100 pc Gaia white dwarf sample within the SDSS footprint.
This sample contains 4016 white dwarfs with $G$ magnitudes ranging from 12.1 to 20.6. The bifurcation in the
white dwarf sequence due to the atmospheric composition \citep{bergeron19} is clearly visible.}
\label{fig1}
\end{figure}

The 100 pc Gaia sample includes more than $10^4$ white dwarfs \citep{gaia18,kilic18,jimenez18,gentile19}, and it is a challenge to obtain
spectroscopy of the entire sample. To make our survey feasible, we limited our follow-up to the SDSS imaging footprint so that we
can take advantage of the prior spectroscopy from the SDSS and other surveys that focused on white dwarfs
\citep[e.g.,][]{bergeron97,bergeron01,mccook99,kilic06a,kilic10,gianninas11,limoges15}. 

To obtain a clean 100 pc white dwarf sample within the SDSS footprint, we used the SDSS Data Release 9 (DR9)
catalogue matched with Gaia DR2. The SDSS DR9 footprint covers 14,555 square degrees of the northern hemisphere sky,
primarily at high $|b|\gtrsim30\arcdeg$ Galactic latitudes, as detailed in \citet{sdssdr9}. We used the
recommendations outlined in \cite{lindegren18} to remove non-Gaussian outliers
in color and absolute magnitude, and limited our sample to objects with $>10\sigma$ significant parallax, $G_{\rm BP}$,
and  $G_{\rm RP}$ photometry. We perform a simple cut in $(G_{\rm BP} - G_{\rm RP}, M_{\rm G})$ space to select our
sample of white dwarfs. Our selection is optimized for reliability rather than completeness; it keeps isolated white dwarfs and
unresolved double degenerates, but removes the majority of the white dwarf + main-sequence binaries.

For completeness, we provide the ADL query that we used to define our sample in the appendix. This query returns 4016
white dwarfs within 100 pc and the SDSS footprint. Table \ref{tabdata} presents the Gaia source ID, coordinates, parallax,
proper motions, and magnitude for each object. 

Figure \ref{fig1} shows these sources in a Gaia color-magnitude diagram. 
The white dwarf sequence extends from $M_G = 8.6$ to 16.5 mag, with the apparent $G$ magnitudes ranging from 12.1 to 20.6. 
The split in the white dwarf sequence due to the atmospheric composition is clearly visible \citep{bergeron19}. 

We cross-matched this sample with the SDSS DR14 spectroscopic catalog \citep{kepler19}, and found 1045 white dwarfs with
SDSS spectra.  We found spectroscopy of 119 additional white dwarfs in \citet{kilic06a,kilic10}, 345 objects in the papers
by the Montr\'eal, group \citep[e.g.,][]{limoges15}, and 26 objects with data from the Supernovae Type Ia Progenitor
Survey survey \citep{napiwotzki01}. We found spectral types for 117 additional  objects in the literature \citep{dufour17}. 

\subsection{Spectroscopic Survey}

Spectral typing of white dwarfs based on low-resolution spectroscopy is relatively easy for stars hotter than about
6000 K. This is because hydrogen lines are clearly visible at these temperatures. Hence, the absence or presence of
H$\alpha$ and/or H$\beta$ is usually sufficient to constrain the atmospheric composition. Helium lines are visible for white
dwarfs hotter than 11,000 K, and additional lines from C$_2$, Ca, and Mg are also visible in DQ and DZ white dwarfs, respectively.
Hydrogen lines disappear below about 5000 K, and the majority of the white dwarfs below this temperature show featureless
spectra; DC spectral type. Hence, optical spectroscopy is not useful for constraining the atmospheric composition of
cooler white dwarfs, unless they are metal-rich. Therefore, we limit our spectroscopic follow-up survey to white dwarfs with
$T_{\rm eff}\geq6000$ K, where our data can reliably constrain the atmospheric composition.

We obtained optical spectroscopy of 711 white dwarfs using the 1.5m Fred Lawrence Whipple Observatory (FLWO)
telescope, the MDM Hiltner 2.4m telescope,  the Apache Point Observatory (APO) 3.5m telescope, the 6.5m Multiple
Mirror Telescope (MMT), and the 8m Gemini South telescope. Table \ref{tab2} presents the details of our observing program,
including the instrument configuration and the number of targets observed at each telescope.

\begin{deluxetable*}{lclcccr} 
\tablecolumns{6} \tablewidth{0pt} 
\tablecaption{Observational Details\label{tab2}} 
\tablehead{\colhead{Telescope} & \colhead{Instrument} & \colhead{Grating} & \colhead{Slit} & \colhead{Resolution} & \colhead{$\lambda$} & \colhead{Targets}\\
 &  & & ($\arcsec$) & (\AA) & (\AA) & }
\startdata 
FLWO 1.5m & FAST     & 300 l mm$^{-1}$ & 1.5 & 3.6 & 3500 - 7400 & 23 \\
                    &               & 600 l mm$^{-1}$ & 1.5 & 2.3  & 3550 - 5530  & 158 \\
MDM 2.4m  & OSMOS & Blue VPH           &  1.2 & 3.3  & 3975 - 6865  & 158 \\
APO 3.5m   & DIS        & R300                  &  1.5 & 6.4  & 5500 - 9200  & 107 \\ 
                    &               & B400                  &  1.5 & 5.2  & 3400 - 5500  & 7 \\ 
6.5m MMT    & Blue Channel & 500 l mm$^{-1}$ & 1-1.25 & 3.8-4.8  & 3700 - 6850  & 233 \\  
                    &                         & 300 l mm$^{-1}$ & 1-1.25 & 6.5-8.1  & 3200 - 7620  & 51 \\ 
Gemini South & GMOS          & B600        &  1.0 & 5.5  & 3670 - 6810 & 21                            
\enddata
\tablecomments{DIS B400 + R300 observations were obtained simultaneously.}
\end{deluxetable*}
 
At the FLWO 1.5m telescope, we used the FAST spectrograph \citep{fabricant98} with either the 600 l mm$^{-1}$ or the 300
l mm$^{-1}$ grating and the $1.5\arcsec$ slit to obtain spectroscopy of 81 targets with spectral resolutions of  2.3 and 3.6 \AA\
over the wavelength ranges 3550 - 5530 \AA\ and 3500-7400 \AA, respectively. These observations were done between 2019 January and 2020 February. We found spectroscopy data for 100 additional targets in the FAST archive \citep{tokarz97}. 

At the MDM 2.4m telescope, we used the OSMOS spectrograph \citep{martini11} with the Blue VPH grism and the $1.2\arcsec$
inner slit to obtain spectroscopy of 158 targets with a spectral resolution of 3.3 \AA\ over the wavelength range 3975-6865 \AA.
These observations were done between 2018 December and 2019 November as part of the MDM OSMOS queue. 

At the APO 3.5m telescope, we used the Dual Imaging Spectrograph (DIS) with the B400 + R300 gratings and a $1.5\arcsec$
slit to obtain spectroscopy of 107 targets with spectral resolutions of 5.2 and 6.4 \AA\ in the blue and red channels, respectively.
The blue and red channels cover the wavelength ranges 3400 - 5500 and 5500 - 9200 \AA, respectively. These observations
were done between 2018 November and 2019 September. Unfortunately, the DIS blue channel suffered from contamination
problems soon after our observing program started, and only seven of our targets have DIS blue channel data available. 

At the 6.5-m MMT, we used the Blue Channel Spectrograph \citep{schmidt89} with the 500 l mm$^{-1}$ grating and a 1
or $1.25\arcsec$ slit to obtain spectroscopy of 233 targets over the wavelength range 3700 - 6850 \AA\ with spectral
resolutions of 3.8 and 4.8 \AA, respectively. We obtained additional observations of 51 stars using the 300 l mm$^{-1}$
grating, which provided spectra from the atmospheric cutoff (3200\AA) to 7620 \AA\ with resolutions of 6.5 and 8.1 \AA\
for the 1 and $1.25\arcsec$ slits, respectively. These observations were done between 2018 November and 2019 October.

At the 8-m Gemini South telescope, we used the Gemini Multi-Object Spectrograph (GMOS) with the B600 grating and
a $1\arcsec$ slit to obtain spectroscopy of 21 targets as part of the queue program GS-2020A-Q-222. Unfortunately,
our program was cut short due to the COVID-19 pandemic, which prohibited observations of additional targets.

Our Gemini observations were done between UT 2020 Jan 21 and March 7. We binned the CCD by $4\times4$ to reduce
the read-out time. This setup provided spectra over the wavelength range 3670 - 6810 \AA, with a resolution of 2 \AA\ per
pixel. GMOS on Gemini has 3 CCDs, with gaps between them. Our observing setup leads to
gaps in spectral coverage between 4677-4703 and 5737-5765 \AA. These gaps can be eliminated by obtaining multiple
exposures with different central wavelength configurations of the grating. Because these gaps have no impact on our spectral
classification, and in order to save telescope time, we did not dither in wavelength to eliminate these gaps. 

We reduced the data from all five telescopes using standard {\sc IRAF} routines. About two dozen targets were observed
on more than one telescope and several spectra with low signal-to-noise ratios were also omitted, resulting in a final sample
of 711 objects with new spectroscopy data. Of these,
477 are DA, 6 are DB, 195 are DC, 7 are DQ, 27 are DZ, and 1 is a dwarf nova. Two targets are likely magnetic white dwarfs with
uncertain spectral types (DAH or DBH), and are included in both the DA and DB samples. 

The use of many different telescope and instrument combinations with different resolving powers can lead
to systematic uncertainties in the physical parameters derived using the spectroscopic method \citep{liebert05}.
Here we do not use the spectroscopic method. We use the photometric method instead (see below), and we rely
only on Gaia parallaxes, and SDSS and Pan-STARRS photometry. Hence, the multiple telescope and instrument
combinations have no impact on the physical parameters derived in our analysis. 
We simply use each spectrum to confirm the spectral type and to constrain the
atmospheric composition of each white dwarf.

\begin{figure}
\includegraphics[width=3.2in, bb=18 144 592 718]{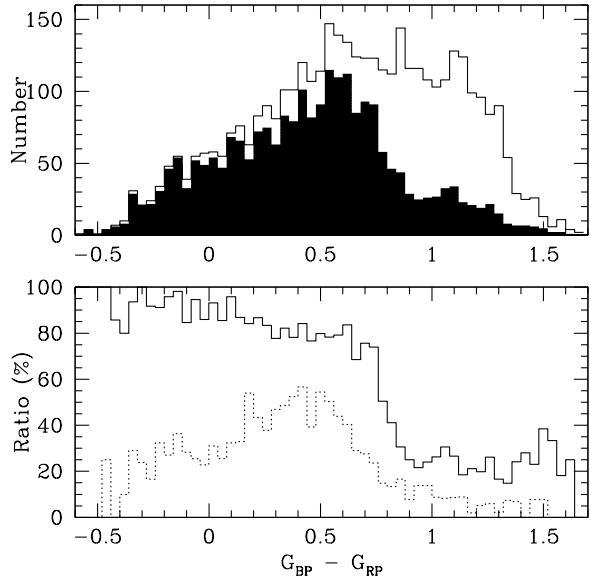} 
\caption{The color distribution of our 100 pc white dwarf sample in the SDSS footprint (solid line, top panel), along with the number of objects
with follow-up spectroscopy (shaded histogram). The ratio of the two, i.e. the completeness level of the spectroscopic follow-up,
is shown as the solid line in the bottom panel. The dotted line shows the completeness of the SDSS spectroscopy only.}
\label{fig2}
\end{figure}

The top panel in Figure \ref{fig2}
shows the color distribution of the entire 100 pc white dwarf sample in the SDSS footprint (solid line) and that of the objects with spectroscopy (shaded histogram). The bottom panel shows the completeness of the spectroscopic follow-up. 

Combining our data with spectroscopy available in the literature, we have spectral classifications for 2361 (59\%) of the 4016
white dwarfs in our sample. Given our survey selection bias for $T_{\rm eff}\geq6000$ K objects,
our spectroscopic follow-up is significantly more complete for hotter stars. A
$\log{g}=8$ pure hydrogen atmosphere white dwarf is predicted to have $G_{\rm BP}-G_{\rm RP}= 0.734$ mag at 6000 K.
There are 2213 white dwarfs bluer than that color in our sample, including 1838 with spectra. Hence, our spectroscopic follow-up
is 83\% complete for white dwarfs hotter than about 6000 K.

\begin{figure*}
\center
\includegraphics[width=3.5in, bb=20 17 592 779]{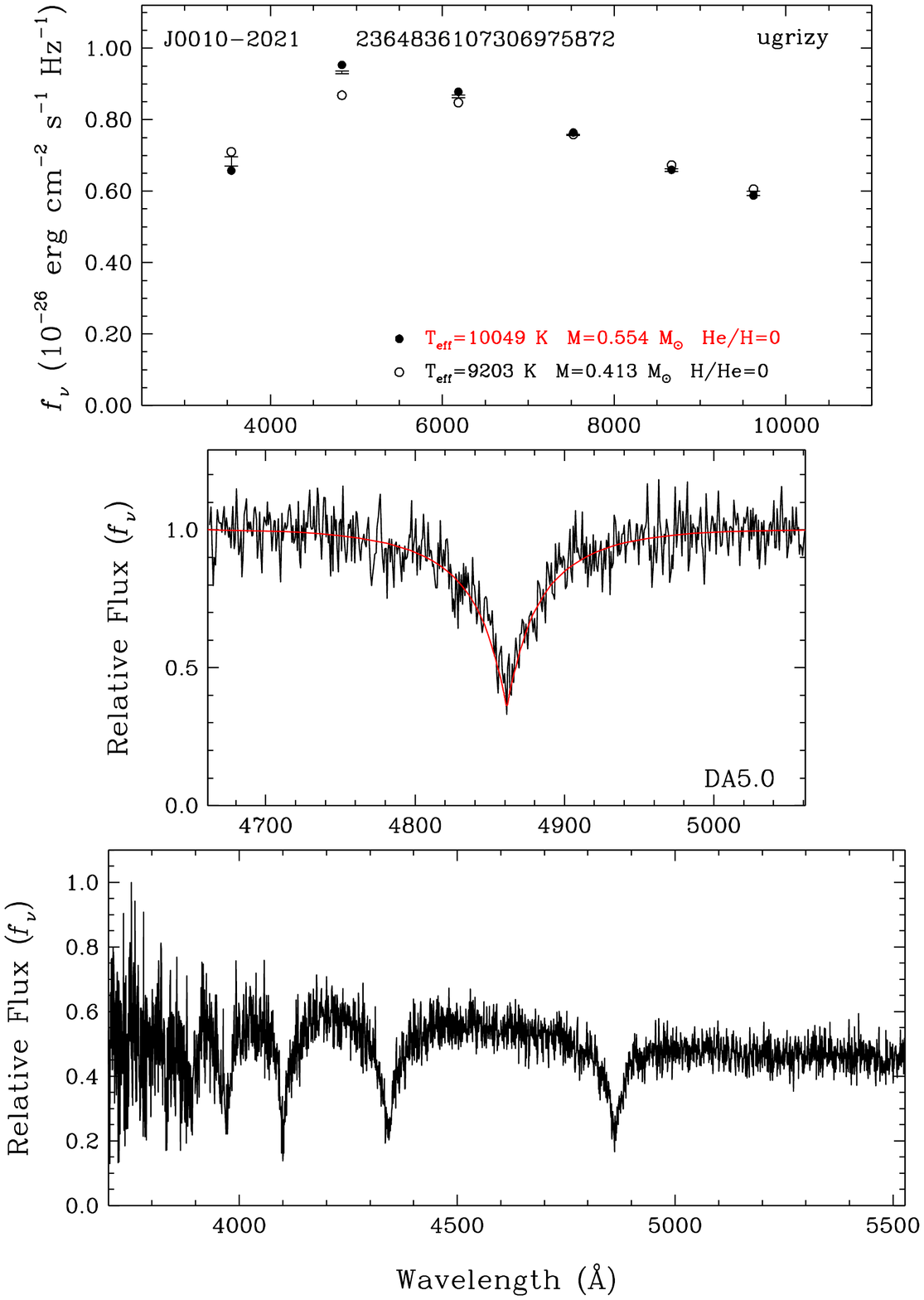}
\includegraphics[width=3.5in, bb=20 17 592 779]{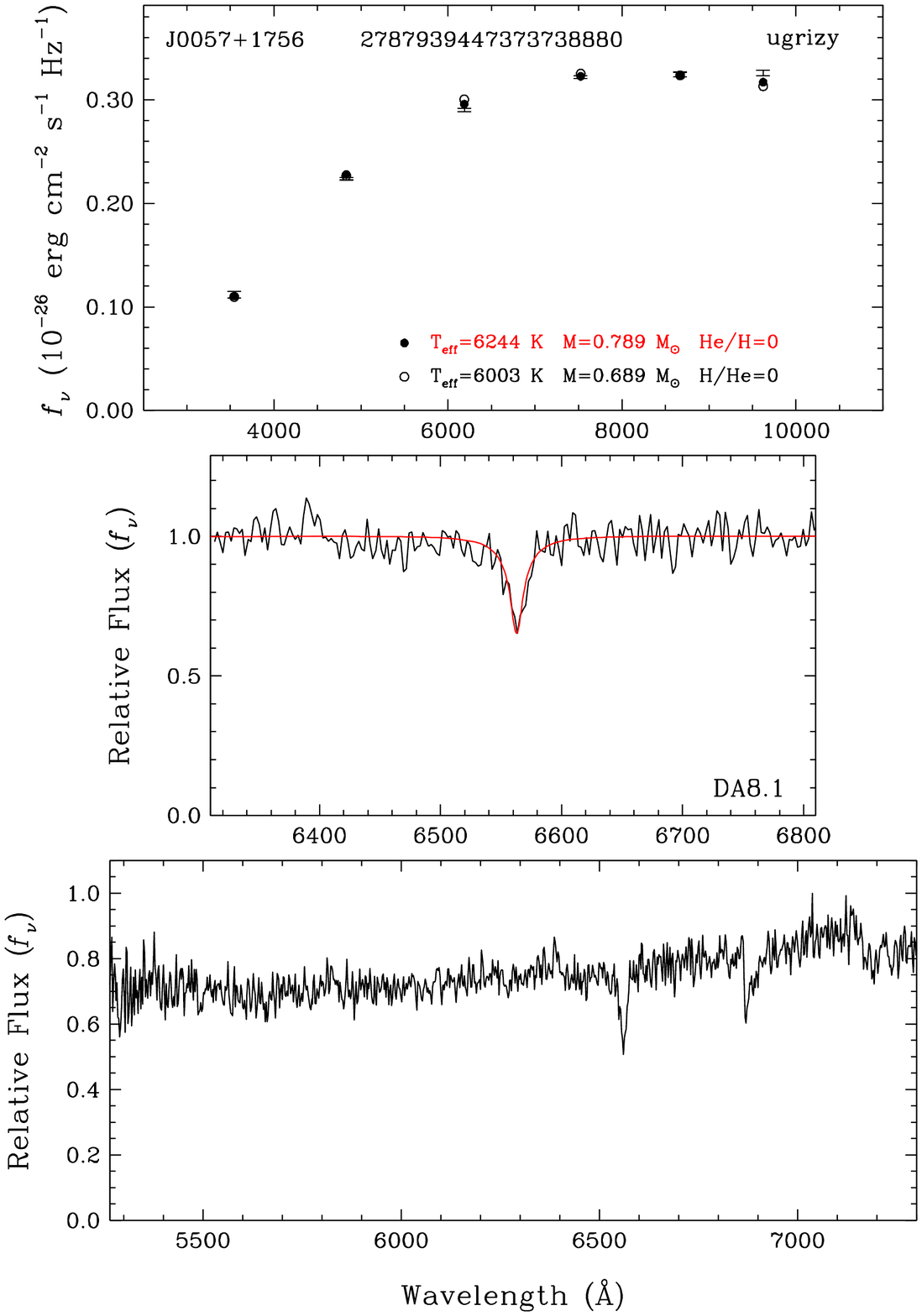}
\caption{Model fits to two new DA white dwarfs observed at the FLWO 1.5m and APO 3.5m telescopes. The top panels show
the best-fitting pure hydrogen (filled dots) and pure helium (open circles) atmosphere white dwarf models to the photometry
(error bars). Each object is labeled based on its Gaia Source ID, object name based on Gaia DR2 coordinates, and the
photometry used in the fitting: $ugrizy$ means SDSS $u$ + Pan-STARRS $grizy$ while $ugriz$ means SDSS $ugriz$.
The atmospheric parameters of the favored solution is highlighted in red. The middle panels, here and in the following figures,
always show the predicted spectrum (red lines) based on the pure hydrogen solution. The bottom panels show a broader
wavelength range for each instrument, revealing additional Balmer lines in the blue for one of these targets.}
\label{figda}
\end{figure*}

\begin{figure*}
\center
\includegraphics[width=3.5in, bb=20 17 592 779]{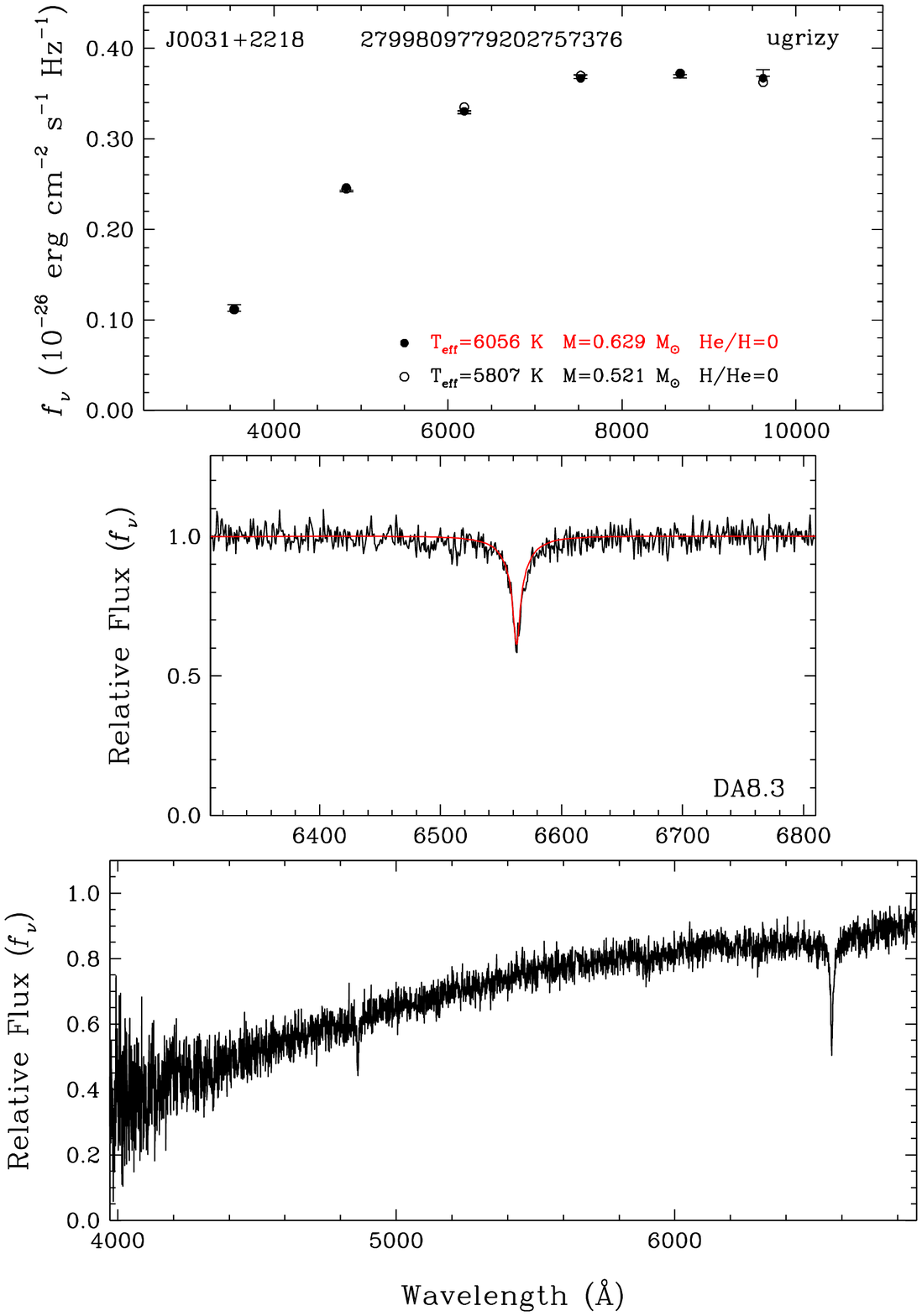}
\includegraphics[width=3.5in, bb=20 17 592 779]{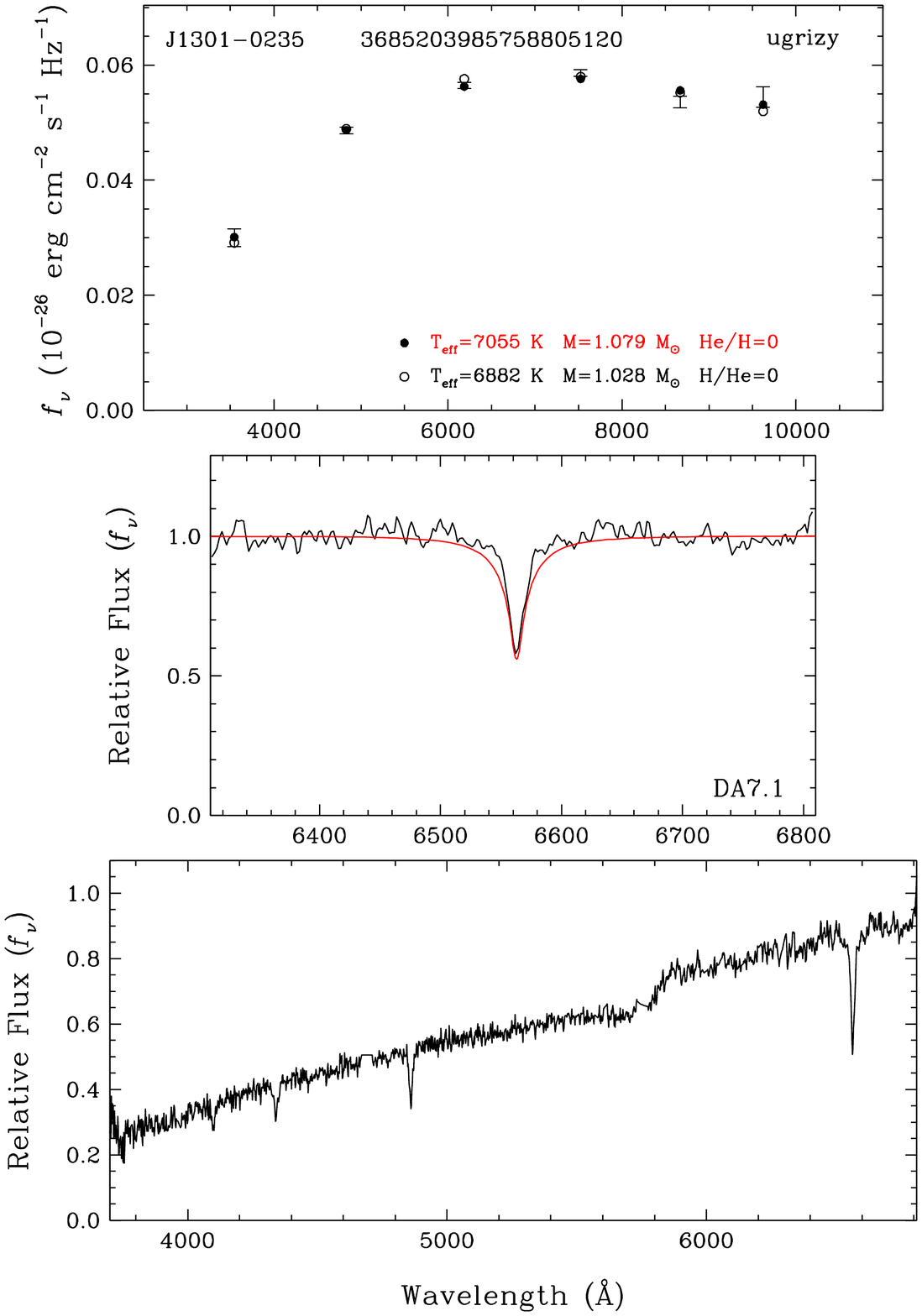}
\caption{Model fits to new DA white dwarfs observed at the MDM 2.4m and Gemini-South telescopes. The symbols and
panels are the same as in Fig. \ref{figda}.}
\label{figda2}
\end{figure*}

\section{Model Atmosphere Analysis}

\subsection{The Fitting Method}

We use the photometric technique as detailed in \citet{bergeron19}, who demonstrated that the SDSS $ugriz$ photometric
system or a combination of SDSS $u$ and the Panoramic Survey Telescope and Rapid Response System
\citep[Pan-STARRS,][]{chambers16}  $grizy$ photometry yield accurate parameters. \citet{bergeron19}
also found that the Pan-STARRS photometry is superior to the SDSS photometry for white dwarf model atmosphere analysis. 
We follow the same approach and use the SDSS $u$ and Pan-STARRS $grizy$ photometry in our analysis.

We convert the observed magnitudes into average fluxes using the appropriate zero points, and compare with the average
synthetic fluxes calculated from model atmospheres with the appropriate chemical composition. A $\chi^2$ value is defined
in terms of the difference between observed and model fluxes over all bandpasses, properly weighted by the photometric
uncertainties, which is then minimized using the nonlinear least-squares method of Levenberg-Marquardt \citep{press86}
to obtain the best fitting parameters. The uncertainties of each fitted parameter are obtained directly from the covariance
matrix of the fitting algorithm, while the uncertainties for all other quantities derived from these parameters are calculated
by propagating in quadrature the appropriate measurement errors.

We fit for the effective temperature and the solid angle, $\pi (R/D)^2$, where $R$ is the radius of the star and $D$ is its
distance. Since the distance is precisely known from Gaia parallaxes, we can constrain the radius of the star directly,
and therefore the mass based on the evolutionary models for white dwarfs. The details of our fitting method, including the
model grids used are further discussed in \citet{bergeron19} and \citet{genest19}. Here we supplement our model grid
with the updated DZ and DQ white dwarf models from \citet{blouin18a} and \citet{blouin19}. Since all of our targets are within 100 pc, we do not correct for reddening.

\subsection{DA White Dwarfs}

Figures \ref{figda} and \ref{figda2} show our methodology for spectral classification based on model fits to four DA white dwarfs
identified in this work. We picked these four stars to also demonstrate the observing setup differences between the different
telescopes.

For each star, the top panel shows the SDSS $u$ and Pan-STARRS $grizy$ photometry (error bars) along with the predicted
fluxes from the best-fitting pure hydrogen (filled dots) and pure helium (open circles) atmosphere models. The labels in
the same panel give the Gaia Source ID, object name based on Gaia DR2 coordinates (for example J0100$-$2021 for the top
left panel in Figure \ref{figda}), and the photometry used in the fitting: $ugrizy$ means SDSS $u$ + Pan-STARRS $grizy$ while
$ugriz$ means SDSS $ugriz$. Some filters were omitted in a few cases where for example they were contaminated by an M
dwarf companion. 

The middle panel, here and in the following figures, always shows the predicted spectrum based on the pure hydrogen
solution, along with the observed H$\alpha$ or H$\beta$ profiles depending on the wavelength coverage of the observations.
Note that we do not fit the spectroscopy data here. Instead, we simply over-plot the predicted hydrogen line profile (red line)
from the photometric fit to see if a given spectrum is consistent with a pure hydrogen atmosphere composition. The spectral
type of each object is also given in the middle panel, along with a temperature index based on 50,400/$T_{\rm eff}$.
The bottom panel shows a broader spectral range for each star revealing the presence of additional Balmer lines in the blue,
if such data are available.

J0010$-$2021(Figure \ref{figda}, left panels) is a relatively warm DA with several Balmer lines visible in its FAST spectrum,
and a significant Balmer jump also visible in the $u$ and $g$ photometry. The photometry and Gaia parallax indicate a pure H
atmosphere with $T_{\rm eff} = 10,049 \pm 57$ K and $M = 0.554 \pm 0.009~M_{\odot}$. The predicted H$\alpha$ line profile
for these parameters provide an excellent match to the observed spectrum, indicating that this is a pure hydrogen atmosphere white dwarf. 

J0057+1756 (Figure \ref{figda}, right panels) is a cooler DA white dwarf with no significant Balmer jump visible in its photometry.
The best-fitting pure hydrogen atmosphere solution has $T_{\rm eff} = 6244 \pm 29$ K and $M = 0.789 \pm 0.011~M_{\odot}$. Even though
our APO data lack blue coverage for this object, the predicted H$\alpha$ line profile from the photometric fit matches the
observed spectrum, indicating that this is also a pure hydrogen atmosphere white dwarf. Note that we did not correct for telluric
features, and the atmospheric B-band (6860-6890 \AA) is visible in our APO data.

Hydrogen lines become weaker with cooler temperatures. J0031+2218 (Figure \ref{figda2}, left panels) is a cool DA white
dwarf with only H$\alpha$ and H$\beta$ lines visible in our MDM spectrum. The best-fitting pure hydrogen atmosphere solution with
$T_{\rm eff} = 6056 \pm 23$ K and $M = 0.629 \pm 0.009~M_{\odot}$ provides an excellent match to the observed H$\alpha$
line profile. 

J1301$-$0235 (Figure \ref{figda2}, right panels) is a slightly warmer and more massive ($M = 1.079 \pm 0.029~M_{\odot}$)
DA white dwarf observed at Gemini. The quantum efficiency correction for the 3 GMOS CCDs is not perfect, resulting
in slightly elevated flux levels for the reddest CCD with coverage beyond 5765 \AA. This has no impact on our spectral typing.
H$\alpha$ through H$\delta$ are visible in our Gemini spectrum of J1301$-$0235, and the photometric solution provides
a good match to the H$\alpha$ line profile, indicating that this is also a pure hydrogen atmosphere white dwarf.

\begin{deluxetable*}{lrlrcc} 
\tablecolumns{6} \tablewidth{0pt} 
\tablecaption{Physical Parameters of the DA White Dwarfs\label{tabda}} 
\tablehead{\colhead{Object} & \colhead{Gaia Source ID} & \colhead{Type} & \colhead{$T_{\rm eff}$} & \colhead{$\log{g}$} & \colhead{$M$} \\
 & & & (K) & (cm s$^{-2}$) & ($M_{\odot}$)}
\startdata   
J0000+0132    & 2738626591386423424 & DA &  $10055 \pm 33$ & $7.975 \pm 0.009$ &  $0.586 \pm 0.007$ \\
J0003$-$0111 & 2449594087142467712 & DA & $ 5371 \pm 36$ & $8.007 \pm 0.043$ & $0.587 \pm 0.036$ \\
J0004$-$0340 & 2447889401738675072 & DA &  $ 7066 \pm 31$ & $7.958 \pm 0.009$ & $0.567 \pm 0.007$  \\
J0006+0755    & 2746037712074342784 & DAH & $ 8610 \pm 74$ & $8.183 \pm 0.026$ & $0.709 \pm 0.023$  \\
J0006$-$0505 & 2444446482939165824 & DA  &  $ 7157 \pm 67$ & $8.156 \pm 0.036$ & $0.689 \pm 0.031$  \\
J0010$-$2021 & 2364836107306975872 & DA &   $10049 \pm 57$ & $7.919 \pm 0.011$ & $0.554 \pm 0.009$  \\
J0011$-$0903 & 2429183303040388992 & DA &   $ 6203 \pm 36$ & $7.931 \pm 0.014$ & $0.548 \pm 0.012$  \\
J0013+0019    & 2545505281002947200 & DA &   $ 9507 \pm 68$ & $7.985 \pm 0.010$ & $0.590 \pm 0.009$  \\
J0013+3246    & 2863526233218817024 & DA &   $10309 \pm 54$ & $7.889 \pm 0.012$ & $0.538 \pm 0.009$   \\
J0015+1353    & 2768116146078155648 & DA &   $ 8439 \pm 56$ & $7.936 \pm 0.016 $ & $0.559 \pm 0.012$                       
\enddata
\tablecomments{This table is available in its entirety in machine-readable format in the online journal. A portion is shown here
for guidance regarding its form and content.}
\end{deluxetable*}

\begin{figure*}
\center
\includegraphics[width=3.5in, bb=20 17 592 779]{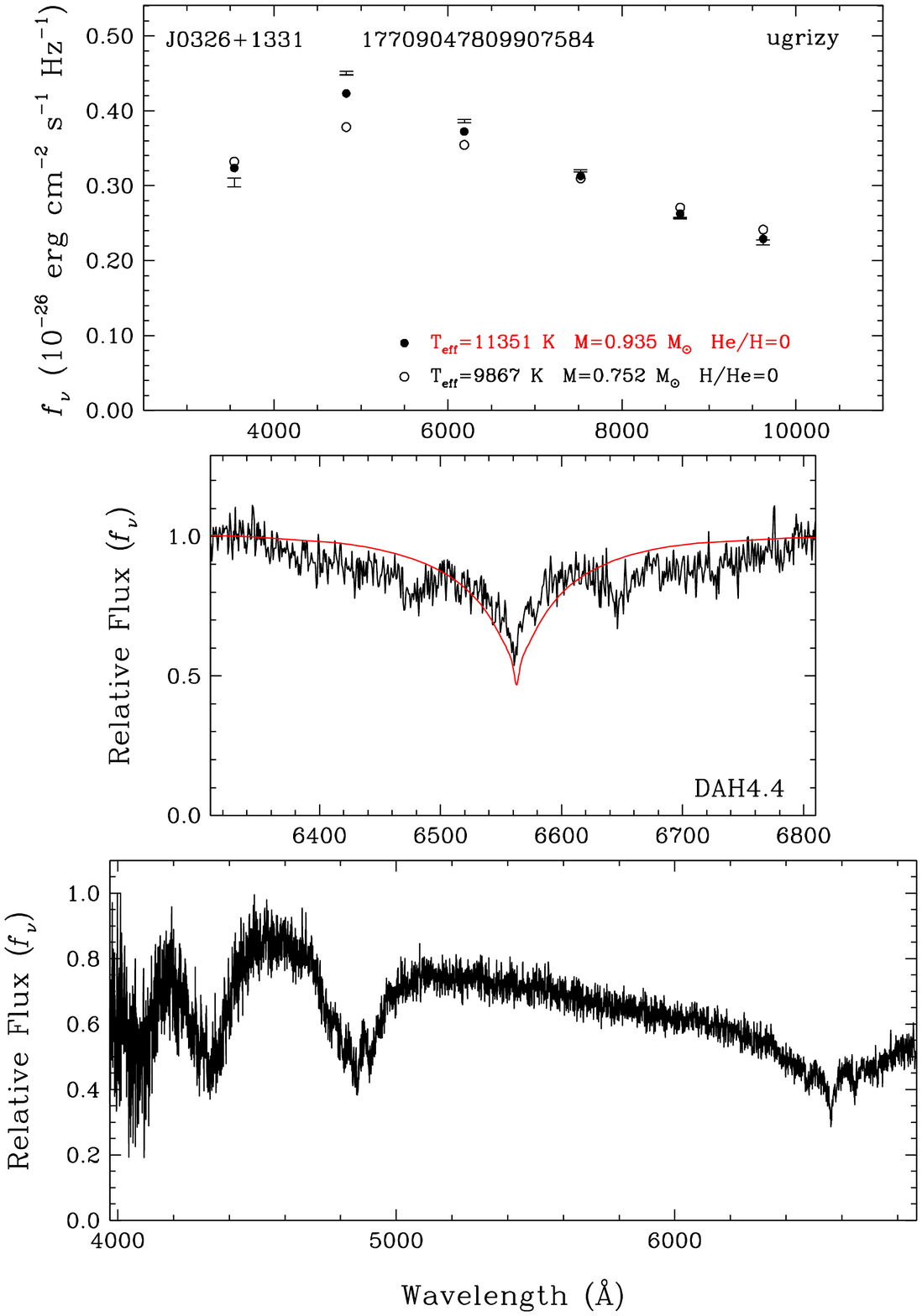}
\includegraphics[width=3.5in, bb=20 17 592 779]{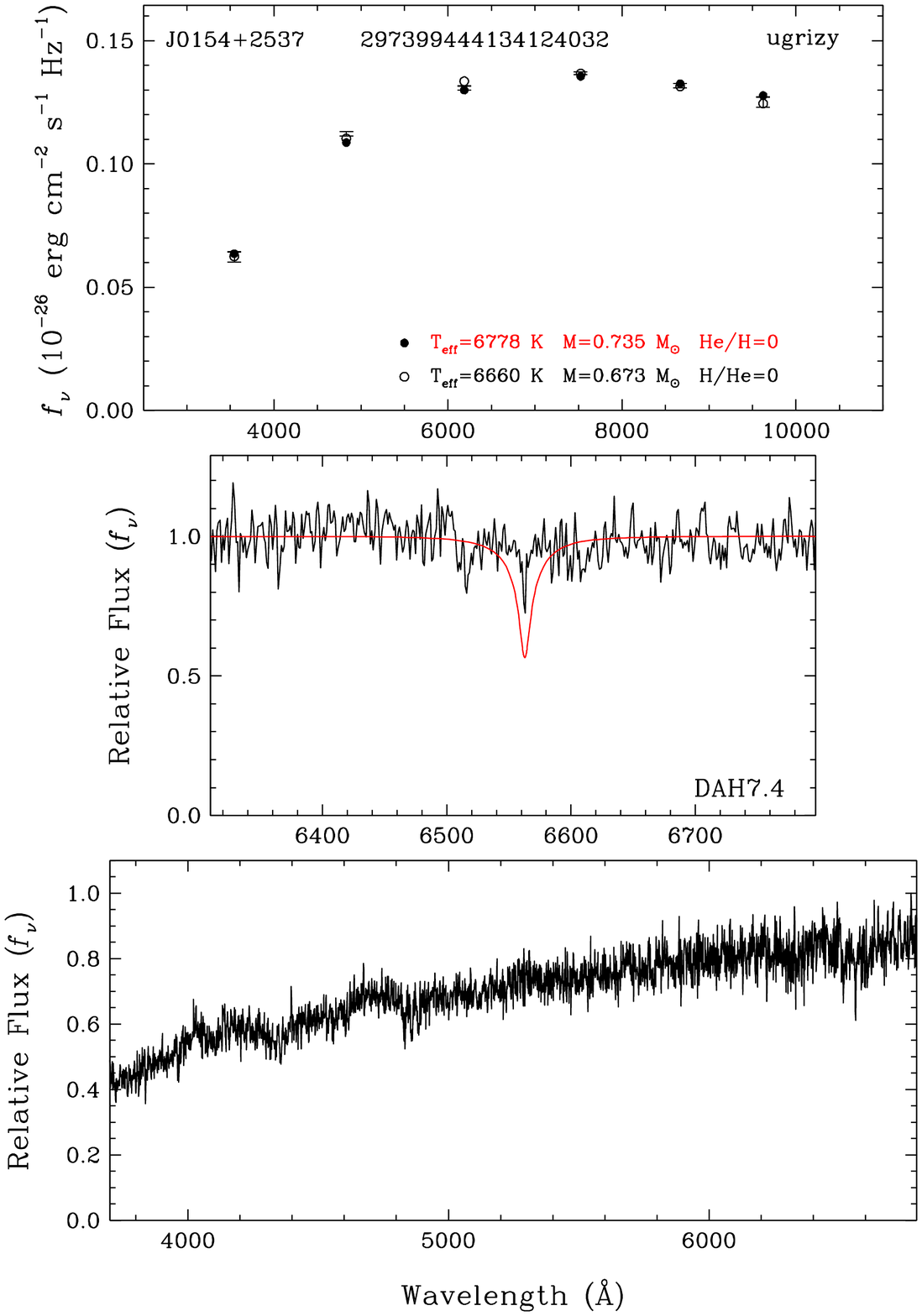}
\vspace{-0.1in}
\caption{Model fits to two new magnetic DA white dwarfs identified in this work. The symbols and panels are the same as in
Figure \ref{figda}. We do not fit for the magnetic field strength here, but simply over-plot the predicted H$\alpha$ line
profile for the best-fitting non-magnetic, pure hydrogen atmosphere model in the middle panel for each star.}
\label{figdah}
\end{figure*}

We present the figures for only four DA white dwarfs here. However, our model fits for all
of the spectroscopically confirmed DA white dwarfs in our sample can be found in the online version of this article. 

We present the physical parameters of all DA white dwarfs in our sample best explained by pure hydrogen atmosphere
models in Table \ref{tabda}. The small errors reported in this table represent the fact that both the photometric and parallax uncertainties are extremely small in most cases. Modern magnitude measurements such as SDSS or Pan-STARRS are
quoted with extremely small uncertainties, sometimes as small as milli-magnitudes. However, as discussed in detail by
\citet{bergeron19}, even though the photometric technique yields precise parameters, the true accuracy of the method
is most likely dominated by the conversion from magnitudes to average fluxes \citep{holberg06}. For example, while the
Pan-STARRS photometric system attempts to be as close as possible to the AB magnitude system, the Pan-STARRS
implementation has an accuracy of only $\sim0.02$ mag according to \citet{tonry12}.

\subsection{Magnetic White Dwarfs}

Figure \ref{figdah} displays our fits to two of the newly identified magnetic DA white dwarfs found in our follow-up survey.
J0326+1331 is a relatively warm and massive DA white dwarf with a significant Balmer jump. For magnetic white dwarfs,
the Balmer jump is not reproduced well by our non-magnetic models. Our MDM spectrum of J0326+1331
reveals broad hydrogen lines that are split into multiple components due to the presence of a strong magnetic field. 
We do not fit for the field strength here, but simply compare the predicted H$\alpha$ line profile from the photometric fit
to the observed spectrum. The observed H$\alpha$ line profile is weaker than expected due to the Zeeman-splitting, indicating
that some of the weak-line objects with noisier spectra may be also magnetic.

J0154+2537 (right panels in Fig. \ref{figdah}) is one such system, where at first glance, our MMT spectrum looks noisy.
However, a closer look at the H$\alpha$ region shows that this line is much weaker than
expected based on the photometric fit, but it is present, and is split into three components due to
a magnetic field. This figure demonstrates the difficulty of identifying magnetic DA white dwarfs at cooler temperatures,
and that some of the DC white dwarfs in our sample may be magnetic white dwarfs with weak lines that might have been
lost in the noise. 

Table \ref{tabda} includes the physical parameters of the magnetic DA white dwarfs, with masses ranging from 0.47
to 1.32 $M_{\odot}$. The median mass of the magnetic DA white dwarfs in our sample is 0.83 $M_{\odot}$. This is similar to the
average mass, 0.87 $M_{\odot}$, of the magnetic white dwarfs presented in \citet{kawka20}. Clearly, the mass distribution
for the magnetic white dwarfs is significantly different than that of the normal DA white dwarfs. This is consistent with the idea that
magnetic white dwarfs form as a result of merging binaries during common envelope evolution \citep{briggs15}.

\subsection{He-rich DA White Dwarfs}

\begin{figure}
\center
\includegraphics[width=3in]{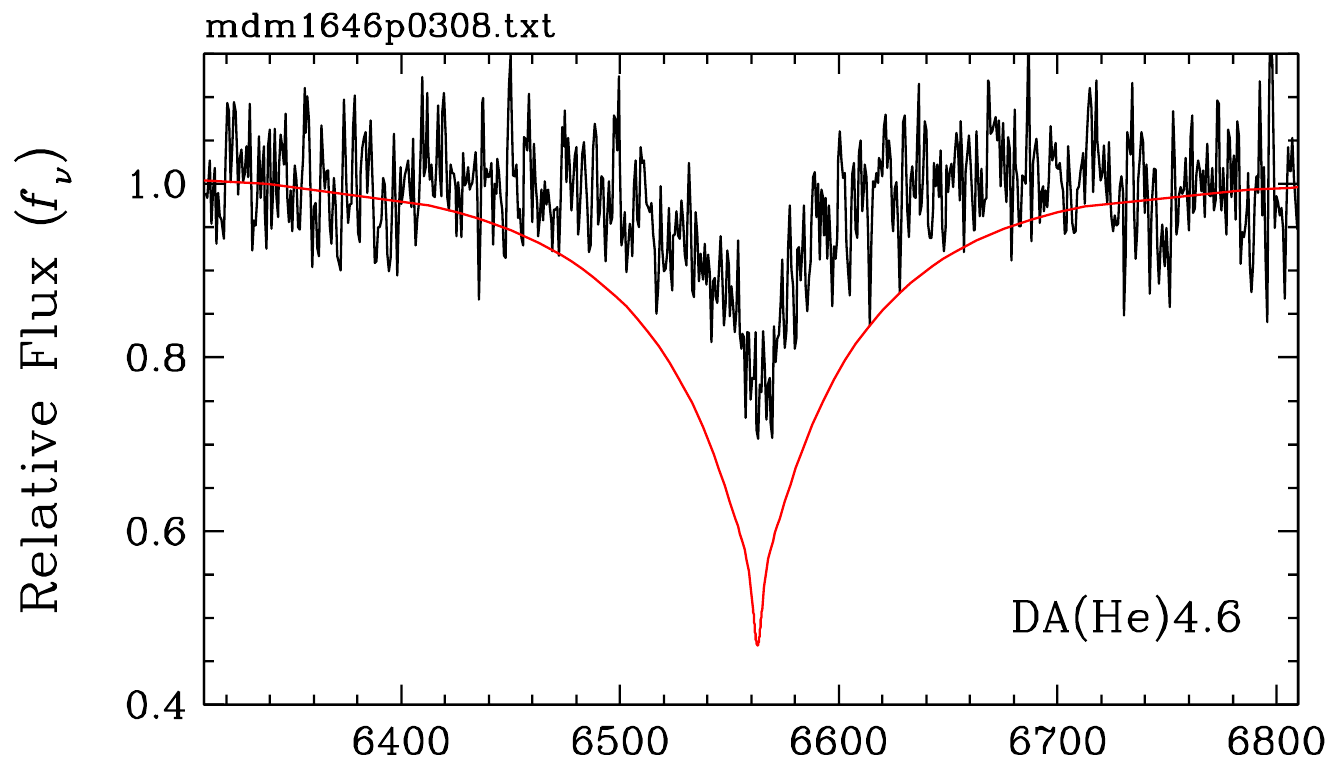}
\includegraphics[width=3.5in, bb=20 17 552 779]{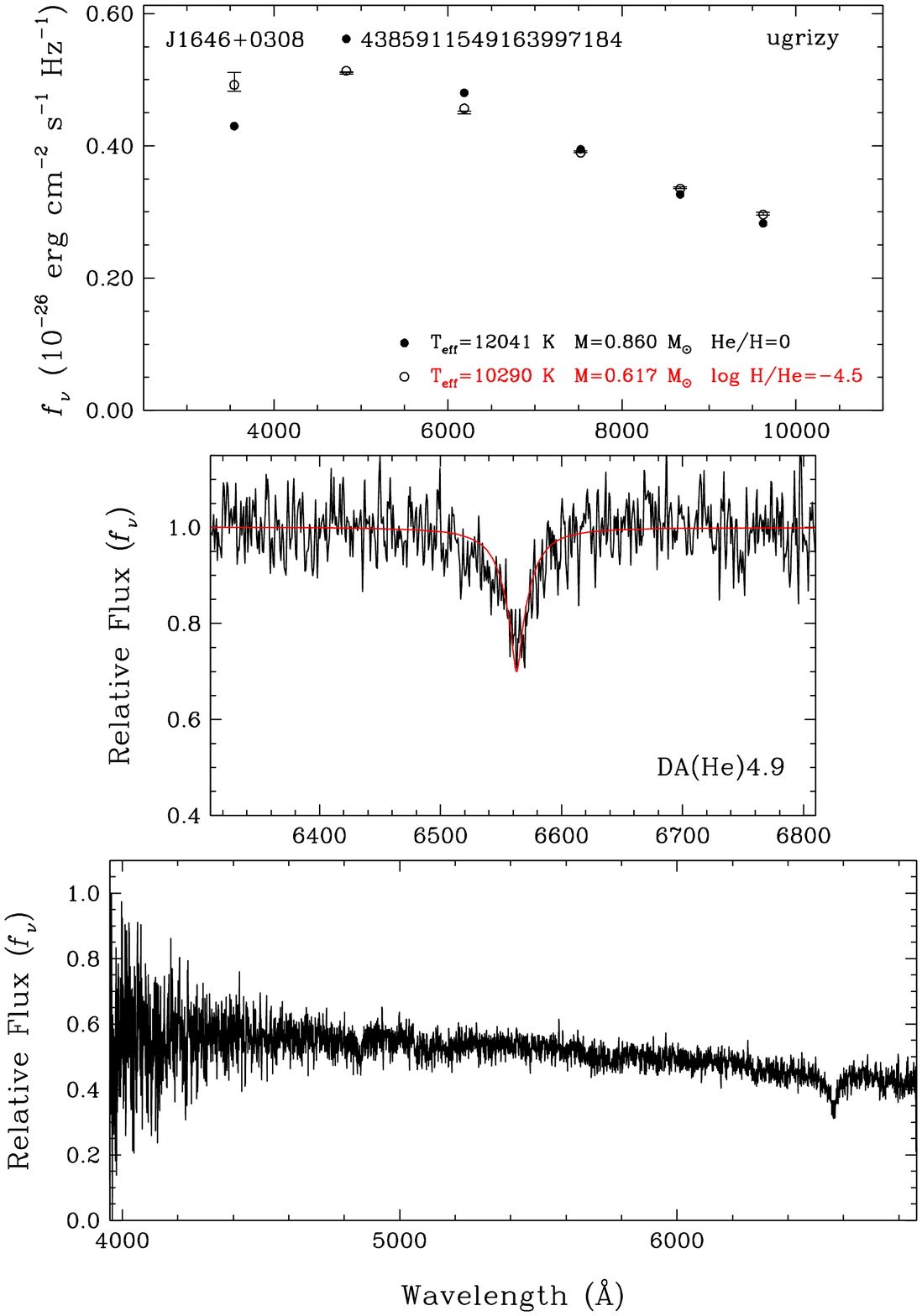}
\caption{Model fits to the He-rich DA white dwarf J1646+0308. The top panel shows the observed (black) and
predicted H$\alpha$ line (red) for the best-fitting photometric solution with a pure hydrogen atmosphere.
The spectral energy distribution clearly favors a He-rich solution (second panel), which
provides an excellent match to the H$\alpha$ line profile (third panel). The bottom panel shows the entire spectral
range of our observations,  revealing H$\alpha$ and H$\beta$ lines much weaker than expected for a typical
DA white dwarf at these temperatures.}
\label{figdahe}
\end{figure}

\begin{figure*}
\center
\includegraphics[width=3.5in, bb=20 17 552 779]{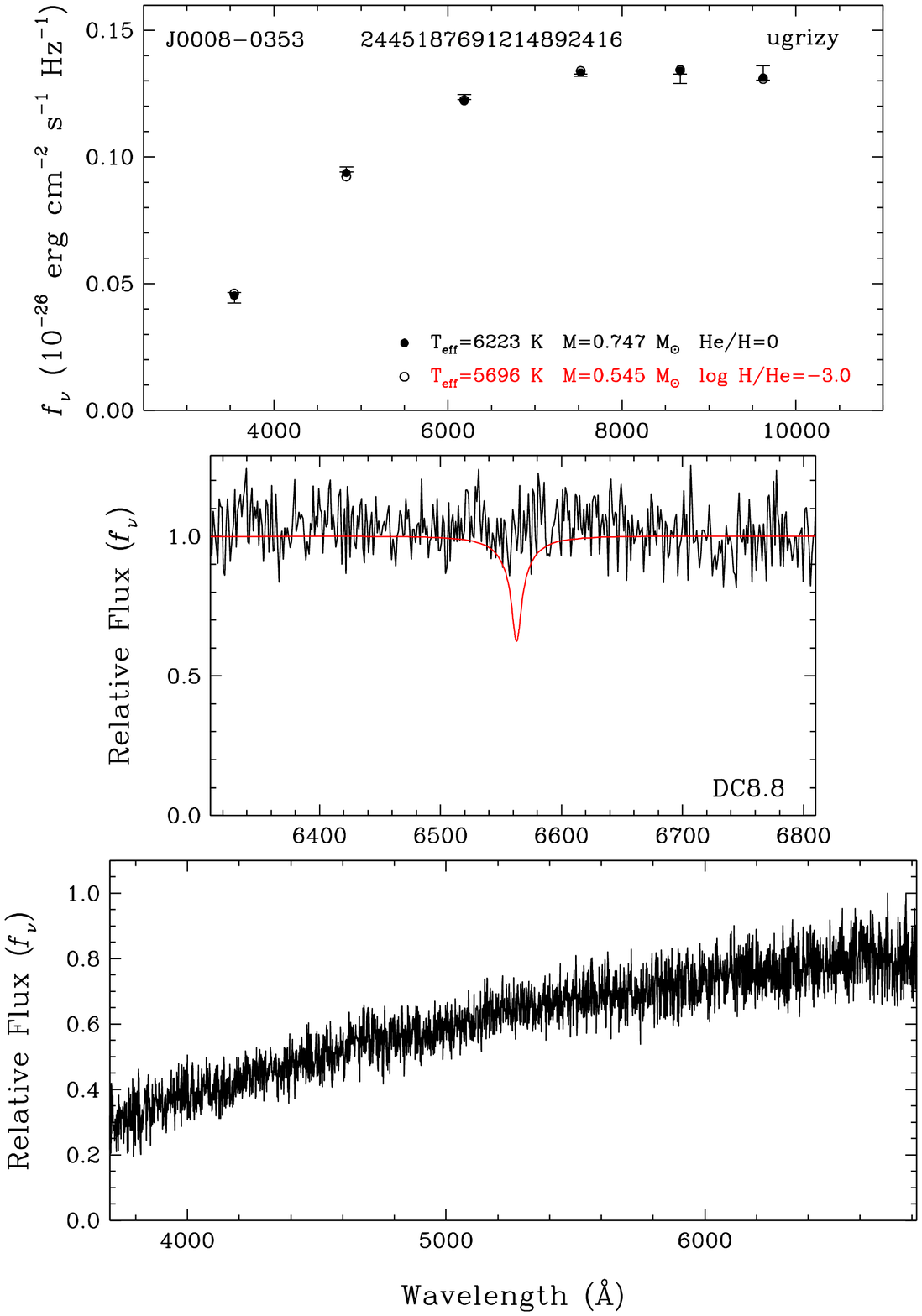}
\includegraphics[width=3.5in, bb=20 17 552 779]{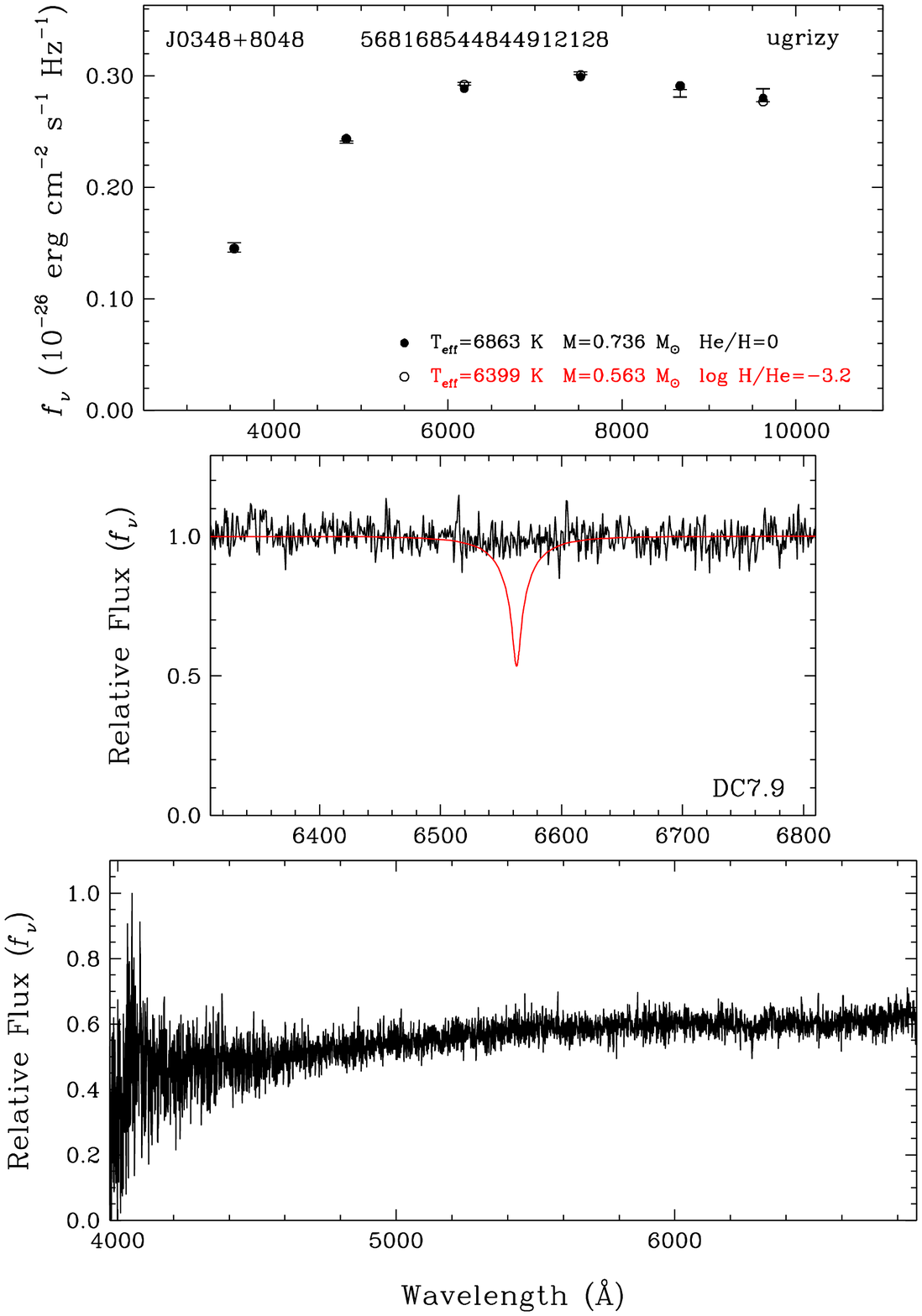}
\caption{Model fits to two of the DC white dwarfs in our sample. The symbols and panels are the same as in Fig. \ref{figda}.
Even though both hydrogen and helium atmosphere solutions match the photometry, pure hydrogen atmosphere models predict significant
H$\alpha$ absorption features that are not observed (middle panels). Hence, these two stars clearly have He-dominated
atmospheres.}
\label{figdc}
\end{figure*}

Ross 640 and L745-46A, are some of the best known examples of DA white dwarfs that have helium-dominated
atmospheres. Hydrogen lines in these stars are heavily broadened through van der Waals interactions in helium dominated
atmospheres \citep{giammichele12,rolland18}. 

There are 13 white dwarfs in our sample that show hydrogen lines, but their spectral energy distributions are best explained by
a helium-rich atmosphere. Figure \ref{figdahe} shows our model fits to one of these DA(He) white dwarfs. J1646+0308 clearly
shows H$\alpha$ and H$\beta$. However, the best-fit photometric solution for a pure hydrogen atmosphere has
$T_{\rm eff} = 12,041$ K and $M=0.86~M_{\odot}$. At that temperature, a pure hydrogen atmosphere white dwarf would display
significantly deeper Balmer lines (top panel) than observed,
and the photometry would also show evidence of a significant Balmer jump between the $u$ and $g$ filters, which is not
observed. However, the spectral energy distribution and the observed H$\alpha$ line profile can both be explained
by a He-dominated atmosphere with $T_{\rm eff}= 10,290$ K, $M=0.617~M_{\odot}$, and trace amounts of hydrogen with
$\log{\rm H/He} = -4.5$. 

Table \ref{tabdahe} presents the best-fit atmospheric parameters for the DA(He) white dwarfs in our sample. 
The model fits for all 13 are available in the online version of this article. Nine of these
are clearly He-dominated DAs, based on the H$\alpha$ line profiles being inconsistent with a pure hydrogen atmosphere solution.
However, for four of these objects, J1611+1322, J1628+1224, J2104+2333, and J2138+2309, our interpretation of a He-dominated
atmosphere depends on the quality of the photometry, in particular in the $u-$band. 

These four stars appear as normal DA stars. Using the spectroscopic method, we find that the spectroscopic solution is also
consistent with the pure hydrogen photometric solution. However, the spectroscopic method has a degeneracy
between a high $\log{g}$ and a high helium content \citep[see Figure 18 in][]{bergeron91}.  Given the temperature range
of 8530 - 11,060 K for these four stars, the observed Balmer jump between the $u$ and $g$ filters is smaller than expected
for pure hydrogen models. The best-fitting mixed atmosphere models have He/H ratios similar to GD 362, which was
confirmed to have a helium-dominated atmosphere through high-resolution spectroscopy \citep{zuckerman07}.
Higher quality $u-$band photometry or higher resolution optical spectroscopy (for the stars with $T_{\rm eff}$ near 11,000 K)
would be useful to confirm the atmospheric compositions for these four targets.

\begin{deluxetable*}{cccccc} 
\tablecolumns{6} \tablewidth{0pt}
\tablecaption{Physical Parameters of the DA White Dwarfs with He-dominated Atmospheres \label{tabdahe}} 
\tablehead{\colhead{Object} & \colhead{Gaia Source ID} & \colhead{$\log{\rm H/He}$} & \colhead{$T_{\rm eff}$} & \colhead{$\log{g}$} & \colhead{$M$} \\
 & & & (K) & (cm s$^{-2}$) & ($M_{\odot}$)}
\startdata 
J0048$-$0124 & 2530629365419780864 & $-$3.5 & $ 9235 \pm 100$ & $8.207 \pm 0.016$ & $0.708 \pm 0.015$ \\
J0851+5426   & 1029968414968983424 & $-$4.0 & $ 8782 \pm  88$ & $8.275 \pm 0.017$ & $0.752 \pm 0.016$ \\
J1024$-$0023 & 3830990156631488128 & $-$4.5 & $ 9487 \pm  75$ & $8.163 \pm 0.019$ & $0.680 \pm 0.017$ \\
J1159+0007   & 3891115064506627840 & $-$2.5 & $ 8566 \pm  44$ & $8.419 \pm 0.007$ & $0.849 \pm 0.006$ \\
J1412+1129   & 1226246251436497152 & $-$2.0 & $ 7360 \pm  56$ & $8.225 \pm 0.025$ & $0.717 \pm 0.022$ \\
J1529+1304   & 1193808772927091712 & $-$4.5 & $ 9254 \pm  53$ & $8.133 \pm 0.011$ & $0.661 \pm 0.009$ \\
J1611+1322   & 4458207634145130368 & $-$2.5 & $ 8529 \pm  50$ & $8.326 \pm 0.007$ & $0.786 \pm 0.007$ \\
J1623+4650   & 1410031887762012800 & $-$4.0 & $ 9206 \pm  82$ & $8.177 \pm 0.012$ & $0.689 \pm 0.011$ \\
J1628+1224   & 4460327252045435648 & $-$1.5 & $11002 \pm 129$ & $7.854 \pm 0.018$ & $0.501 \pm 0.013$ \\
J1646+0308   & 4385911549163997184 & $-$4.5 & $10290 \pm  68$ & $8.060 \pm 0.014$ & $0.617 \pm 0.012$ \\
J2104+2333   & 1840865211187303424 & $-$1.0 & $11059 \pm 107$ & $7.932 \pm 0.014$ & $0.544 \pm 0.011$ \\
J2138+2309   & 1794118516552814336 & $-$1.5 & $ 9695 \pm  33$ & $7.802 \pm 0.007$ & $0.470 \pm 0.005$ \\
J2301+2323   & 2842112183412398336 & $-$5.0 & $ 9974 \pm  52$ & $8.099 \pm 0.009$ & $0.641 \pm 0.008$     
\enddata
\tablecomments{The mixed atmosphere interpretation for J1611+1322, J1628+1224, J2104+2333, and J2138+2309
is based on the significance of the Balmer jump inferred from the $u-$band photometry.}
\end{deluxetable*}

\begin{deluxetable*}{ccccccc} 
\tablecolumns{7} \tablewidth{0pt}
\tablecaption{Physical Parameters of the DB and DC White Dwarfs \label{tabdc}} 
\tablehead{\colhead{Object} & \colhead{Gaia Source ID} & \colhead{Type} &  \colhead{$\log{\rm H/He}$} & \colhead{$T_{\rm eff}$} & \colhead{$\log{g}$} & \colhead{$M$} \\
 & & & & (K) & (cm s$^{-2}$) & ($M_{\odot}$)}
\startdata 
J0001+3237 & 2874216647336589568 & DC & $-$5.0 &  5447 $\pm$ 54 & 7.771 $\pm$ 0.045 & 0.441 $\pm$ 0.033 \\
J0005+3451 & 2876705018245556480 & DC & $-$4.6 & 10138 $\pm$ 73 & 7.965 $\pm$ 0.015 & 0.561 $\pm$ 0.012 \\
J0008$-$0353 & 2445187691214892416 & DC & $-$5.0 &  5714 $\pm$ 52 & 7.953 $\pm$ 0.041 & 0.543 $\pm$ 0.034 \\
J0011+2824 & 2859908324567852416 & DC & $-$3.8 &  8243 $\pm$ 91 & 7.872 $\pm$ 0.021 & 0.503 $\pm$ 0.017 \\
J0011$-$0827 & 2429268309033277184 & DC & $-$4.0 &  8616 $\pm$ 94 & 8.130 $\pm$ 0.028 & 0.658 $\pm$ 0.025 \\
J0014$-$0758 & 2429392661221943040 & DC & $-$5.0 &  5585 $\pm$ 51 & 8.017 $\pm$ 0.048 & 0.581 $\pm$ 0.041 \\
J0019$-$1114 & 2424913628807069056 & DC & $-$3.8 &  8187 $\pm$ 64 & 8.093 $\pm$ 0.025 & 0.633 $\pm$ 0.022 \\
J0021+3827 &  378856389416637056 & DC & $-$5.0 & 11332 $\pm$ 86 & 8.057 $\pm$ 0.011 & 0.617 $\pm$ 0.010 \\
J0038+3409 &  364978005758397952 & DC & $-$4.2 &  9191 $\pm$ 58 & 7.923 $\pm$ 0.018 & 0.534 $\pm$ 0.014 \\
J0049+1727 & 2782037303315690752 & DC & $-$3.0 &  5790 $\pm$ 66 & 7.961 $\pm$ 0.057 & 0.548 $\pm$ 0.047 
\enddata
\end{deluxetable*}

\subsection{DC White Dwarfs}

\ion{He}{1} lines disappear below about 11,000 K and hydrogen lines disappear below about 5000 K.
Hence, the absence of spectral features in DC white dwarfs hotter than 5000 K can be used to rule out a pure H
composition and infer a He-dominated atmosphere. 

Figure \ref{figdc} shows our model fits to two of the DC white dwarfs in our sample, where a pure hydrogen atmosphere can be
safely ruled out. The photometry cannot distinguish between
the pure hydrogen and the helium-dominated solution for J0008$-$0353 (left panels), as both provide acceptable solutions. However,
the best-fitting pure hydrogen atmosphere solution has $T_{\rm eff}=6223$ K, which would display a significant H$\alpha$ line that
is not observed in our MMT spectrum. Hence, J0008$-$0353 clearly has a He-dominated atmosphere. Similarly, the best-fitting
pure hydrogen atmosphere model for J0348+8048 (right panels) requires a H$\alpha$ line that is almost 50\% deep compared to the
continuum. Such a line is clearly absent in our MDM spectrum, also confirming a He-dominated atmosphere for this white dwarf.

\begin{figure*}
\center
\includegraphics[width=3.5in, bb=20 17 552 779]{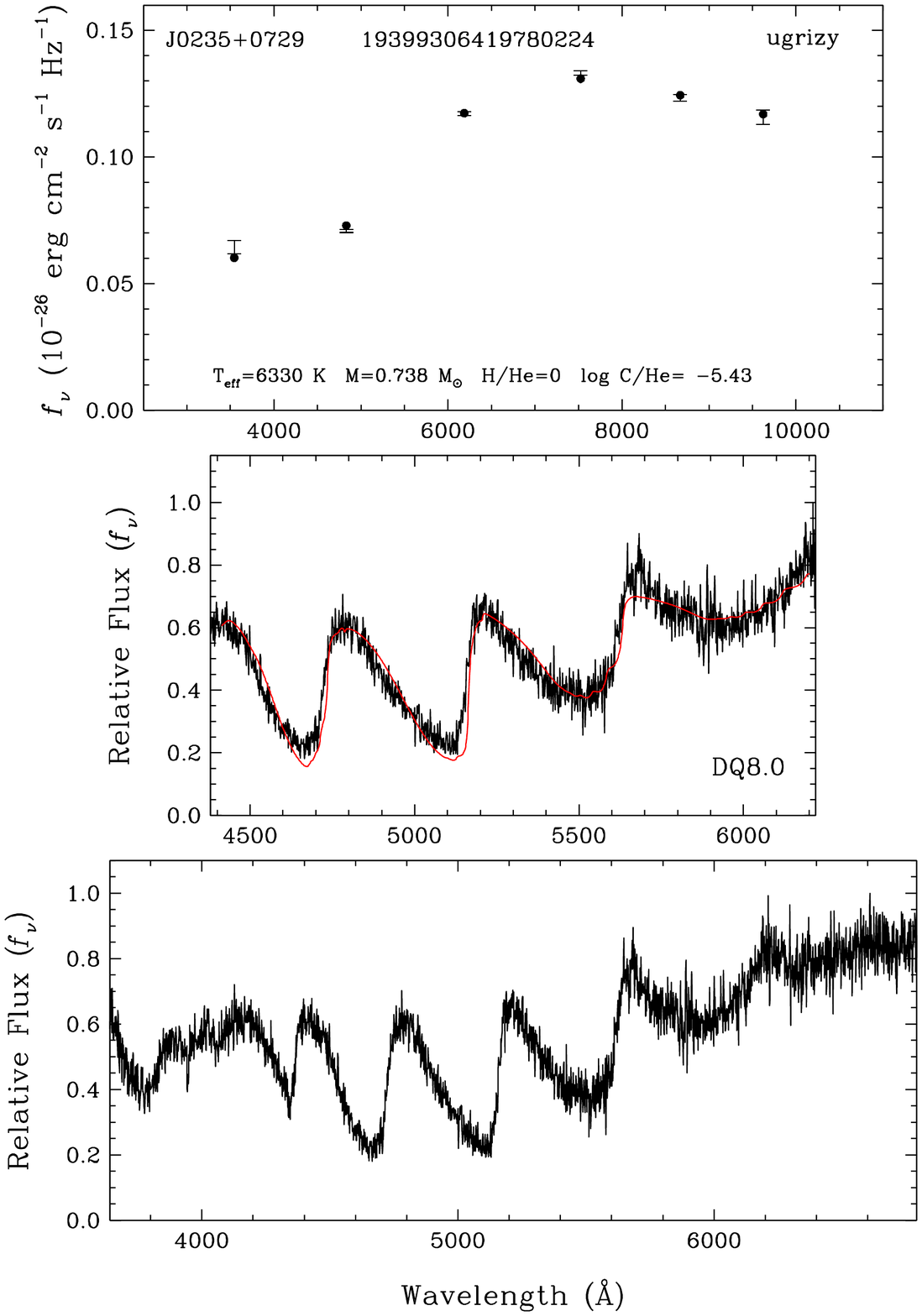}
\includegraphics[width=3.5in, bb=20 17 552 779]{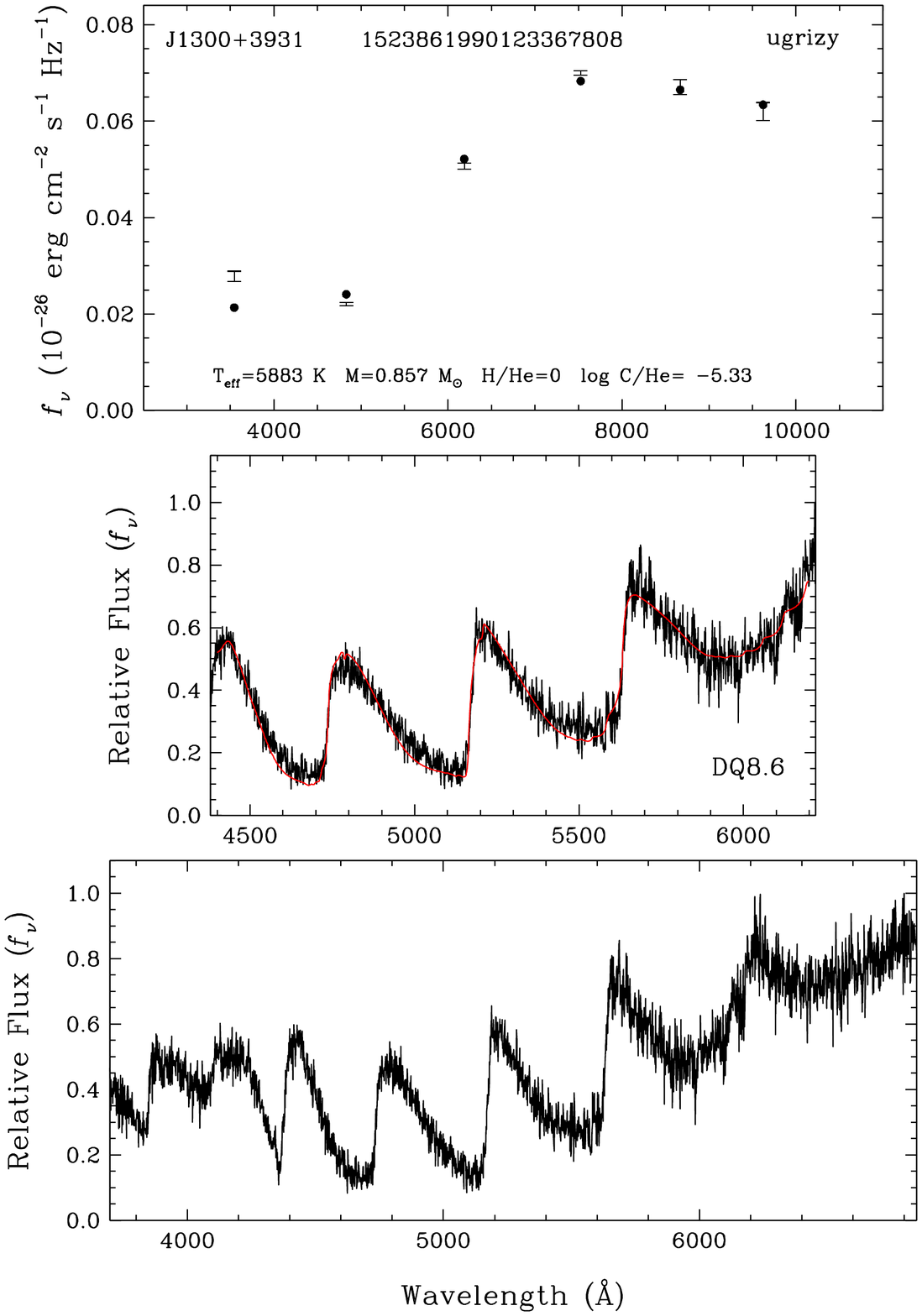}
\caption{Model fits to two DQ white dwarfs observed at the MMT. The top and middle panels
show the photometric and spectroscopic model fits, respectively. The bottom panels display the entire spectral
range of the observations. The atmospheric models provide an excellent match to both photometry and spectroscopy
for these two stars.}
\label{figdq}
\end{figure*}

\begin{figure*}
\center
\includegraphics[width=3.5in, bb=20 17 592 779]{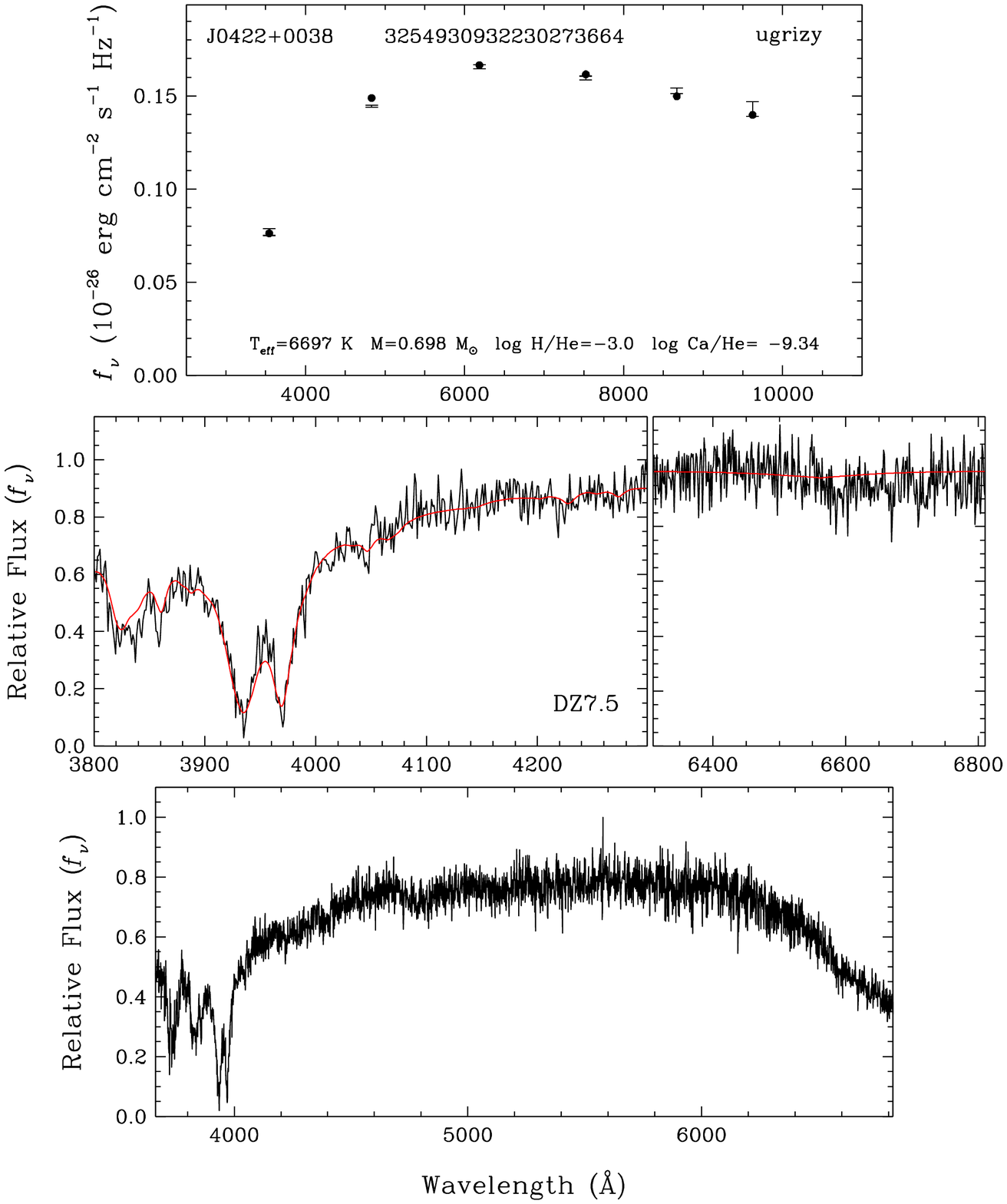}
\includegraphics[width=3.5in, bb=20 17 592 779]{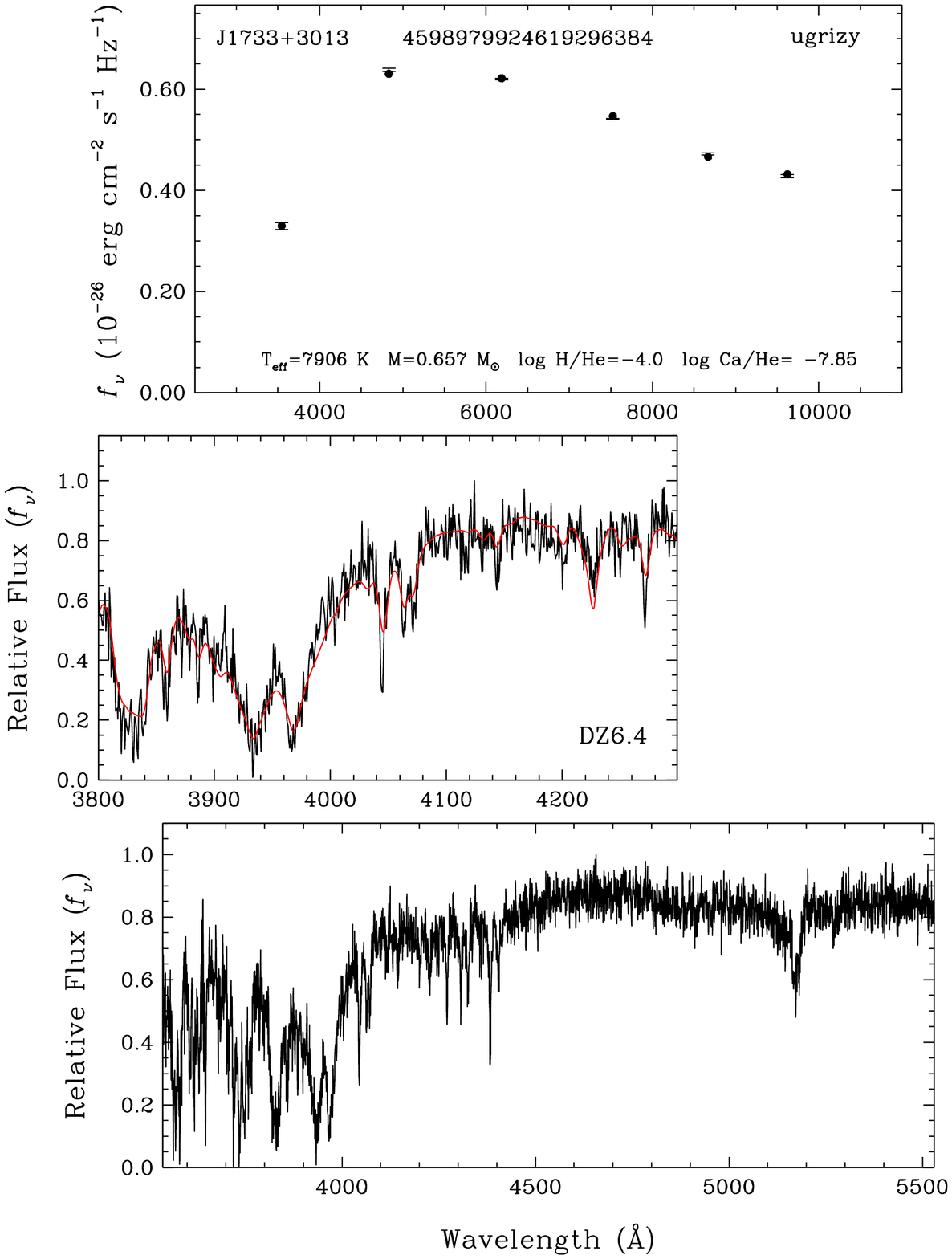}
\caption{Model fits to two DZ white dwarfs observed at the MMT and FAST. The top and middle panels
show the photometric and spectroscopic model fits, respectively. The bottom panels display the entire spectral
range of the observations. The atmospheric models provide an excellent match to both photometry and spectroscopy
for these two stars.}
\label{figdz}
\end{figure*}

Analyzing the mass distribution of non-DA white dwarfs between 11,000 K and 6000 K, \citet{bergeron19} found an almost
complete absence of normal mass non-DA stars based on pure helium atmosphere models, and concluded that pure He
atmosphere white dwarfs must be extremely rare or non-existent in this temperature range. More reasonable mass estimates
require additional electron donors in the form of hydrogen, carbon, or other metals. In fact, they found that about
25\% of the non-DAs in this temperature range are DQ stars contaminated by carbon, and another 25\% are DZ stars contaminated
by metals, and the rest are DC stars. 

\citet{bergeron19} and \citet{rolland18} revisited the suggestion that convective mixing is responsible for the sudden
increase in the number of non-DA white dwarfs below 10,000 K. In this scenario, the relatively small
amount of hydrogen in the upper layers is mixed with the underlying helium convection zone, resulting in extremely
small hydrogen abundances that, in general, do not produce a visible H$\alpha$ line. 

For the analysis of the DC white dwarfs in our sample, we follow \citet{bergeron19} and use mixed hydrogen and helium models
where the hydrogen abundance is adjusted as a function of effective temperature. In the temperature range of DB stars
($T_{\rm eff}\geq11,000$ K) we adopt $\log{\rm H/He} = -5$, and for cooler DC stars, we gradually increase the hydrogen
abundance to $\log{\rm H/He} = -3$ at 6000 K following the predictions of the convective mixing scenario \citep{rolland18}. 
We assume $\log{\rm H/He} = -5$ for the He-rich solution of cooler DC stars.
Table \ref{tabdc} presents the best-fit model parameters for the newly identified DB and DC white dwarfs in our sample.

For DC white dwarfs below 5000 K, there is no way to tell the composition based on optical spectra as both hydrogen and
helium become invisible in their atmospheres. However, pure helium and mixed H/He models result in mass estimates below
0.5 and even 0.4 $M_{\odot}$ for many of the cool DCs. 

\citet{kowalski06} demonstrated that the far red wing of the Lyman $\alpha$ line dominates the opacity in the blue part
of the optical spectra of pure hydrogen atmosphere white dwarfs, and concluded that most cool DC stars should have
pure hydrogen atmospheres. These models provide a much better fit to the $u-g$ colors of the cool DC white dwarfs
(see Figures \ref{fighrdall} and \ref{figugr} below) and provide more reasonable mass estimates of $M>0.5~M_{\odot}$ for the
majority of these stars \citep{bergeron19}. This is why we adopted the pure hydrogen solution for the coolest DCs in our sample.

On the other hand, not all cool white dwarfs have pure hydrogen atmospheres. \citet{blouin19} found that $\approx25$\%
of the DC white dwarfs cooler than 5000 K have helium-rich atmospheres. There are also DZ white dwarfs with clearly
helium-rich atmospheres in the same temperature range. The atmospheric composition of each of the cool DCs should be
constrained based on a detailed analysis of each object, which is outside of the scope of this paper.

There are several DC white dwarfs in our sample that display significant absorption in the Pan-STARRS $z$ and $y-$bands.
These are the so-called IR-faint or ultracool white dwarfs that display collision induced absorption (CIA) from molecular hydrogen.
We discuss these further in section \ref{ultra}.

\subsection{DQ White Dwarfs}

\begin{deluxetable*}{cccccc} 
\tablecolumns{6} \tablewidth{0pt}
\tablecaption{Physical Parameters of the DQ White Dwarfs \label{tabdq}} 
\tablehead{\colhead{Object} & \colhead{Gaia Source ID} & \colhead{$\log{\rm C/He}$} & \colhead{$T_{\rm eff}$} & \colhead{$\log{g}$} & \colhead{$M$} \\
 & & & (K) & (cm s$^{-2}$) & ($M_{\odot}$)}
\startdata 
J0104+4650    &  401215160231429120  & $-$3.21 & $11257 \pm 166$ & $8.646 \pm 0.020$ & $0.997 \pm 0.016$ \\
J0235+0729    &   19399306419780224   & $-$5.43 & $ 6330 \pm  19$ & $8.251 \pm 0.025$ & $0.733 \pm 0.023$ \\
J0441$-$0551 & 3200232157189930880 & $-$5.72 & $ 7619 \pm  49$ & $7.957 \pm 0.013$ & $0.550 \pm 0.011$ \\
J0705$-$1703 & 2935415125246460032 & $-$5.42 & $ 6116 \pm  25$ & $8.156 \pm 0.024$ & $0.670 \pm 0.022$ \\
J1241+2614    & 3961323069532233984 & $-$6.94 & $ 5794 \pm  17$ & $7.946 \pm 0.027$ & $0.539 \pm 0.022$ \\
J1300+3931    & 1523861990123367808 & $-$5.33 & $ 5883 \pm  22$ & $8.429 \pm 0.035$ & $0.852 \pm 0.032$ \\
J1706$-$1238 & 4140966708116861440 & $-$6.83 & $ 6113 \pm  42$ & $7.943 \pm 0.045$ & $0.538 \pm 0.036$  
\enddata
\end{deluxetable*}

\begin{deluxetable*}{ccccccc} 
\tablecolumns{7} \tablewidth{0pt}
\tabletypesize{\scriptsize}
\tablecaption{Physical Parameters of the DZ White Dwarfs \label{tabdz}} 
\tablehead{\colhead{Object} & \colhead{Gaia Source ID} & \colhead{$\log{\rm Ca/He}$} & \colhead{$\log{\rm H/He}$} & \colhead{$T_{\rm eff}$} & \colhead{$\log{g}$} & \colhead{$M$} \\
 & & & & (K) & (cm s$^{-2}$) & ($M_{\odot}$)}
\startdata 
J0115+4733   & 401529105161219968  & $-$10.88 & $-$3 & $ 6436 \pm 63 $ & $ 8.220  \pm 0.054$ & $ 0.712 \pm 0.050 $ \\
J0141+2257   & 290602027028411136  & $-$7.93  & $-$3 & $ 7562 \pm 33 $ & $ 8.044  \pm 0.015$ & $ 0.602 \pm 0.013 $ \\
J0333+0656   & 3276355500412926336 & $-$9.62  & $-$4 & $ 7037 \pm 73 $ & $ 8.149  \pm 0.041$ & $ 0.667 \pm 0.037 $ \\
J0416+2947   & 165677710612522752  & $-$9.72  & $-$4 & $ 6864 \pm 59 $ & $ 8.091  \pm 0.023$ & $ 0.630 \pm 0.021 $ \\
J0422+0038   & 3254930932230273664 & $-$9.34  & $-$3 & $ 6696 \pm 39 $ & $ 8.211  \pm 0.024$ & $ 0.707 \pm 0.022 $ \\
J0425+2834   & 152931141027253888  & $-$10.40 & $-$2 & $ 5988 \pm 31 $ & $ 8.145  \pm 0.020$ & $ 0.662 \pm 0.018 $ \\
J0445+0050   & 3230171064244273024 & $-$10.14 & $-$4 & $ 6071 \pm 35 $ & $ 8.110  \pm 0.027$ & $ 0.640 \pm 0.024 $ \\
J0447+0106   & 3231702860037325824 & $-$10.95 & $-$2 & $ 6164 \pm 43 $ & $ 8.188  \pm 0.017$ & $ 0.691 \pm 0.016 $ \\
J0654+3938   & 950361883331847424  & $-$9.58  & $-$4 & $ 9435 \pm 68 $ & $ 7.961  \pm 0.010$ & $ 0.556 \pm 0.008 $ \\
J0703$-$1656 & 2935446392608812032 & $-$11.27 & $-$4 & $ 7387 \pm 68 $ & $ 8.080  \pm 0.021$ & $ 0.624 \pm 0.018 $ \\
J0717+3630   & 898002379406954752  & $-$9.10  & $-$4 & $ 7133 \pm 61 $ & $ 8.101  \pm 0.029$ & $ 0.637 \pm 0.025 $ \\
J0726+2216   & 866081701427025536  & $-$8.78  & $-$4 & $ 6957 \pm 40 $ & $ 7.991  \pm 0.019$ & $ 0.569 \pm 0.016 $ \\
J0800$-$0840 & 3039671225803688320 & $-$9.93  & $-$4 & $ 6965 \pm 60 $ & $ 8.135  \pm 0.019$ & $ 0.658 \pm 0.017 $ \\
J0805+6624   & 1095169656358227456 & $-$10.94 & $-$4 & $ 6482 \pm 80 $ & $ 8.066  \pm 0.031$ & $ 0.613 \pm 0.027 $ \\
J0902+6503   & 1044511002434919936 & $-$10.34 & $-$4 & $ 6385 \pm 54 $ & $ 8.121  \pm 0.022$ & $ 0.648 \pm 0.019 $ \\
J0912+0808   & 590183142750379136  & $-$10.47 & $-$4 & $ 6338 \pm 47 $ & $ 8.056  \pm 0.019$ & $ 0.607 \pm 0.017 $ \\
J0928+1937   & 634307433430523648  & $-$10.43 & $-$4 & $ 6072 \pm 34 $ & $ 8.050  \pm 0.028$ & $ 0.602 \pm 0.025 $ \\
J1027+5019   & 847272257226467840  & $-$10.71 & $-$2 & $ 5742 \pm 48 $ & $ 8.135  \pm 0.034$ & $ 0.656 \pm 0.031 $ \\
J1114+6546   & 1056003918305358592 & $-$10.40 & $-$4 & $ 5915 \pm 40 $ & $ 8.196  \pm 0.018$ & $ 0.696 \pm 0.016 $ \\
J1137+2005   & 3978988652273088128 & $-$11.17 & $-$3 & $ 6214 \pm 49 $ & $ 8.201  \pm 0.043$ & $ 0.700 \pm 0.039 $ \\
J1502+4933   & 1592242645479324288 & $-$9.79  & $-$4 & $ 6541 \pm 46 $ & $ 8.109  \pm 0.021$ & $ 0.641 \pm 0.018 $ \\
J1552$-$0219 & 4403059253334693248 & $-$9.53  & $-$4 & $ 7666 \pm 59 $ & $ 8.018  \pm 0.021$ & $ 0.586 \pm 0.018 $ \\
J1733+3013   & 4598979924619296384 & $-$7.85  & $-$4 & $ 7908 \pm 38 $ & $ 8.146  \pm 0.008$ & $ 0.667 \pm 0.007 $ \\
J1757+1021   & 4494877446445481472 & $-$8.28  & $-$4 & $ 8358 \pm 46 $ & $ 8.101  \pm 0.008$ & $ 0.639 \pm 0.007 $ \\
J1759+0349   & 4469700519955285760 & $-$9.92  & $-$4 & $11170 \pm 86 $ & $ 7.774  \pm 0.014$ & $ 0.460 \pm 0.010 $ \\
J2025+2839   & 1860318973495642240 & $-$9.03  & $-$4 & $11197 \pm 100$ & $ 7.913  \pm 0.013$ & $ 0.533 \pm 0.010 $ \\
J2058+1657   & 1764481588648685440 & $-$8.38  & $-$1 & $ 6284 \pm 27 $ & $ 8.312  \pm 0.023$ & $ 0.774 \pm 0.022 $ 
\enddata
\end{deluxetable*}

We identify 7 new DQ white dwarfs in our survey. We rely on the photometric technique to determine the best-fit
$T_{\rm eff}$ and $\log{g}$ for these stars, and use the neutral \ion{C}{1} lines or the C$_2$ Swan bands to fit for C/He.
Given the abundances derived from the spectroscopic fit, we repeat our photometric and spectroscopic fits until
a consistent solution is found. Figure \ref{figdq} shows our model fits to two of these stars.
The top and middle panels show our photometric and spectroscopic fits, respectively. The DQ models presented
in \citet{blouin19} provide an excellent match to both the photometry and spectroscopy data for these stars. 

Table \ref{tabdq} presents the best-fit model parameters for all 7 newly identified DQ white dwarfs in our sample.
Six of these are cooler than 8000 K, and display strong C$_2$ Swan bands in their spectra. The remaining object,
J0104+4650, is significantly hotter, and has a relatively noisy MDM spectrum that displays weaker lines. A higher
signal-to-noise ratio spectrum would be helpful in confirming the nature of this system, and constraining its carbon abundance.
The best-fit parameters for J0104+4650, $T_{\rm eff} = 11,257 \pm 166$ K,  $M = 0.997 \pm 0.016~M_{\odot}$, and
$\log{\rm C/He} = -3.21$, place it in the second and more massive DQ white dwarf sequence as discussed in
\citet{blouindufour19}, \citet{koester19}, and \citet{coutu19}. 

\subsection{DZ White Dwarfs}

\begin{figure*}
\center
\includegraphics[width=3.2in, bb=18 144 592 718]{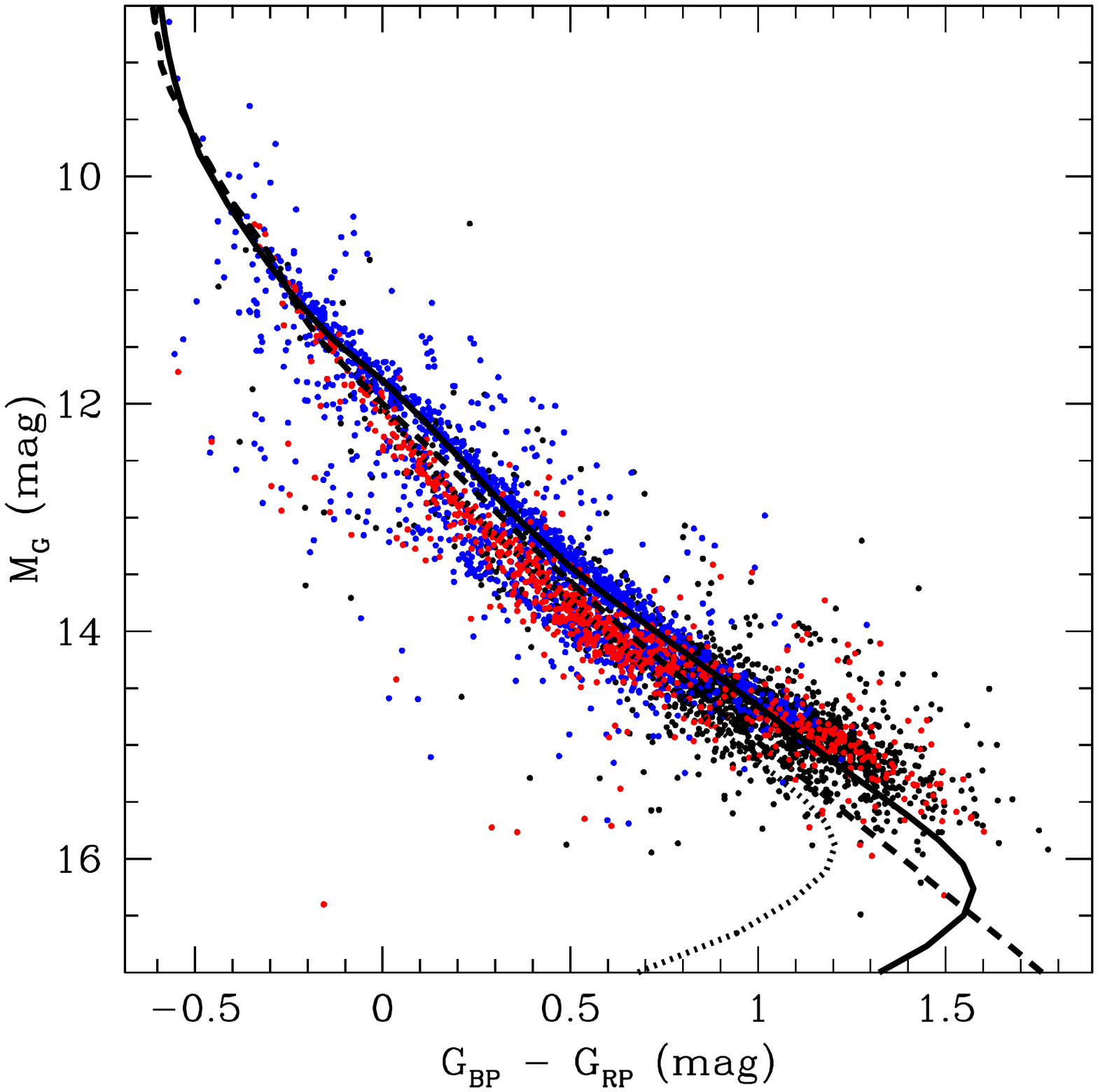} 
\includegraphics[width=3.2in, bb=18 144 592 718]{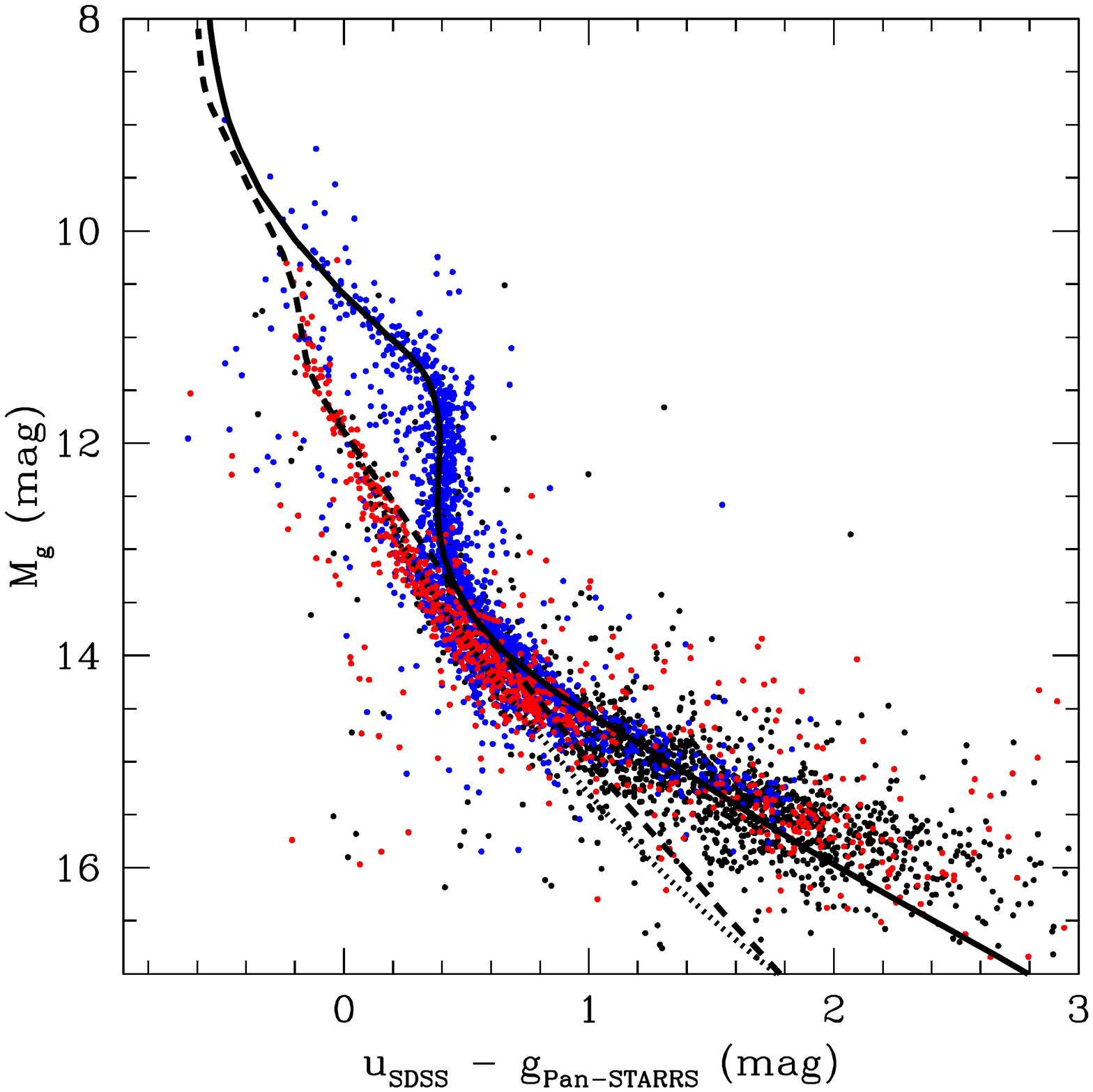}
\caption{Gaia (left) and SDSS + Pan-STARRS (right) color-magnitude diagrams of the 100 pc white dwarf sample
in the SDSS footprint. Blue and red points mark the spectroscopically confirmed DA and non-DA white dwarfs, respectively.
The remaining objects with no follow-up spectroscopy are shown as black dots. The bifurcation between
the DA and non-DA white dwarfs is clearly observed in these diagrams. The solid, dashed, and dotted lines show the cooling
sequences for 0.6 $M_{\odot}$ white dwarfs with pure hydrogen, pure helium, and mixed atmosphere models with
$\log{\rm H/He} = -5$, respectively.}
\label{fighrdall}
\end{figure*}

There are 27 newly identified DZ white dwarfs in our sample that display Ca H and K absorption lines.
We rely on the photometric technique to determine the temperature and surface gravity for these stars, and
fit the blue portion of the spectrum, including the \ion{Ca}{2}  H and K doublet and the \ion{Ca}{1} resonance line, to find Ca/He.
We use the red portion of the spectrum to constrain the H/He abundance ratio.
The abundance ratios of the other heavy elements are scaled to the abundance of Ca to match the abundance ratios of CI
chondrites. Figure \ref{figdz} shows our model fits to two of these stars.  The top and middle panels for each star show our
photometric and spectroscopic fits, respectively. The bottom panels show the entire spectral range of the data.

J0422+0038 (left panels) is a cool DZ white dwarf that shows only Ca absorption features. The absence of a H$\alpha$
absorption feature constrains the H/He ratio in this star, and our model fits provide an excellent match to both the optical
spectra and the photometry of this star. The best-fit model has $\log{\rm H/He} = -3$ and $\log{\rm Ca/He} = -9.34$. 

J1733+3013 (right panels)  is a DZ with both Ca and Mg lines. Our FAST spectrum does not cover H$\alpha$,
but H$\beta$ is clearly absent in this star. Our best-fit model with $\log{\rm Ca/He} = -7.85$ reproduces most
of the observed Ca absorption lines relatively well in the blue. The model fits for the rest of the DZ white dwarfs
are presented in the online version of this article, and the best-fit parameters are presented in Table \ref{tabdz}.
The Ca abundances for these stars range from $\log{\rm Ca/He} = -11.3$ to $-7.9$.

\section{Discussion}

\subsection{The 100 pc White Dwarf Sample in the SDSS Footprint}

With the addition of 711 new spectra from our spectroscopic survey, we have spectral types for 2361 of the
4016 white dwarfs in the 100 pc white dwarf sample in the SDSS footprint. We classify 1539 (or 65\%) as DA, including
85 magnetic white dwarfs, 13 DA(He) stars with He-dominated atmospheres, 1 DAB white dwarf (GD 323), 5 DA + M dwarf systems, and 9 with likely but
uncertain DA: classification. We also identify 53 DB/DBA, 555 (or 24\%) DC, 97 DQ, 115 DZ, 1 cataclysmic variable
(ASASSN-14dx), and 1 dwarf nova (GD 552). 

Figure \ref{fighrdall} shows the distribution of DA and non-DA stars in color-magnitude diagrams based on Gaia (left panel)
and SDSS $u$ and Pan-STARRS $g$ photometry (right panel). The solid, dashed, and dotted lines show the cooling
sequences for 0.6 $M_{\odot}$ white dwarfs with pure hydrogen, pure helium, and mixed atmosphere models with
$\log{\rm H/He} = -5$, respectively. The bifurcation in $u-g$ colors is easily explained by the Balmer jump in pure hydrogen
atmosphere white dwarfs. The Balmer jump suppresses the ultraviolet fluxes of these stars compared to their hydrogen-poor
counterparts, resulting in redder $u-g$ colors. 

The split in the Gaia sequence is also explained by the presence of hydrogen (and/or other electron donors), but in this case
it is the additional opacity from trace amounts of hydrogen in helium-dominated white dwarfs that affects their colors
\citep{bergeron19}. The $\log{\rm H/He} = -5$ mixed atmosphere model sequence, for example, goes through the bulk of
the DC white dwarfs in the Gaia color-magnitude diagram, except the coolest ones. Regardless of the reason (additional
hydrogen or other electron donors), the differences in atmospheric composition can clearly explain the bifurcation seen in
these color-magnitude diagrams.

\begin{figure}
\center
\includegraphics[width=3.2in, bb=18 144 592 718]{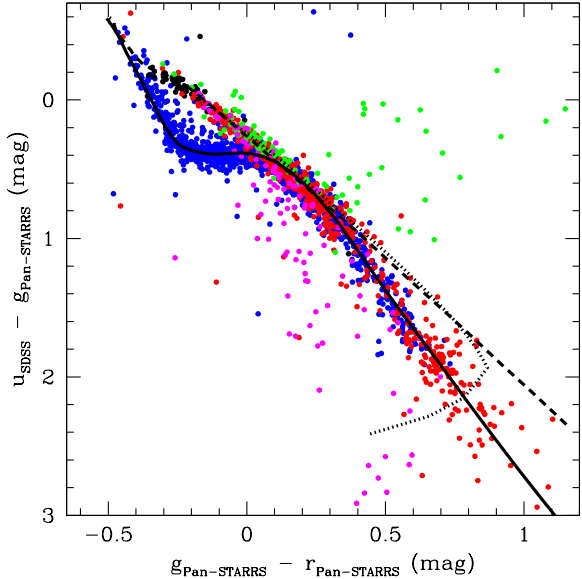} 
\caption{A color-color diagram of the DA (blue), DB (black), DC (red), DQ (green), and DZ (magenta) white dwarfs in
the 100 pc sample and the SDSS footprint. The model sequences are the same as in Figure \ref{fighrdall}. DA and
DB/DC white dwarfs form relatively tight sequences in this diagram, whereas many of the DQ and DZ stars show up as outliers.}
\label{figugr}
\end{figure}

Figure \ref{figugr} shows a color-color diagram of our white dwarf sample based on the SDSS $u$ and Pan-STARRS $g,r$
filters. DA and DB/DC white dwarfs form relatively tight color sequences in this diagram, with the main distinction between them
being the Balmer jump visible in the $u-g$ color for DA white dwarfs. On the other hand, the majority of the DQ (green)
and DZ (magenta) white dwarfs appear as outliers with respect to the DA and DB/DC sequences. 

DZ white dwarfs suffer from metal absorption features mostly in the ultraviolet, which result in redder $u-g$ colors compared to DC stars, whereas
the Swan bands in DQ white dwarfs are strongest in the $g-$filter, resulting in bluer $u-g$ and/or redder $g-r$ colors compared
to the DC white dwarfs. Hence, follow-up spectroscopy of the outliers in similar color-color diagrams may provide an efficient
method to identify additional DQ and DZ white dwarfs in the solar neighborhood.

\begin{figure}
\center
\includegraphics[width=3.2in, bb=18 144 592 718]{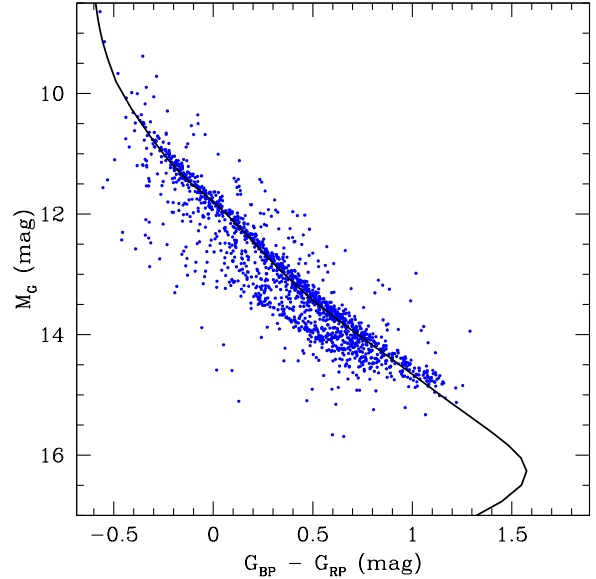}
\caption{Color-magnitude diagram of the spectroscopically confirmed DA white dwarfs in the 100 pc sample and the
SDSS footprint. The solid line shows the cooling sequence for a $0.6~M_{\odot}$ pure hydrogen atmosphere white dwarf.}
\label{fighrdda}
\end{figure}

A striking feature visible in Figures \ref{fighrdall} and \ref{figugr} is that the cooling sequences for 0.6 $M_{\odot}$ pure
helium and mixed H/He atmosphere white dwarfs fail to match the colors of the bulk of the cool DC white dwarfs
with $u-g>0.8$ mag, or $T_{\rm eff}<6000$ K. Hence, a photometric analysis based on helium-rich models produce
mass estimates that are too low \citep[see Figure 12 in][]{bergeron19}. On the other hand, the pure hydrogen atmosphere models
including the red wing of the Lyman $\alpha$ line \citep{kowalski06} provide a much better fit to the $u-g$ colors of the
cool DC white dwarfs in Figures \ref{fighrdall} and \ref{figugr} and provide more normal mass estimates, justifying our
use of pure hydrogen model fits for the coolest ($T_{\rm eff} < 5000$ K) DCs in our sample.

A caveat in our analysis is that some of the cool DCs likely have helium-dominated atmospheres. 
Comparing the pure hydrogen, pure helium, and mixed H/He atmosphere solutions for each star, \citet{blouin19}
found the fraction of helium-rich DCs to be around 25\%. Based on the $u-g$ colors, the fraction of DC stars that
have helium-rich atmospheres appears to be small in our sample. However, a detailed model
atmosphere analysis of individual DC white dwarfs is required for reliable constraints on their atmospheric parameters.

In the following sections, we present the DA white dwarf mass distribution, and then discuss the properties of the DC white
dwarfs, specifically the so-called IR-faint white dwarfs in our sample.

\subsection{The DA Mass Distribution}

Figure \ref{fighrdda} shows a color-magnitude diagram of all spectroscopically confirmed DA white dwarfs in our sample.
The DA white dwarf sequence ends at an absolute magnitude of $M_G \approx 15$, which corresponds to
$T_{\rm eff}=5000$ K for $0.6~M_{\odot}$ white dwarfs. H$\alpha$ disappears below this temperature. Excluding the DA white dwarfs
with weak lines, mixed atmospheres, e.g., DA(He) stars, uncertain spectral types, or M dwarf companions, we have 1508 spectroscopically confirmed DA white dwarfs in our sample that are best-fit by pure hydrogen atmosphere models. Most of these
DA white dwarfs form a tight sequence around $M = 0.6~M_{\odot}$, but a significant contribution from
fainter and more massive white dwarfs, many of which are on the the crystallization sequence
\citep[also referred to as the Q-branch,][]{gaia18,tremblay19}, is also clearly visible. 

\begin{figure}
\center
\includegraphics[width=3.2in, bb=18 144 592 718]{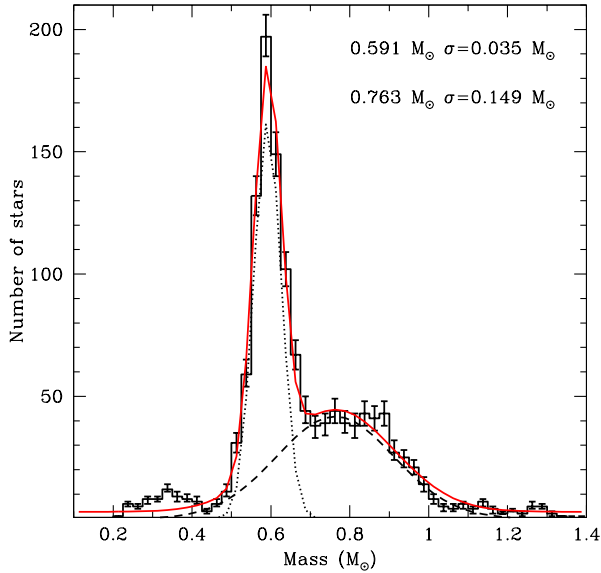} 
\caption{The white dwarf mass distribution (black solid histogram) based on 1337 spectroscopically confirmed DA
white dwarfs hotter than 6000 K in the 100 pc sample and the SDSS footprint. Dotted and dashed lines show the
best-fitting Gaussians to the main peak and the broad shoulder, and the red line shows the composite of the two.}
\label{figmassda}
\end{figure}

\begin{figure*}
\center
\includegraphics[width=3.2in, bb=18 144 592 718]{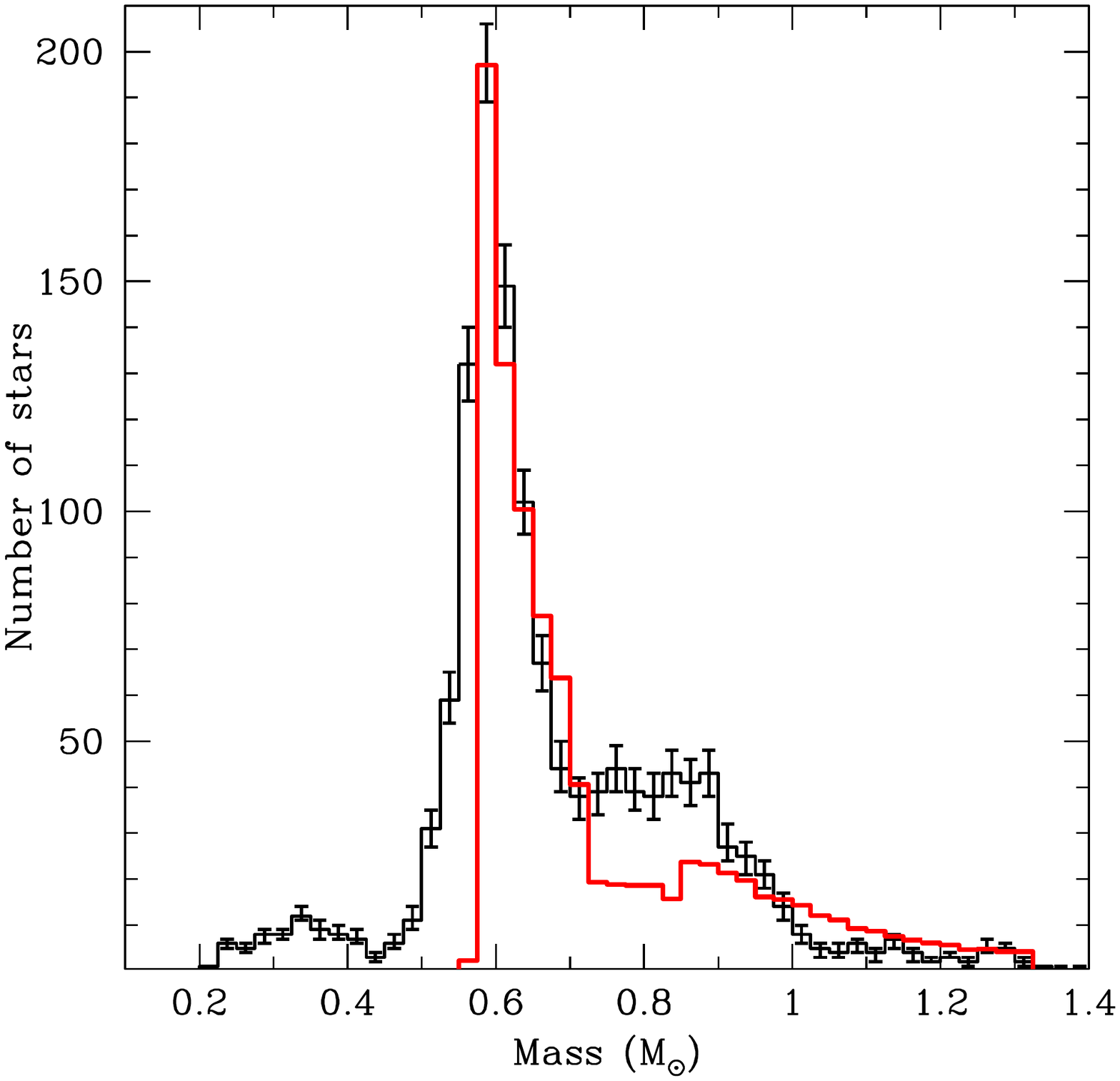} 
\includegraphics[width=3.2in, bb=18 144 592 718]{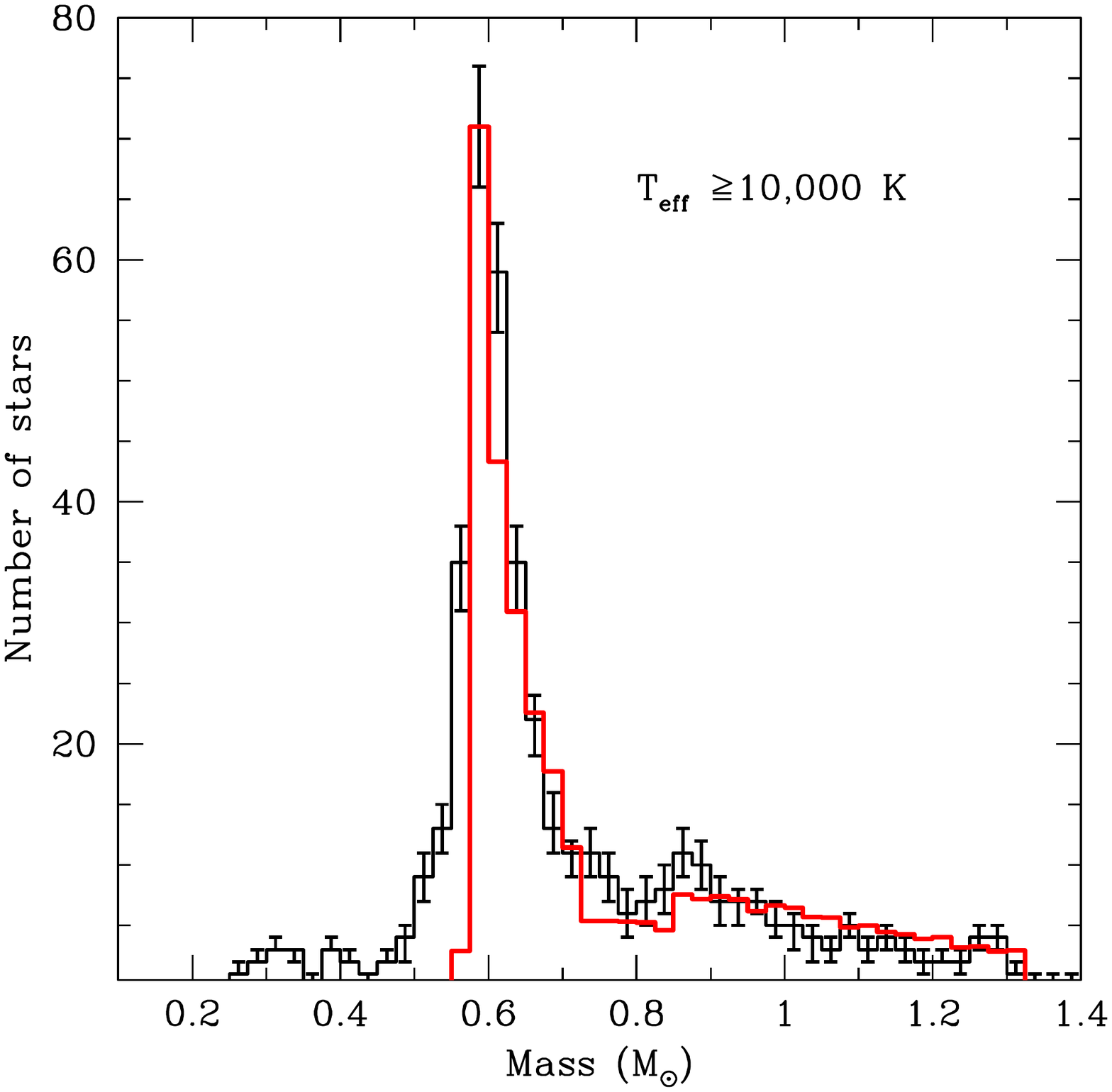}
\caption{{\it Left panel:} The DA white dwarf mass distribution (black histogram) compared to the
predicted mass distribution for a 10 Gyr old disk population that evolved in isolation
(red line). {\it Right panel:} The same as in the left panel, but for a more restrictive sample of white dwarfs with
$T_{\rm eff}\geq 10,000$ K.}
\label{figmassmodel}
\end{figure*}

Figure \ref{figmassda} shows the mass distribution of our DA white dwarf sample. Since our spectroscopic completeness
falls significantly beyond 6000 K, we limit our mass distribution to 1337 stars with $T_{\rm eff}\geq 6000$ K.
We estimate the uncertainties in each
mass bin based on a Monte Carlo analysis where we replace the mass of each white dwarf $M$ with $M + g~\delta M$,
where $\delta M$ is the error in mass and $g$ is a Gaussian deviate with zero mean and unit variance.   
For each of 1000 sets of modified masses for the entire sample, we calculate the mass distribution, and
we take the range that encompasses 68\% of the probability distribution function as the $1\sigma$ uncertainties for each
mass bin. 

Remarkably, the mass distribution has an extremely narrow peak and a broad shoulder. Fitting two Gaussians to the observed
distribution, we find that the mean peak is centered at 0.59 $M_{\odot}$ with a $1\sigma$ spread of only $0.035~M_{\odot}$,
whereas the contribution from massive white dwarfs is best-fit with a Gaussian at 0.76 $M_{\odot}$ and a $1\sigma$ spread
of $0.15~M_{\odot}$.There is also a low mass peak around 0.35 $M_{\odot}$ in the mass distribution, but it is smaller than previously observed in magnitude-limited samples like the PG survey \citep{liebert05}. This is not surprising as low-mass white
dwarfs are over-represented in magnitude-limited surveys due to their larger radii and luminosities. A caveat here is that
our mass estimates for low-mass white dwarfs may be erroneous since they were based on the assumption of a single star,
and that low-mass white dwarfs are usually found in binary systems \citep{marsh95,kilic20}. Hence, further observations
would be useful for constraining the binary nature of these systems and the mass distribution below 0.5 $M_{\odot}$.

Figure \ref{figmassmodel} compares the DA mass distributions for $T_{\rm eff}\geq 6000$ K (left panel) and a more restrictive
$T_{\rm eff}\geq 10,000$ K (right panel) sample of stars against the predictions from a 10 Gyr old disk population with
a constant star-formation rate (SFR, solid line) and a Gaia magnitude limit of $G = 20$. We use a Salpeter mass function
\citep{salpeter55}, assume solar metallicity, and that the stars evolve in isolation. We use the main-sequence + giant-branch
lifetimes from \citet{hurley00} to decide if a star evolves into a white dwarf within 10 Gyr, and if so, to estimate its cooling age.

We calculate the final white dwarf masses based on the latest MIST-based piece-wise initial-final mass relation (IFMR)
from \citet{cummings18}, and use the evolutionary models\footnote{See http://www.astro.umontreal.ca/$\sim$bergeron/CoolingModels.} similar to those described in \citet{fontaine01} with (50/50) C/O-core compositions,
$q({\rm He})\equiv M_{\rm He}/M_{\star}=10^{-2}$, and $q({\rm H})=10^{-4}$, to estimate their temperatures.
Just like in the observed
samples, we only include white dwarfs with $T_{\rm eff}\geq 6000$ K (left panel) or 10,000 K (right panel) in these simulations.
We normalize the disk model to match the peak of the observed mass distribution at $0.59~M_{\odot}$.
Our simulations fail to account for the relatively large numbers of low-mass ($M<0.5~M_{\odot}$) white dwarfs. This is not
surprising, as the majority of these low mass white dwarfs are likely in binary white dwarf systems \citep{marsh95}, which
are excluded from our synthetic populations. 

This figure reveals remarkable differences between the mass distributions for the hot DAs and the overall population.
The mass distribution for the hot DAs (right panel) shows a roughly linear decline in the number of massive white dwarfs beyond
0.7 $M_{\odot}$, and the 10 Gyr old disk model of single stars with constant star formation provides an excellent match
to this mass distribution. Note that we simply scaled the model predictions to match the peak of the observed mass distribution,
hence the excellent match between the observations and the predictions demonstrate that our choices of a Salpeter IMF,
a constant SFR, the IFMR, and the white dwarf cooling models are appropriate for simulating white dwarf populations in the
solar neighborhood. At 10,000 K, only white dwarfs more massive than $1~M_{\odot}$ would have crystallized. Hence,
the uncertainties in the cooling models due to crystallization and its associated effects (see below) are mostly avoided
for hot white dwarfs.

Even though both the hot DA and the overall DA mass distributions peak at $0.59~M_{\odot}$, the latter (left panel) is
significantly different: it shows a nearly constant number of massive white dwarfs between 0.7 - 0.9 $M_{\odot}$, and
then a sudden decrease to nearly zero at 1 $M_{\odot}$ and beyond.  The synthetic populations fail to match this distribution.
This figure demonstrates that for cool white dwarfs there is a mechanism that leads to a pile-up
of massive white dwarfs with $M \sim 0.8~M_{\odot}$ and a sudden disappearance of more massive white dwarfs.

\begin{figure}
\center
\includegraphics[width=3in, bb=18 144 592 718]{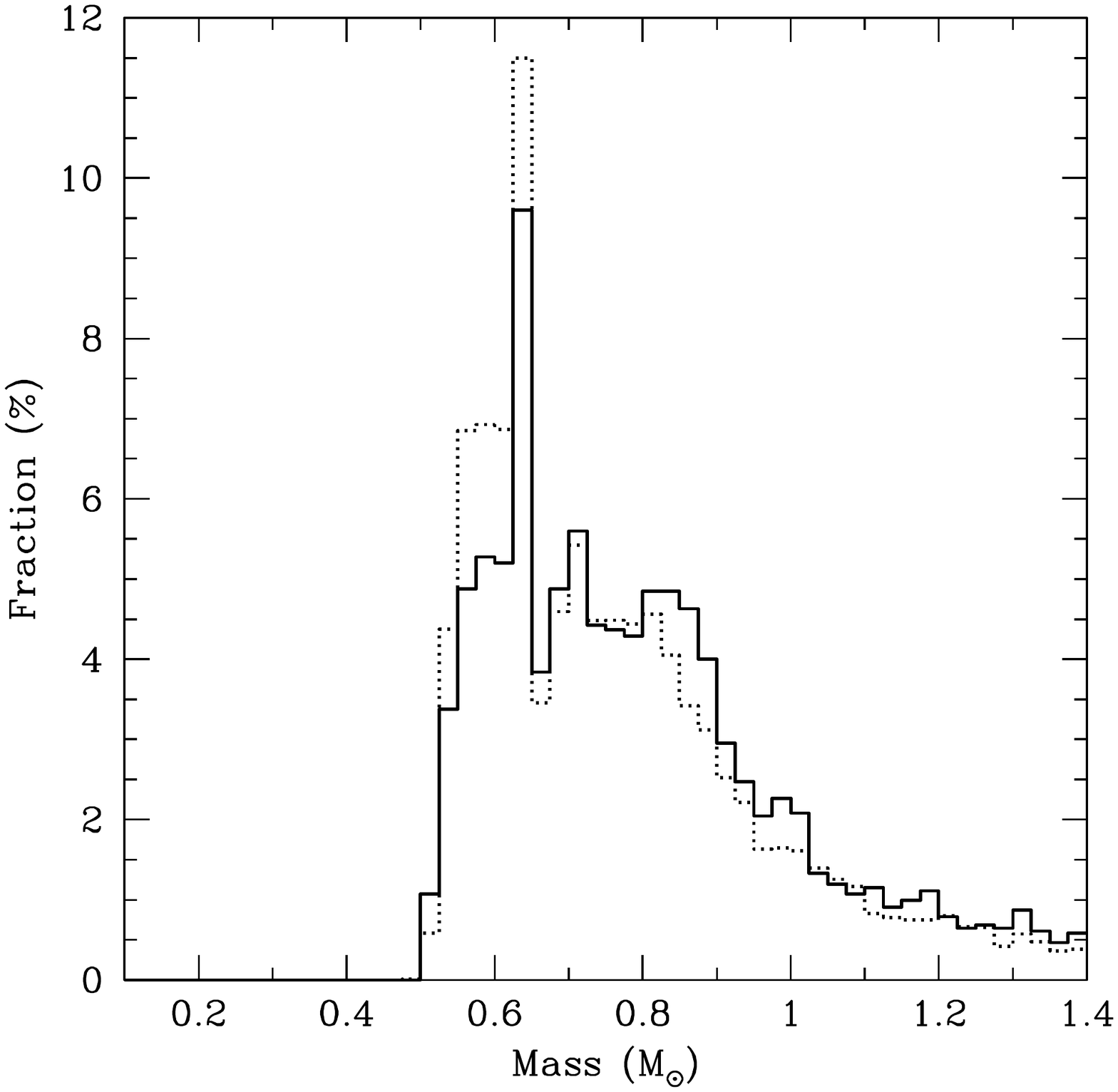}
\includegraphics[width=3in, bb=18 144 592 718]{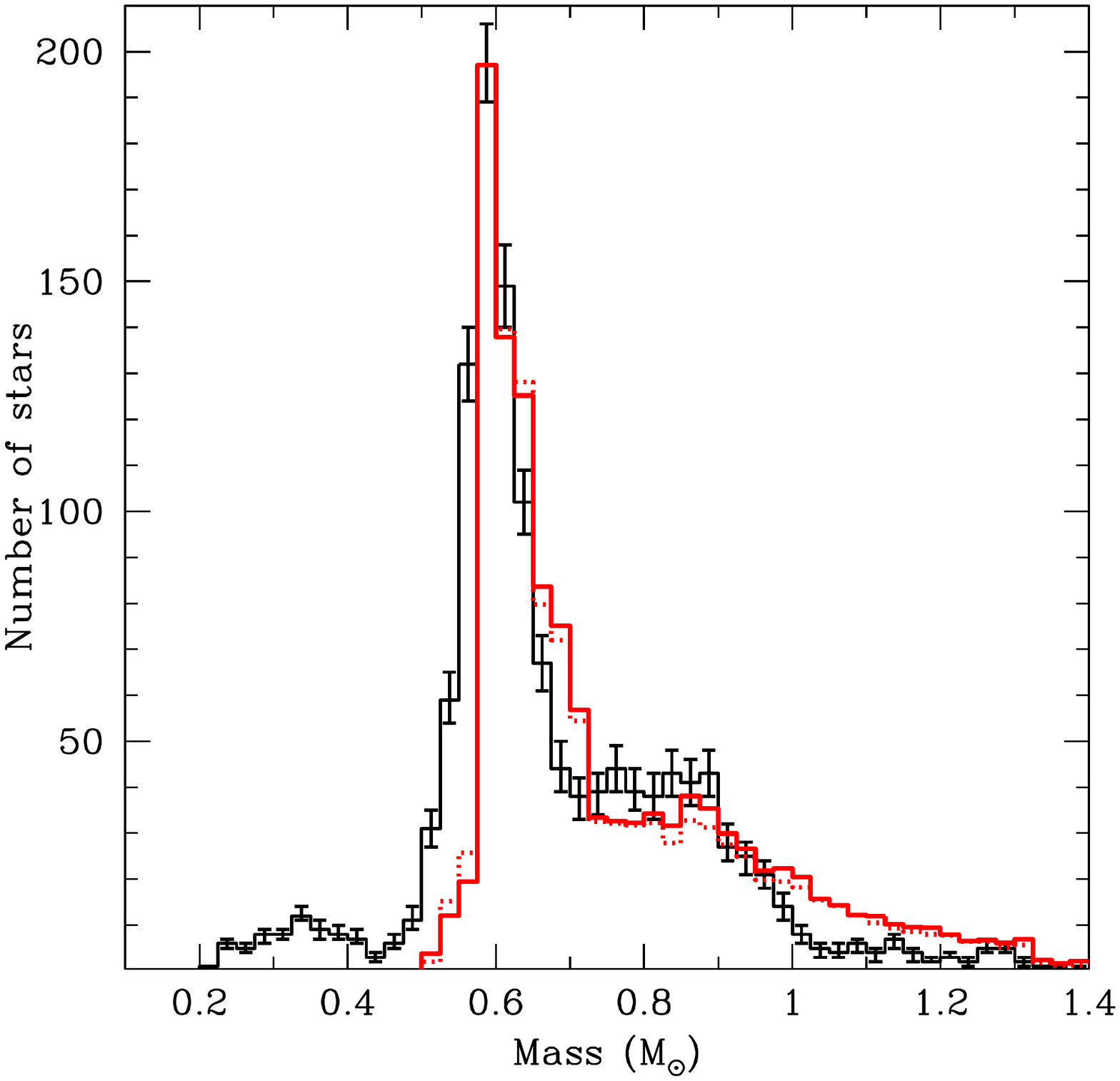} 
\caption{{\it Top panel:} The predicted mass distribution for single white dwarfs that form through mergers during their
main-sequence or post-main-sequence evolution. The solid and dotted lines show the results assuming an
initial lognormal and flat period distribution for the progenitor main-sequence stars, respectively.
{\it Bottom panel:} The DA white dwarf mass distribution (black histogram) compared to the predicted
mass distribution with 30\% contribution from mergers (red lines, the solid and dashed lines represent
the same period distributions as in the top panel). Both observations and simulations shown here are restricted
to white dwarfs with $T_{\rm eff}\geq 6000$ K.}
\label{figmerger}
\end{figure}

\begin{figure*}
\center
\includegraphics[width=3.2in, bb=18 144 592 718]{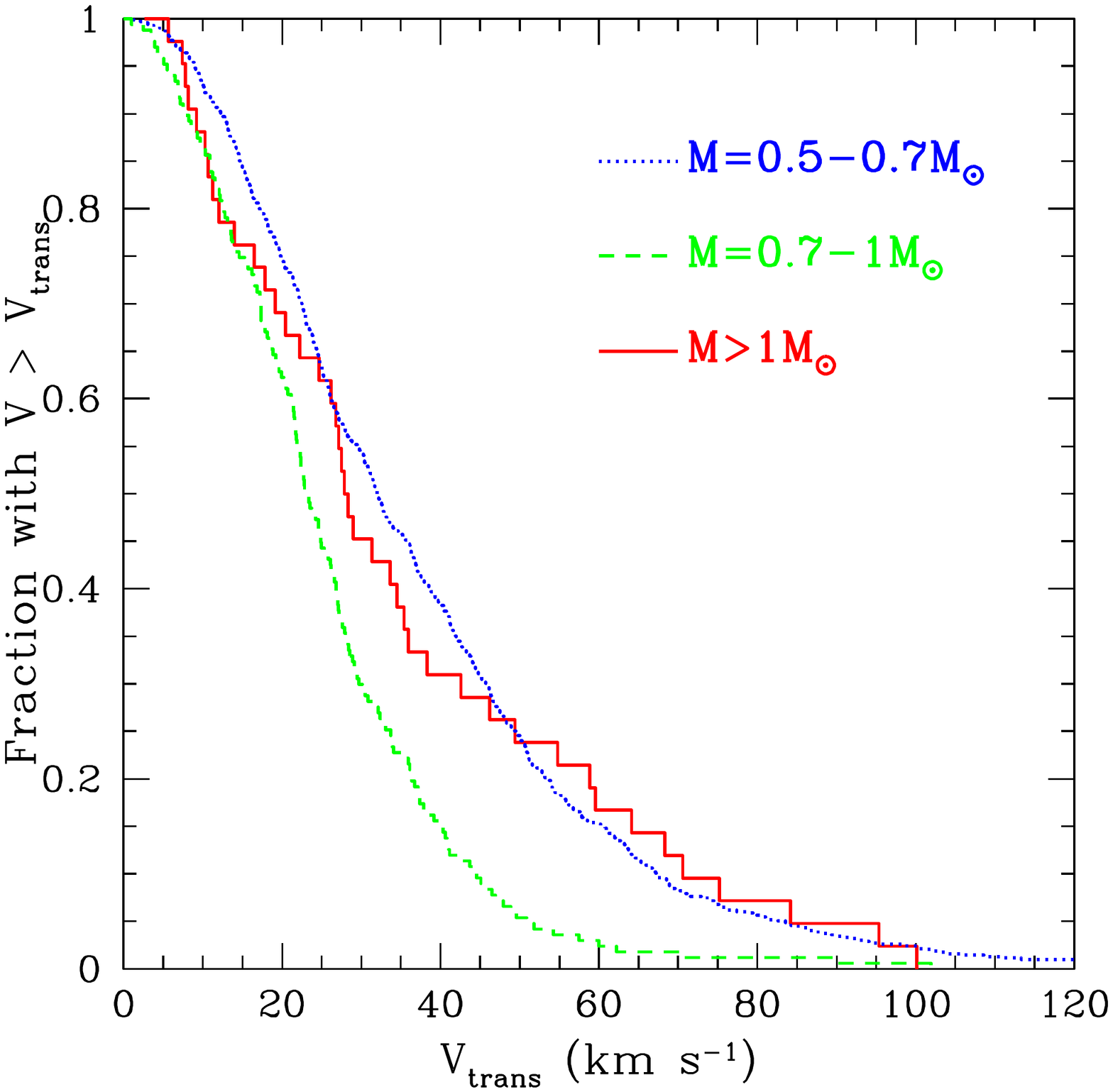}
 \includegraphics[width=3.2in, bb=18 144 592 718]{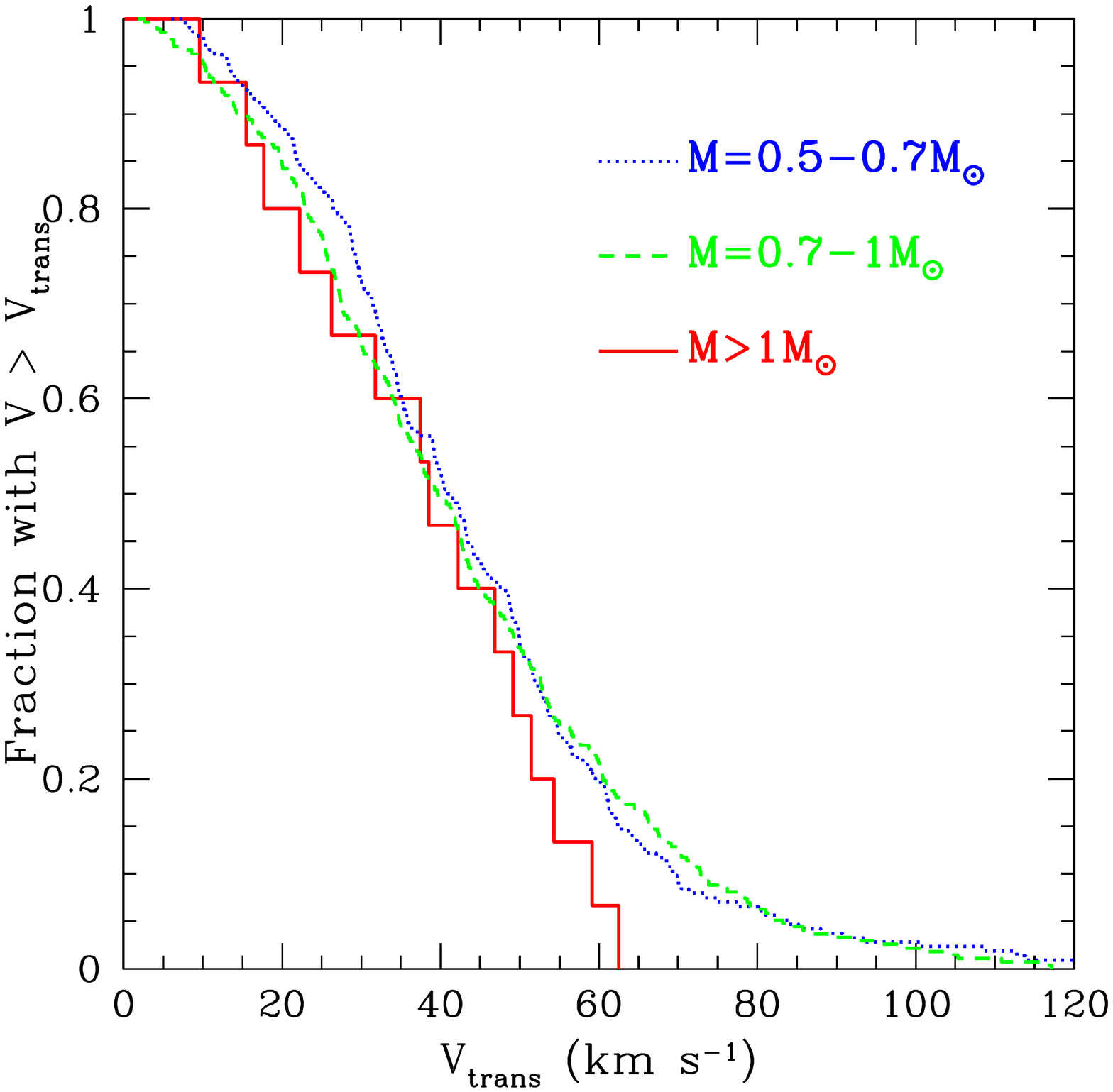}
\caption{Cumulative distribution of transverse velocities for DA white dwarfs with cooling ages (assuming single star evolution)
younger than (left panel) and older than (right panel) 2 Gyr and with $M=$ 0.5-0.7, 0.7-1, and $>1~M_{\odot}$.}
\label{figvtan}
\end{figure*}

\subsubsection{The Effects of Mergers on the Mass Distribution}

\citet{liebert05} interpreted the excess of massive white dwarfs in the Palomar-Green Survey of DA white dwarfs as
evidence of binary evolution. \citet{toonen17} performed binary population synthesis calculations to show that up to
23\% of the white dwarfs in the solar neighborhood might have descended from binary mergers during (post-)main-sequence
evolution. Recently, \citet{temmink20} found that binary mergers can account for 10-30\% of all single white dwarfs.
However, they found that these mergers should not significantly alter the shape of the white dwarf mass distribution.

Now that we have a precise mass distribution based on 1337 spectroscopically confirmed white dwarfs with
$T_{\rm eff}\geq 6000$ K in a volume-limited
sample, we can directly compare the predictions from the binary population synthesis calculations to see if mergers can
explain the over-abundance of massive white dwarfs in our sample. 

We use the binary-star evolution (BSE) algorithm of
\citet{hurley02} to track the evolution of $10^5$ main-sequence binaries with primary masses randomly drawn between
0.8 and $8~M_{\odot}$ from a Salpeter mass function and the secondary masses drawn from a uniform mass ratio distribution
between 0 and 1. We assume constant star formation rate, and track the evolution of these systems over 10 Gyr. We perform
simulations with two different period distributions; lognormal and flat \citep{toonen17}. We use the evolutionary times from
BSE to estimate the cooling age of each single white dwarf that formed through mergers, and estimate its current temperature
based on the white dwarf cooling models.  

Figure \ref{figmerger} top panel shows the predicted mass distribution for single white dwarfs that form as a result
of mergers. The results from the lognormal and flat period distributions are shown as solid and dotted lines, respectively.
The overall shapes of the predicted mass distributions are similar; both predict a dominant peak around 0.6 $M_{\odot}$
with a tail towards higher masses. The main difference between the simulations using lognormal and flat period distributions is the
number of systems that form through mergers \citep[see,][for a detailed discussion]{toonen17}.

To see if mergers can help explain a significant portion of the massive white dwarfs in our sample, we compare our DA
mass distribution against the predictions for blended populations of 70\% white dwarfs from a 10 Gyr old disk population
that evolved in isolation + 30\% single white dwarfs that formed through mergers. We picked 30\% contribution from
mergers, as this is the upper limit in the binary population synthesis calculations from \citet{temmink20}.

Figure \ref{figmerger} bottom panel shows a comparison between our 100 pc DA white dwarf mass distribution against
these blended populations, the solid red line for the lognormal period distribution and the dotted line for the flat period distribution.
Note that both observations and simulations shown here are restricted to white dwarfs with $T_{\rm eff}\geq 6000$ K.
Just like in Figure \ref{figmassmodel}, we normalize the synthetic populations to match the peak of the
observed mass distribution at 0.59 $M_{\odot}$.  

Even though the blended populations help increase the number of $M\sim0.8~M_{\odot}$ white dwarfs produced in
the simulations and provide a better fit to the observational data in that mass range, the overall fit to the mass distribution
is still problematic as these blended populations also produce more of the $M\sim0.7$ and $>1~M_{\odot}$ white dwarfs. 
The observed sample shows significant deficits for both $M\sim0.7$ and $>1~M_{\odot}$ white dwarfs compared to the
model predictions.  

\subsubsection{The Transverse Velocity Distribution}

Merger populations can also reveal themselves through their kinematics. Double white dwarf merger products should
have higher velocity dispersion because they would be part of an older population that is kinematically heated up
\citep{coutu19,cheng20}.  Studying the kinematics of hot ($T_{\rm eff}>13,000$ K) DA white dwarfs from the PG survey
and the SDSS DR4 white dwarf catalog \citep{eisenstein06}, \citet{wegg12} found that the transverse velocity dispersion
decreases from $\sigma \sim 50$ km s$^{-1}$ for 0.5 $M_{\odot}$ white dwarfs to $\sim$20 km s$^{-1}$ for 0.8 $M_{\odot}$
white dwarfs. This is consistent with the expectations from single star evolution as the progenitors of lower mass white
dwarfs belong to an older population where the disc heating is significant. 

\citet{wegg12} did not find any high mass white dwarfs traveling at $>50$ km s$^{-1}$ in their young DA sample, and
concluded that the observed kinematics are consistent with the majority
of high mass white dwarfs forming through single star evolution. This is similar to the conclusions reached by \citet{cheng20},
who estimate  that the fraction of massive white dwarfs that form through double white dwarf mergers is around 10\% in the
0.8-0.9 $M_{\odot}$ mass range. The exception to this is the massive DQ white dwarfs that display significantly larger transverse
velocities, with 45\% displaying velocities in excess of 50 km s$^{-1}$ \citep{coutu19}. Hence, many of the massive DQ
white dwarfs are likely produced through double white dwarf mergers. 

Figure \ref{figvtan} shows the cumulative distribution of transverse velocities for our DA white dwarf sample split into three
mass ranges, which correspond to the main peak in the mass distribution (0.5-0.7 $M_{\odot}$), the broad shoulder at 0.7-1
$M_{\odot}$, and very massive white dwarfs with $M > 1~M_{\odot}$. The left and right panels show the distributions for
the young and old DA samples with cooling ages (assuming single star evolution) younger than and older than 2 Gyr,
respectively. The transverse velocities are based on Gaia DR2 parallaxes and proper motions. 

Looking at the cumulative transverse velocity distribution for the younger DAs (left panel), the velocity dispersion decreases
for 0.7-1 $M_{\odot}$ white dwarfs compared to the 0.5-0.7 $M_{\odot}$ population. This is similar to what is seen in
\citet{wegg12}, which suggests that the mergers are unlikely to explain the number of 0.7-1 $M_{\odot}$ white dwarfs.
On the other hand, the velocity distribution for the most massive white dwarfs with $M > 1~M_{\odot}$ shows an excess
of higher velocity objects compared to the 0.7-1 $M_{\odot}$ sample: 10 of the 44 hot DA white dwarfs with $M > 1~M_{\odot}$
have $V_{\rm trans}>50$ km s$^{-1}$. This would indicate a substantial contribution from mergers for the most massive white
dwarfs, and is consistent with the results from \citet{temmink20} and \citet{cheng20}, who found that the fraction of
$> 1~M_{\odot}$ white dwarfs that come from mergers is $\sim$30\%. 

\begin{figure}
\center
\includegraphics[width=3.2in, bb=18 144 592 718]{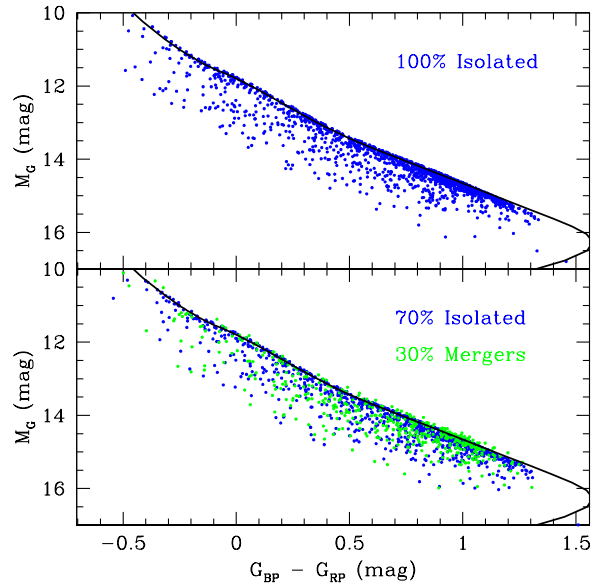} 
\caption{Synthetic color-magnitude diagrams for pure hydrogen atmosphere white dwarfs for a 10 Gyr old disk population and assuming a constant star formation rate. The top panel is for stars that evolved in isolation, whereas the bottom panel includes
single white dwarfs that formed through mergers. Each panel displays the same number of stars as in Figure \ref{fighrdda}, for
a fair comparison. The solid line shows the cooling sequence for a $0.6~M_{\odot}$ pure hydrogen atmosphere white dwarf.}
\label{figcmd}
\end{figure}

\begin{figure*}
\center
\includegraphics[width=8in,bb=70 120 800 500]{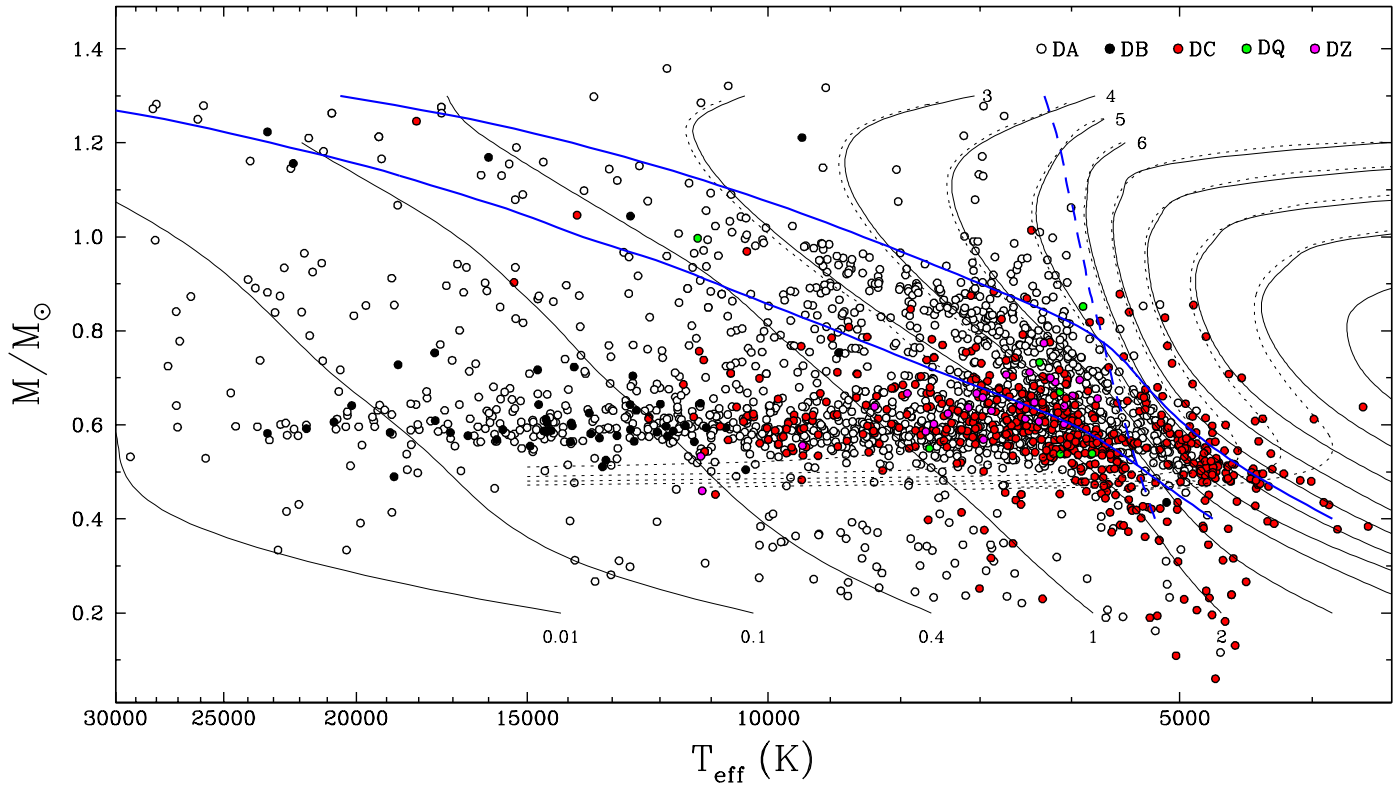} 
\caption{Stellar masses as a function of effective temperature for all spectroscopically confirmed white dwarfs in the
100 pc sample and the SDSS footprint. The parameters have been determined using photometric fits to the SDSS $u$ and
Pan-STARRS $grizy$ photometry and Gaia DR2 parallaxes. Solid curves are theoretical isochrones, labeled
in units of $10^9$ years, obtained from cooling sequences with C/O-core compositions,
$q({\rm He}) \equiv M({\rm He})/M_{\star} = 10^{-2}$, and $q$(H) = $10^{-4}$, while the dotted curves correspond to isochrones
with the main sequence lifetime taken into account. The lower blue solid curve indicates the onset of crystallization at the
center of evolving models, while the upper one indicates the locations where 80\% of the total mass has solidified.
The dashed curve indicates the onset of convective coupling. Note that the low-mass DA white dwarfs in this diagram are
most likely in unresolved degenerate binary systems, and their mass estimates may be erroneous since they were based on the
assumption of a single star.}
\label{figtm}
\end{figure*}

The transverse velocity distribution for the older DAs (right panel) is remarkably different. Here the distributions
for the 0.5-0.7 and 0.7-1 $M_{\odot}$ white dwarfs are almost identical, but the velocity distribution for the
$M>1~M_{\odot}$ cool DA sample in Figure \ref{figvtan} is significantly different: it is missing the high velocity tail
visible in the velocity distributions of the $M<1~M_{\odot}$ stars. This means that we are missing the oldest
(and therefore the coolest) massive DA white dwarfs. Based on the velocity distributions for
both hot and cool DAs, we can safely conclude that mergers cannot explain the relatively large numbers of
$M\sim 0.8~M_{\odot}$ white dwarfs, and we must look for an alternative explanation for these systems.

Another way to demonstrate this is through color-magnitude diagrams. Figure \ref{figcmd} displays synthetic color-magnitude
diagrams for pure hydrogen atmosphere white dwarfs for a 10 Gyr old disk population with constant SFR and a Gaia magnitude
limit of $G = 20$. The top panel assumes evolution in isolation, whereas the bottom panel includes single white dwarfs that
form through mergers. A comparison with Fig. \ref{fighrdda} shows that none of the simulated populations, including the one
with the mergers, can explain the observations. 

\subsubsection{Crystallization}

\citet{tremblay19} identified a pile up in the cooling sequence of white dwarfs in Gaia color-magnitude
diagrams, and demonstrated that the latent heat of crystallization causes this pile up. The release of latent heat results in
a slowdown in the evolution of a white dwarf. Since massive white dwarfs crystallize first, the effects of crystallization is
most significant in their number distribution \citep[see also][]{bergeron19}. 

\citet{bergeron19} note that the most significant effect of crystallization is
actually the Debye cooling phase, where the transition from the classical regime to the quantum regime occurs. 
In the quantum regime, the specific heat decreases with cooling, rapidly depleting the thermal energy reservoir of the star,
and making it essentially disappear from view as a black dwarf. Again, this is most significant for massive white dwarfs.

Our sample contains 44 DA white dwarfs with $T_{\rm eff}>10,000$ K and cooling ages less than 2 Gyr, and only
15 DAs with $T_{\rm eff}=$ 6000-10,000 K and cooling ages more than 2 Gyr. A remaining question is whether the
near absence of $M>1~M_{\odot}$ DA white dwarfs at cooler temperatures is due to those stars turning into massive
DC stars or not.  

Figure \ref{figtm} shows the stellar masses as a function of temperature
for all spectroscopically confirmed DA and non-DA white dwarfs in the 100 pc sample and the SDSS footprint. 
We restrict this figure to the objects with $T_{\rm eff} <$ 30,000 K, where the photometric technique is the most reliable
\citep{genest19}. Theoretical isochrones, with and without the main sequence lifetime taken into account \citep{bergeron01},
are shown as solid and dotted curves, respectively. The nearly horizontal isochrones for massive white dwarfs at cooler
temperatures indicate the Debye cooling regime, where the thermal reservoir of the star is rapidly depleted. This occurs
below about $T_{\rm eff}=$ 6000 K for $M\geq 1~M_{\odot}$ white dwarfs.

The lower solid blue curve marks the region where crystallization starts at the center of the
white dwarf, and with further cooling, the solidification front moves upward. By the time $\sim80$\% of the star has solidified,
most of the latent heat would have been spent, and the upper solid blue curve marks this limit of 80\% crystallization.
The dashed curve marks the onset of convective coupling, when the convection zone first reaches into the degenerate interior,
where all of the thermal energy of the star resides. Convective coupling significantly changes the cooling rates of white dwarfs, as
evidenced by the sudden change in slope of the isochrones shown here.

Figure \ref{figtm} demonstrates that the pile up of massive white dwarfs seen in our 100 pc sample is clearly due to
delays in cooling from the latent heat of crystallization. We see an enhanced number of white dwarfs between the
two solid blue curves, where most of the latent heat of crystallization would have been released. 

Just like the DA white dwarfs, the mass distribution for the non-DA stars is also dominated by 0.6 $M_{\odot}$ white dwarfs,
with a small number of massive DC white dwarfs also present. Including trace amounts of hydrogen in our DC white dwarf
fits leads to normal mass estimates for these stars, with no bias towards mass estimates lower than 0.5 or 0.4 $M_{\odot}$
at the cool end of the distribution. 

Interestingly, there are only four massive DC white dwarfs
with masses near 1 $M_{\odot}$. Three of these, J1621+0432, J1039$-$0325, and J1105+5225, are warmer than 13,000 K.
Hence, to be classified DC, they must be strongly magnetic DAH or DBH white dwarfs. The remaining object, J0930+0628,
has $T_{\rm eff} = 6418 \pm 54$ K and $M=1.01 \pm 0.02~M_{\odot}$ assuming a composition of $\log{\rm H/He}=-3.2$.
J0930+0628 appears to be the only genuine DC white dwarf more massive than 1 $M_{\odot}$. 

Note that there are a number of cool and massive DA stars in the 40 pc sample of \citet[][see their Fig. 21]{limoges15}.
However, the masses for those objects are mostly determined spectroscopically. For those in common (18 objects cooler
than 8000 K and with $M >0.8~M_{\odot}$ based on spectroscopy),  we find that the photometric mass is much smaller
than the spectroscopic mass for all objects with spectroscopic masses above $0.9~M_{\odot}$. This suggests that these
objects are either double degenerate systems or He-rich DAs. The fact that the photometric and spectroscopic temperature
estimates agree for most of these objects would favor the latter explanation. Alternatively, a relatively weak magnetic
field can also broaden the lines without displaying explicit Zeeman splitting in the low-resolution spectra, leading to erroneously
high mass estimates. However, two of these objects, J1022+4600 and J1420+5322, have photometric mass
estimates below $0.5~M_{\odot}$, and therefore they are likely in binary systems.

\begin{figure}
\center
\includegraphics[width=4in,bb=60 40 630 740]{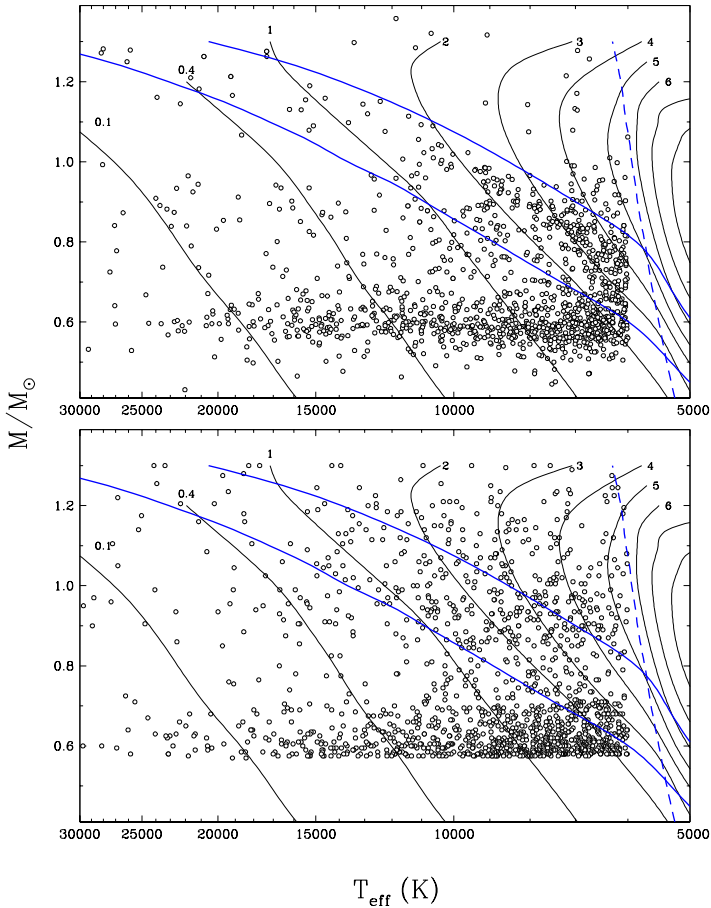}
\caption{Observed (top, similar to Fig. \ref{figtm}) and predicted (bottom) white dwarf mass and temperature distribution
for a 10 Gyr old disk population with a constant star formation rate based on the current evolutionary models. Also shown
are the same isochrones, crystallization, and convective coupling curves as in Figure \ref{figtm}. We limit this figure
to stars hotter than 6000 K, since our follow-up spectroscopy is severely incomplete below that temperature.}
\label{figber}
\end{figure}

\subsubsection{Evolutionary Model Predictions}

To check if the current evolutionary models \citep{fontaine01} provide a reasonable match to the significance of the pile up
of $M\sim0.8~M_{\odot}$ white dwarfs in the 100 pc sample, we show the observed (top panel, similar to Figure \ref{figtm})
and the predicted (bottom panel) DA white dwarf mass and temperature distribution for a 10 Gyr old disk population
with a constant star formation rate in Figure \ref{figber}. These are the same
evolutionary models used in \citet{bergeron19} and \citet{tremblay19}. 

Since our spectroscopic follow-up is severely incomplete beyond 6000 K, here we restrict our comparison to stars
with $T_{\rm eff}\geq6000$ K. Gaia's limiting magnitude also comes into play below this temperature. A 1 $M_{\odot}$ pure
hydrogen atmosphere white dwarf is predicted to have $M_{\rm G} = 15$ mag at 6000 K. Hence, Gaia DR2 catalog suffers from incompleteness issues, even within 100 pc, for massive white dwarfs below this temperature. We limit our simulations to
objects brighter than $G=20$ mag to account for this bias.

A complication in the comparison and simulations shown in Figure \ref{figber} is the spectral evolution of white dwarfs; the
atmospheric composition
of a white dwarf changes throughout its evolution due to several different mechanisms
\citep[see][for a detailed discussion]{blouin18a}. Given the white dwarf luminosity function, our sample is dominated by
stars cooler than 10,000 K, where we also observe an increase in the ratio of non-DA to DA stars due to convective mixing.
This means that a fraction of the DA white dwarfs turn into non-DAs below 10,000 K. We account for spectral evolution
in our simulations by assuming 1/3 of all DAs below 10,000 K turn into non-DAs, essentially removing them from our
simulated sample.

A comparison of the observed and the predicted mass and temperature distributions of our 100 pc white dwarf sample shows
that  the synthetic population is significantly different from the observed one in two major ways: 1- There is no significant
pile-up of $M\sim0.8~M_{\odot}$ white dwarfs, and 2- Massive white dwarfs with $M> 1~M_{\odot}$ do not disappear quickly
enough. 

\begin{figure*}
\center
\includegraphics[width=3.2in, bb=100 450 592 779]{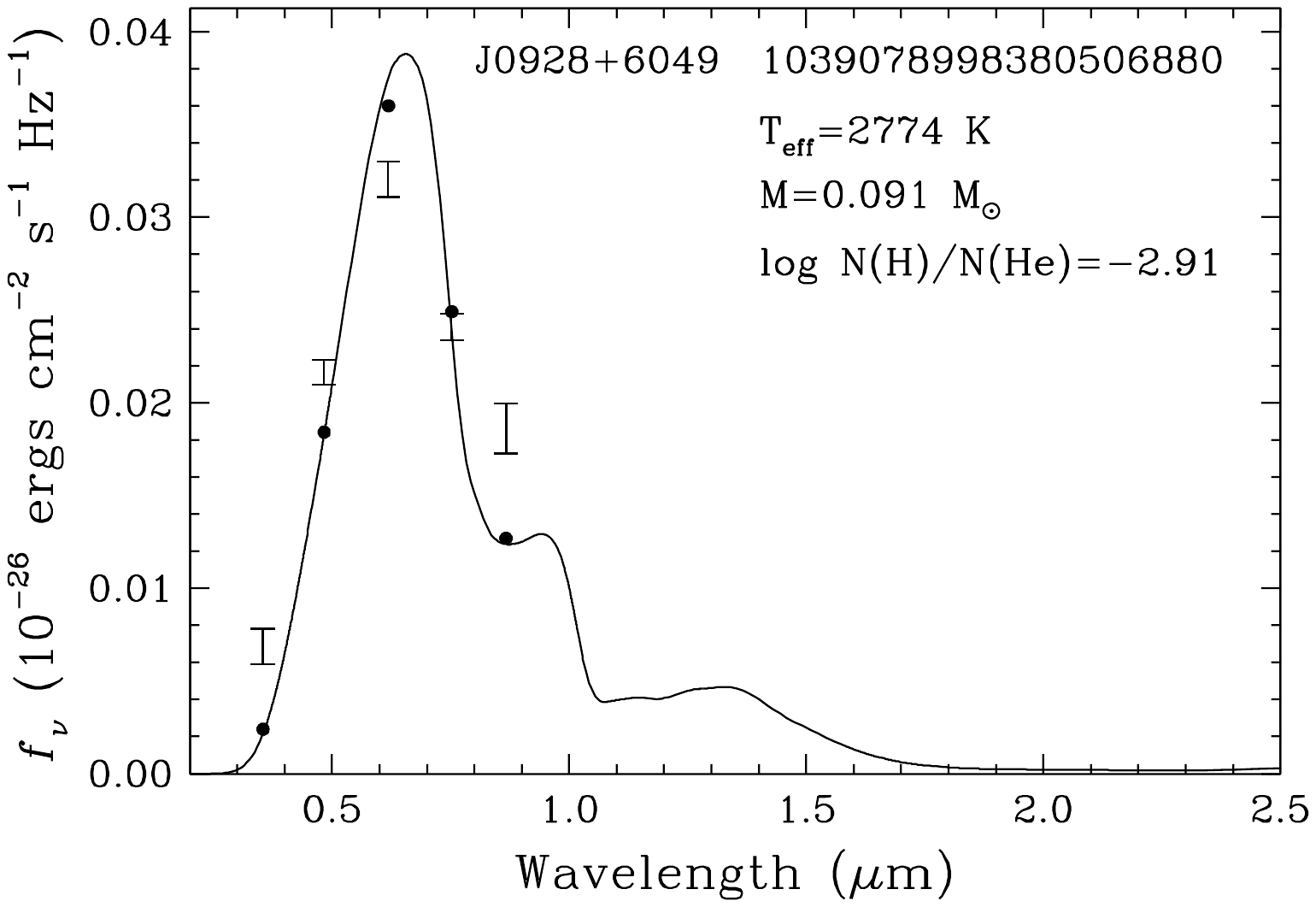}
\includegraphics[width=3.2in, bb=100 450 592 779]{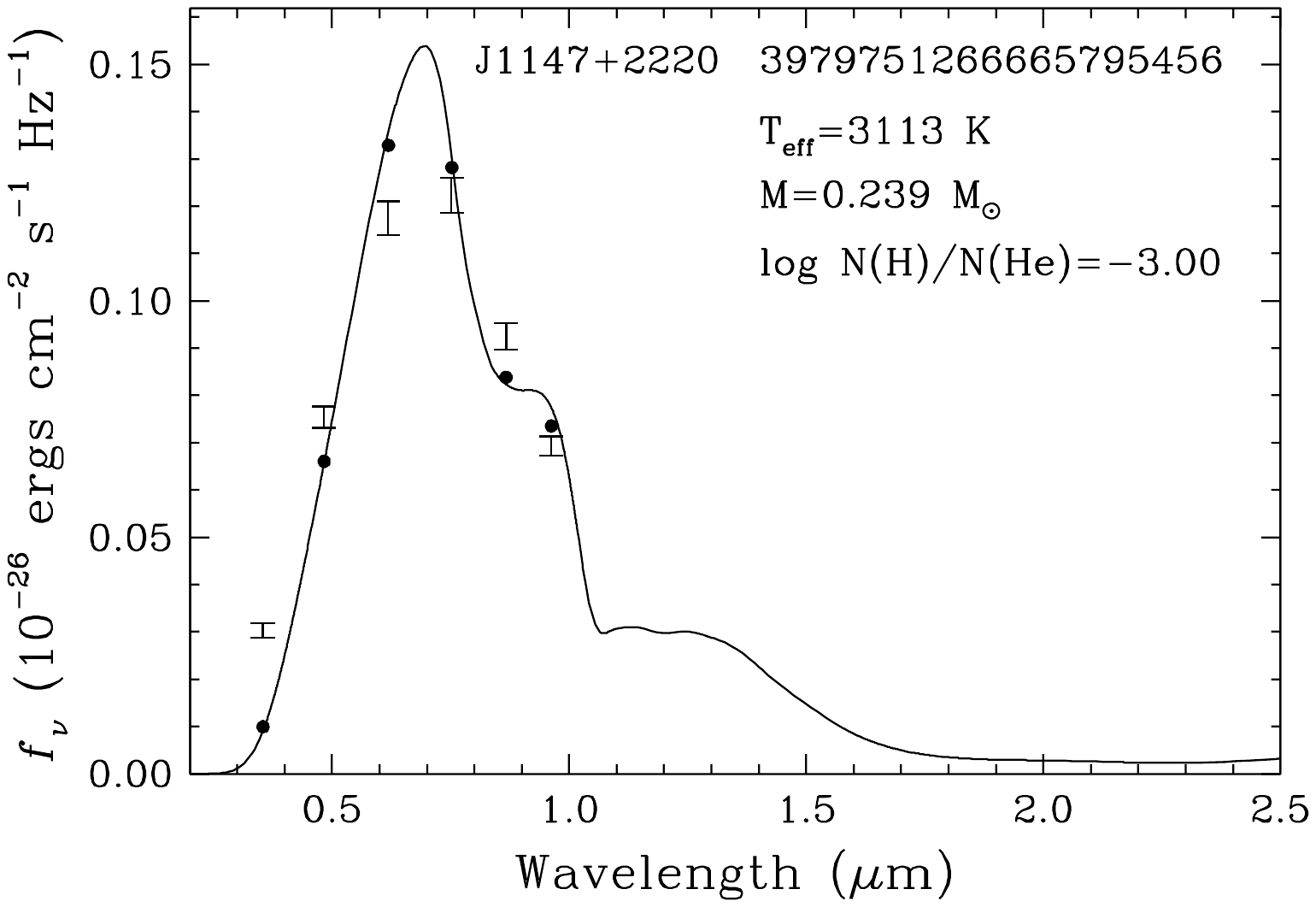}
\includegraphics[width=3.2in, bb=100 400 592 779]{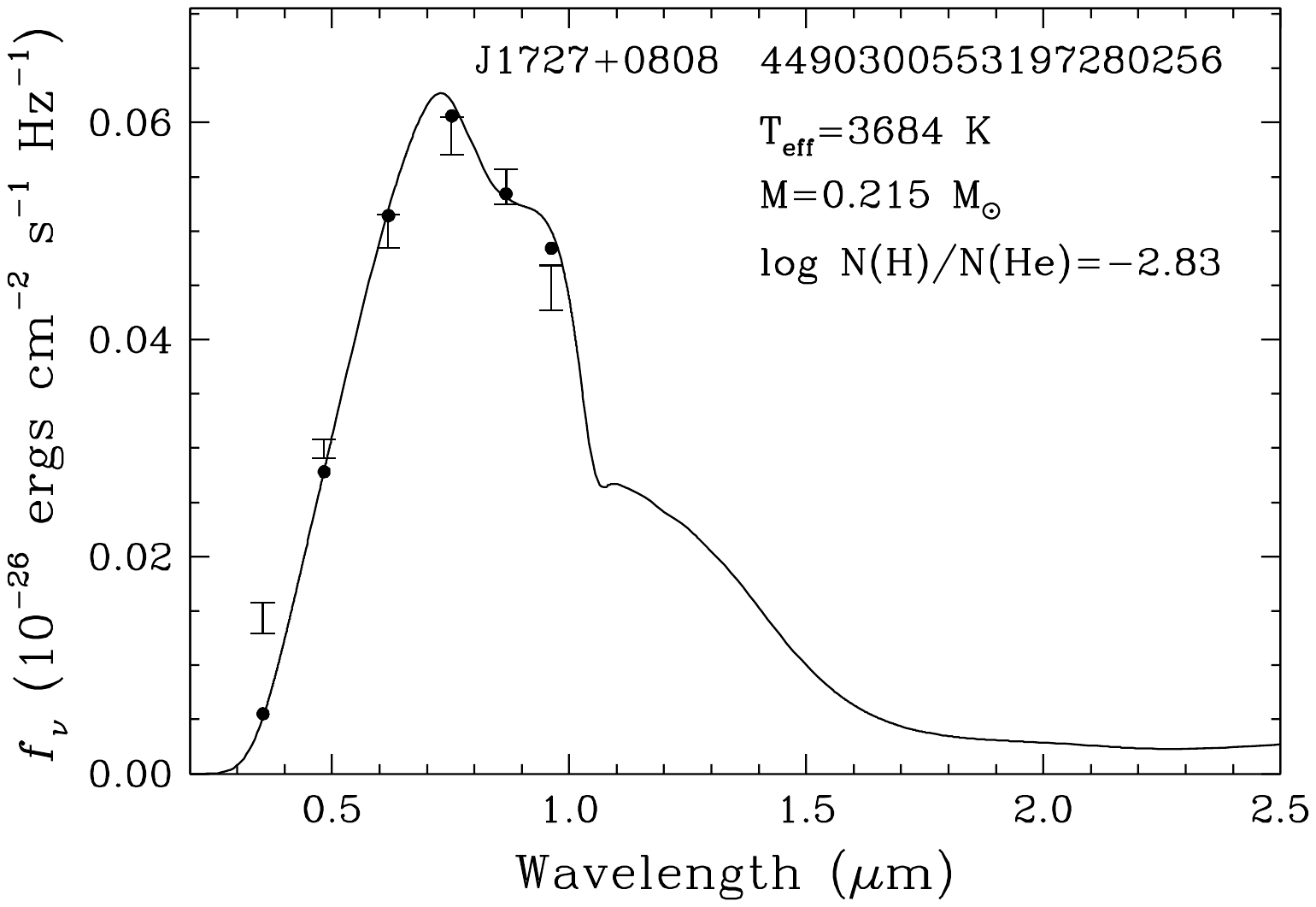}
\includegraphics[width=3.2in, bb=100 400 592 779]{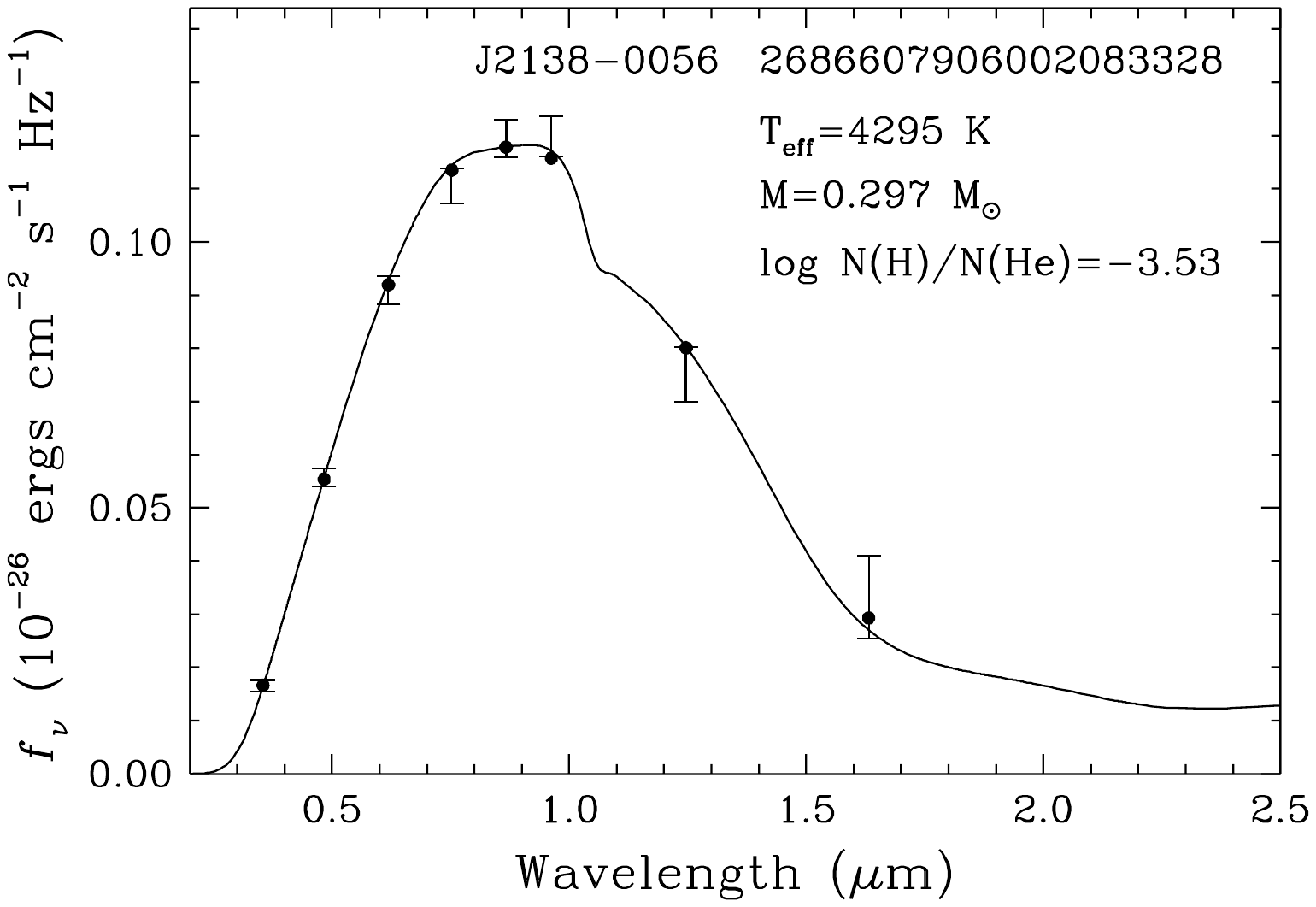} 
\caption{Fits to the spectral energy distributions of four new IR-faint white dwarfs identified in the 100 pc sample and the SDSS
footprint. Solid lines show the monochromatic fluxes for the best-fit model for each star, and dots show the synthetic photometry
of those models in each filter. Three of these objects, J0928+6049, J1147+2220, and J1727+0808, are spectroscopically
confirmed DC white dwarfs.
The remaining target, J2138$-$0056, currently lacks optical spectroscopy. All four targets show significant absorption in the
near-infrared bands, $z,y, J$ or $H$, and are best explained by mixed H/He atmospheres.}
\label{figultra}
\end{figure*}

Even though the evolutionary models that include crystallization predict a slow down in the
cooling at the right location in the mass versus effective temperature diagram, where a pile up of $\sim 0.8~M_{\odot}$ white
dwarfs is observed \citep{tremblay19,bergeron19}, the estimated delay in evolutionary times is clearly insufficient to
produce any significant pile-up in the synthetic populations. In addition, the synthetic populations predict a significant
population of massive white dwarfs at cooler temperatures that are simply absent in the observational sample. 

Studying the number density and velocity distribution of massive white dwarfs on the Q branch, \citet{cheng19} arrived at a
similar conclusion, and found that the cooling delay from crystallization alone is not sufficient and suggested
$^{22}$Ne settling as a source of extra cooling delay. In addition, the effects of phase separation and $^{16}$O
sedimentation are currently not included in the evolutionary models \citep{fontaine01}. Additional factors such as the uncertain
Coulomb plasma parameter at the liquid-solid phase transition, the C/O profile in the core, and the thickness of the surface hydrogen and helium layers
also play a role in the cooling rates of these objects \citep{tremblay19}. Hence, our observational data clearly point to
the onset of crystallization and its associated effects to explain the properties of the observed white dwarf mass distribution, but
the current evolutionary models are missing important physics related to this process. 

\subsection{IR-faint WDs}
\label{ultra}

Among the 555 DC white dwarfs identified in our sample, there are a handful of known IR-faint white dwarfs
that show optical and near-infrared flux deficits, presumably due to the CIA from molecular hydrogen. 
Even though we tend to refer to them as ultracool, implying they have temperatures below 4000 K, it is unclear if the majority
of them are actually this cool, as mixed H/He atmosphere model fits indicate temperatures as high as $\sim5500$ K for some of them \citep{kilic10,gianninas15}. 

For example, \citet{blouin18b} performed a detailed model atmosphere analysis of J0804+2239, a DZ white dwarf
with significant absorption in the near-infrared. J0804+2239 is within 100 pc, and included in our sample.
\citet{blouin18b} find best-fit parameters of $T_{\rm eff} = 4970 \pm 100$ K, $\log{g}=7.98 \pm 0.05$, and
$\log{\rm H/He} = -1.6 \pm 0.2$. Hence, the term ``IR-faint'' is more appropriate for these objects. 

\begin{figure*}
\center
\includegraphics[width=3.2in, bb=18 144 592 718]{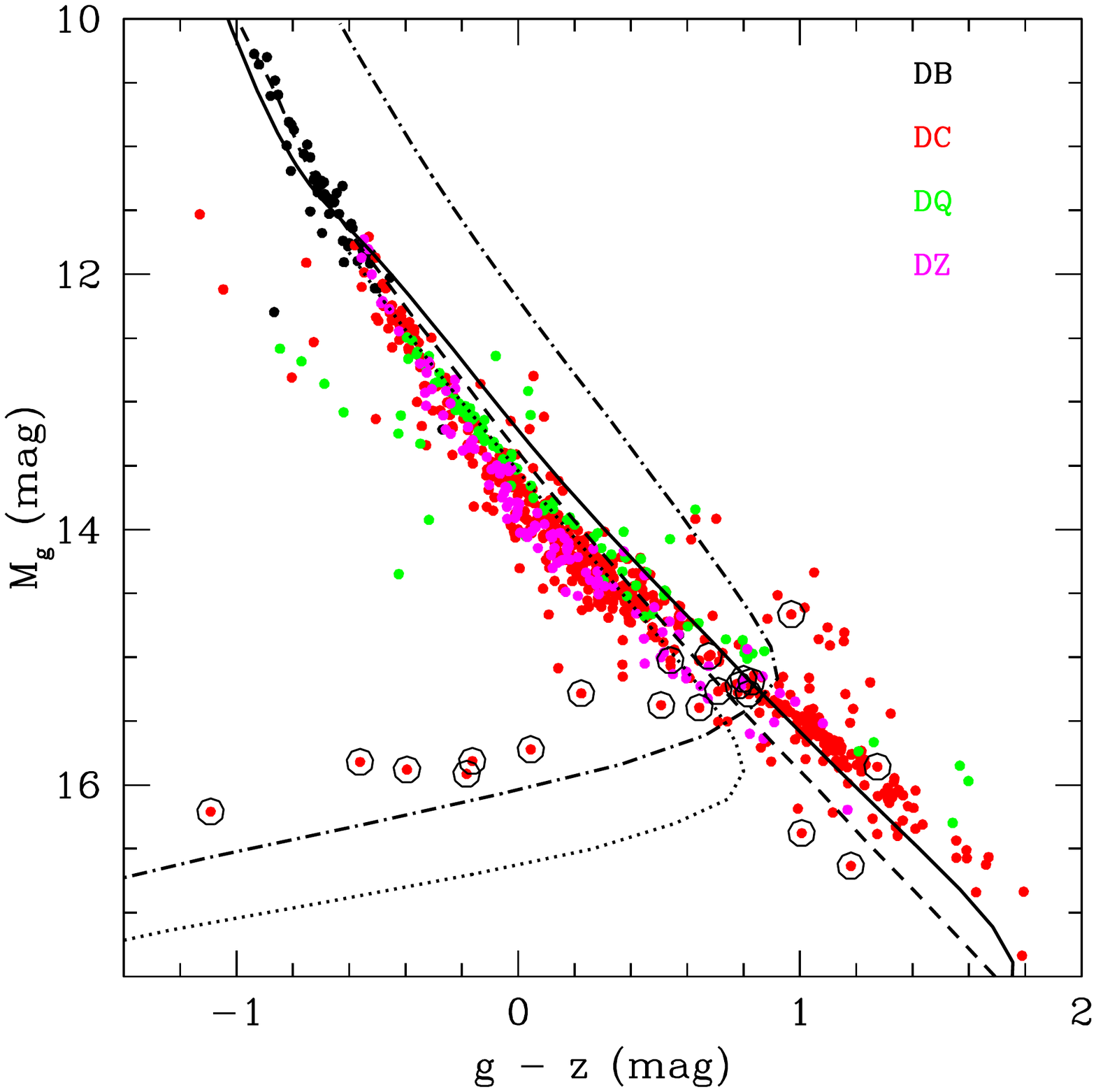} 
\includegraphics[width=3.2in, bb=18 144 592 718]{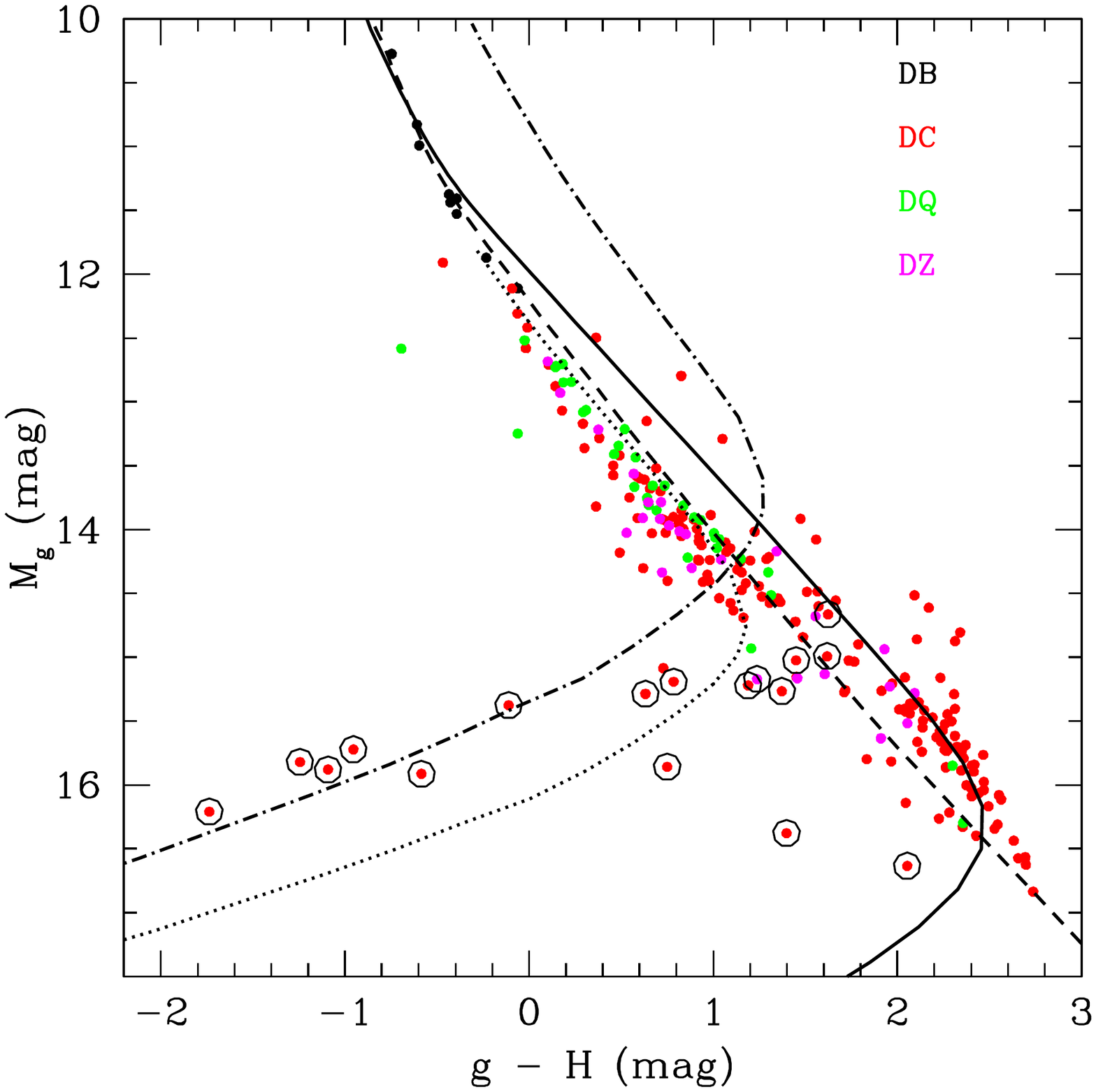} 
\caption{Near-infrared color-magnitude diagrams for the spectroscopically confirmed non-DA white dwarfs in the 100 pc
sample and the SDSS footprint. IR-faint white dwarfs are marked by open circles.
The solid, dashed, and dotted lines show the cooling sequences for 0.6 $M_{\odot}$ white
dwarfs with pure H, pure He, and mixed H/He models with $\log{\rm H/He} = -3$, respectively. The dashed-dotted line
shows the same mixed atmosphere models for a 0.2 $M_{\odot}$ white dwarf.}
\label{figgz}
\end{figure*}

We identify four additional white dwarfs with optical and near-infrared flux deficits in the 100 pc sample and the SDSS footprint.
Figure \ref{figultra} shows the spectral energy distributions for these four stars. Solid lines and dots show the monochromatic
fluxes and the synthetic photometry for the best-fit model, respectively. 

Three of these objects, J0928+6049, J1147+2220, and J1727+0808 are spectroscopically confirmed DC white dwarfs with
SEDs that peak in the $r-$ or $i-$band and with flux deficits visible in the Pan-STARRS $z$ and $y$ bands. The remaining
object, J2138$-$0056 does not have follow-up spectroscopy, but it clearly shows a significant flux deficit in the near-infrared
based on the UKIRT Infrared Deep Sky Survey \citep[UKIDSS,][]{ukidss} $JH$ photometry. If available, we also include the
near-infrared photometry in our model fits. These fits indicate temperatures as low as 2774 K and hydrogen abundances
of $\log{\rm H/He} \approx -3$. Note that this abundance corresponds to roughly the maximum intensity of the H$_2$-He CIA. 

The intensity of CIA is controlled by the photospheric density and the H$_2$ abundance. The former decreases and the latter
increases with an increase in the H/He ratio, which implies that there is a degeneracy between high and low hydrogen
abundances below and above $\log{\rm H/He} \approx -3$ \citep{blouin18b}. For example, instead of the solution shown in
Figure \ref{figultra}, the spectral energy distribution of J2138$-$0056 can also be fitted with a model that has
$T_{\rm eff}= 4630$ K, $M = 0.429~M_{\odot}$, and $\log{\rm H/He} = -2$. The model degeneracy is not an issue for the other
three stars; the infrared flux deficit is much stronger in these stars, including the Pan-STARRS $z-$ and $Y-$band photometry.
Their spectral energy distributions require the maximum possible CIA to reproduce the flux deficit, so there is a unique
solution for these objects. 

Our mixed H/He atmosphere models barely replicate the broad-band photometry for IR-faint white dwarfs,
and no model can currently match the observed spectra. In addition, the relatively
low temperature estimates force our model solutions to have relatively large radii to match the observed fluxes, which
in turn require $M\sim0.2~M_{\odot}$ and $\log{g}\sim7$ for these objects. In fact, many of the previously
known IR-faint white dwarfs with parallax measurements seem over-luminous and require relatively large radii
and small masses \citep[e.g.,][]{bergeron02,gianninas15}. 

Using the clean sample of extremely low-mass (ELM, $M\sim0.2~M_{\odot}$) white dwarfs from \citet{brown20}, we estimate
that the local space density of ``observed'' ELM white dwarfs is 100 to 200 kpc$^{-3}$ depending on the choice of Galactic
scale length parameters \citep[see][for a detailed discussion]{brown16}. This means that there should be 0.4-0.8 ELM white
dwarfs with $T_{\rm eff}=$ 9000-22,000 K within 100 pc of the Sun. Limiting the sample to the SDSS footprint lowers this estimate
by about $3 \times$. To estimate the number of cool ELM white dwarfs,
we use the evolutionary models for solar-metallicity progenitors from \citet{istrate16}. We predict $N\sim1$ ELM white
dwarf with $4000 < T_{\rm eff} < 6000$ K, but we find 20 IR-faint white dwarfs in our sample. Hence, we can safely conclude
that the majority of the IR-faint white dwarfs cannot be extremely low mass.

Figure \ref{figgz} shows near-infrared color-magnitude diagrams for the non-DA white dwarfs in our sample.
The left panel is solely based on the Pan-STARRS data, whereas the right panel uses $H-$band data from the UKIDSS
Large Area Survey, and \citet{kilic10} for the majority of the IR-faint white dwarfs. We matched our list of non-DA white dwarfs with the
Large Area Survey from the UKIDSS Data Release 11, and found 244 objects with near-infrared observations.
IR-faint white dwarfs, including the four newly identified systems presented here, are marked with open circles.
We also show the cooling sequences for pure hydrogen, pure helium, and mixed H/He atmosphere white dwarfs, for reference. 

Strikingly, 17 of the 20 IR-faint white dwarfs shown in this figure form a sequence in these color-magnitude diagrams.
This is the first time such a sequence is clearly observed in a color-magnitude diagram. The cooling sequence for 0.2
$M_{\odot}$ white dwarfs with $\log{\rm H/He}=-3$ comes closest to matching this sequence, but the match is not perfect
in either diagram. 

The IR-faint white dwarf sequence is connected to a region in the $M_g$ versus $g-z$ color-magnitude
diagram where the number of DC white dwarfs appears to be low. There are a significant number of DC white dwarfs
blueward of $g-z=0.5$ mag, and there are also a large number of them redward of $g-z=0.9$ mag, but the DC white
dwarf population is significantly depleted in between these two colors. This color range corresponds to
$\approx5000$-6000 K for 0.6 $M_{\odot}$ pure hydrogen white dwarfs. This is the same temperature range where \citet{bergeron97,bergeron01} found depleted numbers of helium-rich stars compared to their hydrogen-rich counterparts
\citep[see also][]{blouin19}. This is likely related to the spectral evolution of white dwarfs and 
transformation of helium-rich white dwarfs to hydrogen-rich composition, and vice-versa, through some yet unexplained phenomenon \citep{malo99,chen12}. 

We interpret the observed sequence of IR-faint white dwarfs in Figure \ref{figgz} as a cooling sequence of
white dwarfs with mixed H/He atmospheres. Given that the sequence is so tight indicates that these stars probably have
similar hydrogen abundances, around $\log{\rm H/He} =-3$. In addition, the analysis by \citet{bergeron19} indicates that
some of the warmer DC stars (above 6000 K) have such trace amounts of hydrogen in their atmospheres. Hence, it is possible
that such an IR-faint white dwarf sequence due to CIA is unavoidable. 

Clearly, a large fraction of the warmer DC white dwarfs turn into hydrogen-rich DC stars below 5000 K, otherwise we would see a lot more objects with CIA
absorption, but a few of those likely retain their helium-dominated atmospheres, and end up on the IR-faint white dwarf cooling
sequence instead.

In summary, we see a mix of hydrogen- and helium-atmosphere white dwarfs above 6000 K. These are the DA and DC/DQ/DZ
white dwarfs, respectively. Since hydrogen lines disappear below 5000 K, we have to rely on other indicators for composition,
like the presence of metals, the red wing of the Ly $\alpha$ line, and CIA. There are three possible outcomes for cool DC white
dwarfs below 5000 K: 1- The color-magnitude diagram suggests that most cool white dwarfs evolve as pure hydrogen atmospheres through some unknown process. 2- Some evolve as IR-faint white dwarfs on a cooling sequence with hydrogen abundances
that are typical of some of the hotter DC and DZ white dwarfs, whose origin is likely convectively mixed DAs. 3- Some also
evolve as DC white dwarfs with almost pure helium atmospheres, with hydrogen abundances low enough to produce no visible
CIA signatures. The progenitors of the latter likely have very little hydrogen, potentially DQ stars or cooled off DB stars. 

\section{Conclusions}

We present the results from a follow-up spectroscopy survey of the 100 pc white dwarf sample in the SDSS footprint.
Our follow-up is complete for 83\% of the white dwarfs hotter than 6000 K. We perform a detailed model atmosphere analysis
of each white dwarf, and use this sample
to constrain the DA white dwarf mass distribution, which displays an extremely narrow peak at 0.59 $M_{\odot}$ and broad
shoulder up to 0.9 $M_{\odot}$. 

Surprisingly, the number of massive white dwarfs remains roughly constant in the
0.7-0.9 $M_{\odot}$ range. The predictions from single and binary population synthesis studies fail to match this over-abundance
of massive white dwarfs. The location of the pile-up of $\sim0.8~M_{\odot}$ white dwarfs in the mass versus temperature
diagrams and the disappearance of $M>1~M_{\odot}$ white dwarfs from these samples are consistent with the expectations for
crystallization and related effects. 

The former is explained by
a delay in cooling due to the release of latent heat of crystallization, whereas the latter is explained by enhanced cooling
in the quantum regime where massive white dwarfs quickly disappear from the sample. However, we find that the current
evolutionary models are unable to match the significance of the pile-up of $\sim0.8~M_{\odot}$ and the sudden disappearance
of the more massive stars. 

We suggest that additional cooling delays from $^{16}$O sedimentation upon crystallization and $^{22}$Ne gravitational
settling likely play a significant role in the evolution of white dwarfs, and that convective coupling may
lead to even faster cooling than currently predicted.

We also discuss the prevalence of IR-faint white dwarfs in the 100 pc sample and the SDSS footprint. We find 20 IR-faint
white dwarfs that show significant flux deficits in the optical and/or near-infrared. We demonstrate that these white dwarfs form a sequence in
color-magnitude diagrams. In addition, the IR-faint white dwarf sequence seems to be connected to a region in the color-magnitude
diagrams where the number of helium-dominated atmosphere white dwarfs is low. This suggests that the transition of helium-rich white dwarfs into hydrogen-rich atmospheres, and
vice-versa, may
also explain the appearance of IR-faint white dwarfs, and that IR-faint white dwarfs likely have mixed H/He atmospheres.

\acknowledgements
This work is supported in part by the NSF under grant AST-1906379, the NSERC Canada, the Fund FRQ-NT (Qu\'ebec),
and the Smithsonian Institution. SB acknowledges support from the Laboratory Directed Research and Development program
of Los Alamos National Laboratory under project number 20190624PRD2.

We thank B. Kunk, E.\ Martin, and A.\ Milone for their assistance with observations obtained
at the MMT Observatory, a joint facility of the Smithsonian  Institution and the University of Arizona.

This work is based on observations obtained at the MDM Observatory, operated by Dartmouth College,
Columbia University, Ohio State University, Ohio University, and the University of Michigan.
We thank J.\ Rupert for obtaining the MDM data as part of the OSMOS queue.

This research is based in part on  observations obtained with the Apache Point Observatory 3.5-meter
telescope, which  is owned and operated by the Astrophysical Research Consortium.  

Based on observations obtained at the Gemini Observatory, which is operated by the Association of Universities for Research in Astronomy, Inc., under a cooperative agreement with the NSF on behalf of the Gemini partnership: the National Science Foundation (United States), National Research Council (Canada), CONICYT (Chile), Ministerio de Ciencia, Tecnolog\'{i}a e Innovaci\'{o}n Productiva (Argentina), Minist\'{e}rio da Ci\^{e}ncia, Tecnologia e Inova\c{c}\~{a}o (Brazil), and Korea Astronomy and Space Science Institute (Republic of Korea).

\facilities{MMT (Blue Channel spectrograph), FLWO:1.5m (FAST spectrograph),
Gemini (GMOS spectrograph), APO (DIS), MDM (OSMOS)}

%\bibliographystyle{aasjournal} 
%\bibliography{ref}

\appendix

The ADL query used to select our sample of 100 pc white dwarfs in the SDSS footprint is given below.

\begin{tt}
\noindent SELECT g.source\_id, g.ra, g.dec, parallax, parallax\_over\_error, pmra, pmra\_error, pmdec, pmdec\_error, phot\_g\_mean\_mag, phot\_g\_mean\_flux\_over\_error, phot\_bp\_mean\_mag, phot\_bp\_mean\_flux\_over\_error, phot\_rp\_mean\_mag, phot\_rp\_mean\_flux\_over\_error, phot\_g\_mean\_mag + 5.0*log10(parallax/100.0) AS M\_G, s.ra,s.dec, u\_mag, u\_mag\_error, g\_mag, g\_mag\_error, r\_mag, r\_mag\_error, i\_mag, i\_mag\_error, z\_mag, z\_mag\_error, clean\_flag \\
FROM   gaiadr2.gaia\_source AS g, gaiadr1.sdssdr9\_original\_valid AS s, gaiadr2.sdssdr9\_best\_neighbour AS xs \\
WHERE g.source\_id = xs.source\_id AND s.sdssdr9\_oid = xs.sdssdr9\_oid AND parallax > 10.0 \\
  AND parallax\_over\_error > 10.0 
  AND phot\_bp\_mean\_flux\_over\_error > 10 \\
  AND phot\_rp\_mean\_flux\_over\_error > 10 
  AND astrometric\_n\_good\_obs\_al > 5 \\
  AND (SQRT(astrometric\_chi2\_al/(astrometric\_n\_good\_obs\_al - 5.0)) < 1.2 OR \\
    SQRT(astrometric\_chi2\_al/(astrometric\_n\_good\_obs\_al - 5.0)) < 
       1.2*exp(-0.2*(phot\_g\_mean\_mag - 19.5))) \\
  AND phot\_bp\_rp\_excess\_factor BETWEEN 1.0 + (0.03*POWER(phot\_bp\_mean\_mag - phot\_rp\_mean\_mag, 2.0))  
                                   AND 1.3 + (0.06*POWER(phot\_bp\_mean\_mag - phot\_rp\_mean\_mag, 2.0)) \\
  AND phot\_g\_mean\_mag + 5.0*log10(parallax/100.0) > (3.333333*(phot\_bp\_mean\_mag - phot\_rp\_mean\_mag) + 8.333333)
\end{tt}

\end{document}